\documentclass{aa}
\usepackage{graphicx}
\usepackage{txfonts}
\usepackage{natbib}
\usepackage{hyperref}
\usepackage{caption}
\usepackage{amssymb}

\hypersetup{pagebackref=true, colorlinks=true, linkcolor=red, citecolor=blue, urlcolor=blue, bookmarks=true} 

\def\gsimeq{\hbox{\raise0.5ex\hbox{$>\lower1.06ex\hbox{$\kern-1.07em{\sim}$}$}}} 
\def\lsimeq{\hbox{\raise0.5ex\hbox{$<\lower1.06ex\hbox{$\kern-1.07em{\sim}$}$}}} 

\bibpunct{(}{)}{;}{a}{}{,} 

\begin{document}

\title{The infrared-radio correlation of star-forming galaxies is strongly M$_{\star}$-dependent but nearly redshift-invariant since z$\sim$4}

\author{I.~Delvecchio\inst{1,2}\thanks{email: ivan.delvecchio@inaf.it}
\and E.~Daddi\inst{2} 
\and M. T. Sargent\inst{3}
\and M. J. Jarvis\inst{4,5}
\and D. Elbaz\inst{2}
\and S. Jin\inst{6,7}
\and D. Liu\inst{8}
\and I. H. Whittam\inst{4,5}
\and H. Algera\inst{9}
\and R. Carraro\inst{10}
\and C. D'Eugenio\inst{2}
\and J. Delhaize\inst{11}
\and B. S. Kalita\inst{2}
\and S. Leslie\inst{9}
\and D. Cs. Moln{\'a}r\inst{12}
\and M. Novak\inst{8}
\and I. Prandoni\inst{13}
\and V. Smol{\v{c}}i{\'c}\inst{14}
\and Y. Ao\inst{15,16}
\and M. Aravena\inst{17}
\and F. Bournaud\inst{2}
\and J. D. Collier\inst{18,19}
\and S. M. Randriamampandry\inst{20,21}
\and Z. Randriamanakoto\inst{20}
\and G. Rodighiero\inst{22}
\and J. Schober\inst{23}
\and S. V. White\inst{24}
\and G. Zamorani\inst{25}
}

\institute{INAF - Osservatorio Astronomico di Brera, via Brera 28, I-20121, Milano, Italy \& via Bianchi 46, I-23807, Merate, Italy
\and  CEA, Irfu, DAp, AIM, Universit\`e Paris-Saclay, Universit\`e de Paris, CNRS, F-91191 Gif-sur-Yvette, France
\and Astronomy Centre, Department of Physics \& Astronomy, University of Sussex, Brighton, BN1 9QH, England
\and Astrophysics, Department of Physics, Keble Road, Oxford, OX1 3RH, UK
\and Department of Physics \& Astronomy, University of the Western Cape, Private Bag X17, Bellville, Cape Town, 7535, South Africa 
\and Instituto de Astrof\'isica de Canarias (IAC), E-38205 La Laguna, Tenerife, Spain
\and Universidad de La Laguna, Dpto. Astrof\'isica,  E-38206 La Laguna,Tenerife, Spain
\and MPI for Astronomy, Königstuhl 17, D-69117 Heidelberg, Germany
\and Leiden Observatory, Leiden University, P.O. Box 9513, 2300RA Leiden, the Netherlands
\and Instituto de F\'\i{}sica y Astronom\'\i{}a, Universidad de Valpara\'\i{}so, Gran Breta\~{n}a 1111, Playa Ancha, Valpara\'\i{}so, Chile
\and Department of Astronomy, University of Cape Town, Private Bag X3, Rondebosch 7701, South Africa
\and INAF - Osservatorio Astronomico di Cagliari, Via della Scienza 5, I-09047 Selargius (CA), Italy
\and INAF - Istituto di Radioastronomia, Via P. Gobetti 101, 40129 Bologna, Italy 
\and Department of Physics,  University of Zagreb,  Bijeni{\v{c}}ka cesta 32, 10002 Zagreb, Croatia
\and Purple Mountain Observatory \& Key Laboratory for Radio Astronomy, Chinese Academy of Sciences, Nanjing, China
\and School of Astronomy and Space Science, University of Science and Technology of China, Hefei, Anhui, China
\and N\'ucleo de Astronom\'ia, Facultad de Ingenier\'ia y Ciencias, Universidad Diego Portales, Av. Ej\'ercito 441, Santiago, Chile
\and The Inter-University Institute for Data Intensive Astronomy (IDIA), Department of Astronomy, University of Cape Town, Private Bag X3, Rondebosch, 7701, South Africa
\and School of Science, Western Sydney University, Locked Bag 1797, Penrith, NSW 2751, Australia
\and South African Astronomical Observatory, P.O. Box 9, Observatory 7935, Cape Town, South Africa 
\and A\&A, Department of Physics, Faculty of Sciences, University of Antananarivo, B.P. 906, Antananarivo 101, Madagascar 
\and University of Padova, Department of Physics and Astronomy, Vicolo Osservatorio 3, I-35122, Padova, Italy
\and Laboratoire d’Astrophysique, EPFL, CH-1290 Sauverny, Switzerland
\and Department of Physics and Electronics, Rhodes University, PO Box 94, Makhanda, 6140, South Africa
\and INAF - Osservatorio Astronomico di Bologna, via P. Gobetti 93/3, 40129 Bologna, Italy
}

   \date{Received }

  \abstract{
  Several works in the past decade have used the ratio between total (rest 8-1000$\mu$m) infrared and radio (rest 1.4~GHz) luminosity in star-forming galaxies (q$_{IR}$), often referred to as the ''infrared-radio correlation'' (IRRC), to calibrate radio emission as a star formation rate (SFR) indicator. Previous studies constrained the evolution of q$_{IR}$ with redshift, finding a mild but significant decline, that is yet to be understood. For the first time, we calibrate q$_{IR}$ as a function of \textit{both} stellar mass (M$_{\star}$) and redshift, starting from an M$_{\star}$-selected sample of $>$400,000 star-forming galaxies in the COSMOS field, identified via (NUV-r)/(r-J) colours, at redshifts 0.1$<$z$<$4.5. Within each (M$_{\star}$,z) bin, we stack the deepest available infrared/sub-mm and radio images. We fit the stacked IR spectral energy distributions with typical star-forming galaxy and IR-AGN templates, and carefully remove radio AGN candidates via a recursive approach. We find that the IRRC evolves primarily with M$_{\star}$, with more massive galaxies displaying systematically lower q$_{IR}$. A secondary, weaker dependence on redshift is also observed. The best-fit analytical expression is the following: q$_{IR}$(M$_{\star}$,z)=(2.646$\pm$0.024)$\times$(1+z)$^{(-0.023\pm0.008)}$-(0.148$\pm$0.013)$\times$($\log~M_{\star}$/M$_{\odot}$-10). Adding the UV dust-uncorrected contribution to the IR as a proxy for the total SFR, would further steepen the q$_{IR}$ dependence on M$_{\star}$. We interpret the apparent redshift decline reported in previous literature as due to low-M$_{\star}$ galaxies being progressively under-represented at high-redshift, as a consequence of binning only in redshift and using either infrared or radio-detected samples. The lower IR/radio ratios seen in more massive galaxies are well described by their higher observed SFR surface densities. Our findings highlight that using radio-synchrotron emission as a proxy for SFR requires novel M$_{\star}$-dependent recipes, that will enable us to convert detections from future ultra deep radio surveys into accurate SFR measurements down to low-SFR, low-M$_{\star}$ galaxies. 
  }

  \titlerunning{The IRRC evolves primarily with M$_{\star}$}
  \authorrunning{I. Delvecchio et al.}

  \maketitle


\section{Introduction} \label{intro}

For nearly fifty years astronomers have studied the observed correlation between total infrared (IR; rest-frame 8-1000~$\mu$m, i.e. L$_{IR}$) and radio (e.g. rest-frame 1.4 GHz, i.e. L$_{1.4~GHz}$) spectral luminosity arising from star formation, usually referred to as the ``infrared-radio correlation'' (IRRC, e.g. \citealt{Helou+85}; \citealt{deJong+85}). This tight (1$\sigma \sim$0.16~dex, e.g. \citealt{Molnar+20} submitted) correlation is often parametrized by the IR-to-radio luminosity ratio q$_{\rm IR}$, defined as (e.g. \citealt{Helou+85}; \citealt{Yun+01}):
\begin{equation}
 q_{IR} = \log \Bigg(\frac{L_{IR}~[W]}{3.75\times10^{12} [Hz]} \Bigg) - \log (L_{1.4~GHz}~[W~Hz^{-1}])
\label{eq:qir}
 \end{equation}
where 3.75$\times10^{12}$~Hz represents the central frequency over the far-infrared (FIR, rest-frame 42-122$~\mu$m) domain, usually scaled to IR in the recent literature. In the local Universe, the IRRC (or its parametrization $q_{IR}$) appears to hold over at least three orders of magnitude in both $L_{IR}$ and $L_{1.4~GHz}$ (e.g. \citealt{Helou+85}; \citealt{Condon92}; \citealt{Yun+01}). Broadly speaking, this is because the infrared emission comes from dust heated by fairly massive ($\gtrsim$5~M$_{\odot}$) OB stars, while radio emission arises from relativistic cosmic ray electrons (CRe) accelerated by shock waves produced when massive stars ($\gtrsim$8~M$_{\odot}$) explode as supernovae. Nevertheless, CRe are also subject to different cooling processes, as they propagate throughout the galaxy, which are mainly caused by inverse Compton, bremsstrahlung and ionization losses (e.g. \citealt{Murphy09}).

Surprisingly enough, despite all such processes at play, infrared and radio emission are observed to correlate, both in local star-forming late-type galaxies and even in merging galaxies (e.g. \citealt{Condon+93}, \citeyear{Condon+02}; \citealt{Murphy13}). This has been a strong motivator for using radio-continuum emission as a dust-unbiased star formation rate (SFR) tracer also in the faint radio sky (e.g. \citealt{Condon92}; \citealt{Bell03}; \citealt{Murphy+11}, \citeyear{Murphy+12}). Moreover, measuring the offset from the IRRC has been widely used to indirectly identify radio-excess active galactic nuclei (AGN; e.g. \citealt{Donley+05}; \citealt{DelMoro+13}; \citealt{Bonzini+15}; \citealt{Delvecchio+17}). 

These applications, however, deeply rely on a proper understanding of whether and how the IRRC evolves over cosmic time and across different types of galaxies. Despite its extensive application in extragalactic astronomy, the detailed physical origins of the IRRC and the nature of its cosmic evolution have long been debated (e.g. \citealt{Harwit+75}; \citealt{Rickard+84}; \citealt{deJong+85}; \citealt{Helou+85}; \citealt{Hummel+88}; \citealt{Condon92}; \citealt{Garrett02}; \citealt{Appleton+04}; \citealt{Murphy+08}; \citealt{Jarvis+10};  \citealt{Sargent+10}; \citealt{Ivison+10a},\citeyear{Ivison+10b}; \citealt{Bourne+11}; \citealt{Smith+14}; \citealt{Magnelli+15}; \citealt{CalistroRivera+17}; \citealt{Delhaize+17}; \citealt{Gurkan+18}; \citealt{Molnar+18}; \citealt{Algera+20b}).

For example, some studies of local star-forming galaxies (SFGs), ranging from dwarf (e.g. \citealt{Wu+08}) to ultra-luminous infrared galaxies (ULIRGs; L$_{IR} >$10$^{12}$~L$_{\odot}$; e.g. \citealt{Yun+01}) concluded that the IRRC remains linear across a wide range of L$_{IR}$. Conversely, other studies have argued that at low luminosities the IRRC may break down, consistent with a non-linear trend of the form L$_{IR}\propto$L$_{1.4~GHz}^{0.75-0.90}$ (e.g. \citealt{Bell03}; \citealt{Hodge+08}; \citealt{Davies+17}; \citealt{Gurkan+18}), which might be partly induced by dust heating from old stellar populations \citep{Bell03}.

Several models have attempted to explain this non linearity. On the one hand, \textit{calorimetric models} assume that galaxies are optically thick in the ultraviolet (UV), so that UV emission is fully re-emitted in the IR, likewise CRe radiate away their total energy through synchrotron emission before escaping the galaxy (e.g. \citealt{Voelk89}). These conditions might hold in the most massive (stellar mass M$_{\star}\gtrsim$10$^{10}$~M$_{\odot}$) SFGs, because of their increasing compactness (i.e. the size-mass relation R$_{e}\propto M_{\star}^{0.22}$, \citealt{vanderWel+14}), that might enhance their ability to retain the gas ejected by stars. However, this is likely to break down towards lower M$_{\star}$ galaxies, due to smaller sizes and lower obscuration (e.g. \citealt{Bourne+12}). On the other hand, \textit{non-calorimetric} models or the optically thin scenario (\citealt{Helou+93}; \citealt{Niklas+97}; \citealt{Bell03}; \citealt{Lacki+10a}), argue that several physical mechanisms cancel each other out, creating a sort of \textit{conspiracy} that keeps the IRRC unexpectedly tight and linear. Indeed, both IR and radio luminosities should underestimate the total SFR in low M$_{\star}$ and low SFR surface density galaxies \citep{Bell03}, inducing a departure of the IRRC from linearity. This is however not observed. Radio synchrotron models postulate that such small galaxies are not able to prevent CRe from escaping, causing a global deficit of radio emission at fixed SFR. Similarly, the IR domain becomes less sensitive to SFR in low-M$_{\star}$ galaxies (e.g. \citealt{Madau+14}), generating an IR deficit of a similar amount that might counter-balance the radio and keep the IRRC linear. Understanding the discrepancy between model predictions and observations is crucial, since the linearity (or not) of the IRRC has direct implications for using radio emission as a SFR tracer.

From an observational perspective, it is widely recognized that a tight relation links SFR and M$_{\star}$ in nearly all SFGs, namely the ``main sequence'' of star formation (MS, scatter$\sim$0.2--0.3~dex). This relation holds from z$\sim$5 down to the local Universe (e.g. \citealt{Brinchmann+04}; \citealt{Noeske+07}; \citealt{Elbaz+11}; \citealt{Whitaker+12}; \citealt{Speagle+14}; \citealt{Schreiber+15}; \citealt{Lee+15}), showing a flattening at high M$_{\star}$ and an evolving normalization with redshift. Because the SFR is directly linked to L$_{IR}$, especially in massive SFGs \citep{Kennicutt98}, the existence of the MS gives an additional argument that studying q$_{IR}$ as a function of M$_{\star}$ could be of the utmost importance for our understanding of what drives the IRRC in galaxies.

Recent studies have corroborated the idea that the IRRC slightly, but significantly, declines with redshift (\citealt{Ivison+10b}; \citealt{Magnelli+15}; \citealt{CalistroRivera+17}; \citealt{Delhaize+17}), in the form of q$_{\rm IR} \propto$(1+z)$^{[-0.2:-0.1]}$, although the physical explanation for such evolution is still uncertain. Somewhat different conclusions were reached by other works (e.g. \citealt{Garrett02}; \citealt{Appleton+04}; \citealt{Ibar+08}; \citealt{Jarvis+10}; \citealt{Sargent+10}; \citealt{Bourne+11}) which ascribe this apparent evolution to selection effects. For instance, these include comparing flux-limited samples, each with a different selection function. 

In this regard, we note that any selection method is sensitive to brighter, i.e. more massive galaxies towards higher redshifts. By binning in redshift, only a restricted range in galaxy M$_{\star}$ will be detectable at each redshift for any flux limited sample, thus inducing a bias as a function of z. Therefore, it is timely to examine the evolution of the IRRC as a function of M$_{\star}$ and redshift \textit{simultaneously}. We emphasize that our approach is fully empirical. However, a possible M$_{\star}$ dependence of the IRRC is expected from some synchrotron emission models (e.g. \citealt{Lacki+10b}; \citealt{Schober+17}), and might reflect some combination of the underlying physics originating the IRRC (see Sect.~\ref{discussion}).
 
The main goal of the present paper is to calibrate the IRRC for the first time as a function of \textit{both} M$_{\star}$ and redshift over a wide range. To this end, we start from an M$_{\star}$-selected sample of $>$400,000 galaxies at 0.1$<$z$<$4.5 collected from deep UltraVISTA images in the Cosmic Evolution Survey \citep{Scoville+07} (centered at RA=+150.11916667; Dec=+2.20583333 (J2000)). Then we leverage the new de-blended far-IR/sub-mm data \citep{Jin+18} recently compiled in COSMOS, which allow us to circumvent blending issues due to poor angular resolution and measure L$_{IR}$ for typical MS galaxies out to z$\sim$4. In addition, we exploit the deepest radio-continuum data taken from the VLA-COSMOS 3 GHz Large Project \citep{Smolcic+17}. Individual detections will be combined with stacked flux densities of non-detections, at both IR and radio frequencies to assess the average q$_{IR}$ as a function of M$_{\star}$ and redshift. 

The layout of this paper is as follows. A description of the sample selection and multi-wavelength ancillary data is given in Sect.~\ref{sample}. We describe the stacking analysis in Sect.~\ref{method}, including measurements of L$_{IR}$ (Sect.~\ref{ir_stacking}) and L$_{1.4~GHz}$ (Sect.~\ref{radio_stacking}). The average q$_{IR}$ as a function of M$_{\star}$ and redshift is presented in Sect.~\ref{results}, where we perform a careful subtraction of radio AGN at different M$_{\star}$ via a recursive approach. Our main results are discussed and interpreted in Sect.~\ref{discussion} in the framework of previous observational studies and theoretical models. The main conclusions are summarized in Sect.~\ref{summary}. In addition, we test our total 3~GHz flux densities in Appendix \ref{Appendix_radio}. A detailed comparison between radio stacking results is presented in Appendix \ref{Appendix_ancillary}. In Appendix~\ref{Appendix_AGN} we discuss how the final IRRC is sensitive to our AGN subtraction method. Finally, in Appendix.~\ref{Appendix_comparison} we quantify how different assumptions from the literature would change our main results.

Throughout this paper, magnitudes are given in the AB system \citep{Oke74}. We assume a \citet{Chabrier03} initial mass function (IMF) and a $\Lambda$CDM  cosmology with $\Omega_{\rm m}$ = 0.30, $\Omega_{\rm \Lambda}$ = 0.70, and H$\rm _0$ = 70 km s$^{-1}$ Mpc$^{-1}$ \citep{Spergel+03}.

\section{Multi-wavelength data and sample selection} \label{sample}

In this Section we describe the creation of a $K_s$ prior catalogue that we used to select our parent sample in the COSMOS field. 

\begin{figure}
\centering
     \includegraphics[width=\linewidth]{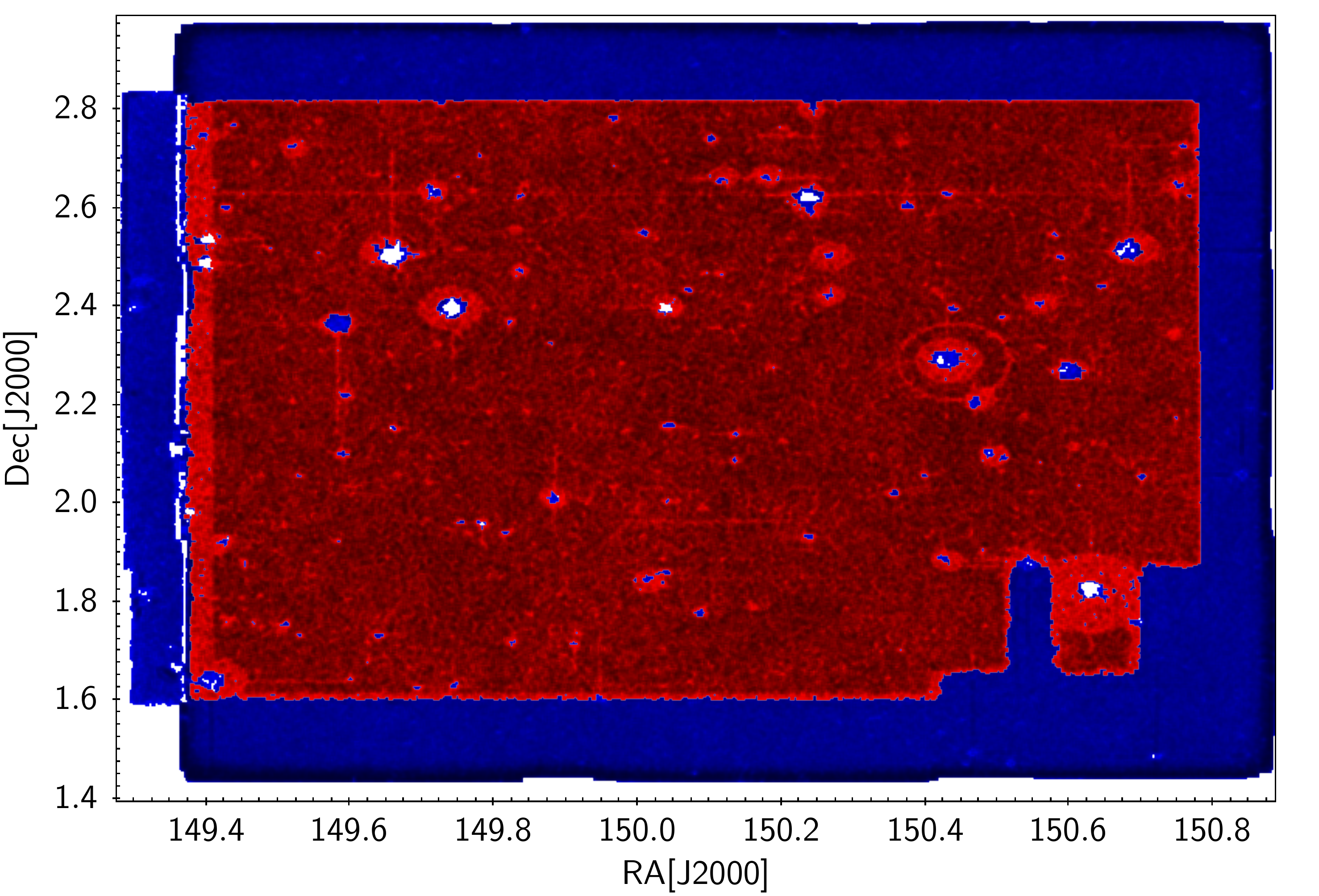}
 \caption{\small Distribution of the full COSMOS2015 \citep{Laigle+16} source list over the COSMOS area (blue dots). The subset of 413,678 $NUVrJ$--based star-forming galaxies analyzed in this work (red dots) includes sources from \citet{Laigle+16} and \citet{Muzzin+13} within the UltraVISTA area, with the exception of masked regions due to saturated or contaminated photometry. See Sect.~\ref{sample} for details.
 }
   \label{fig:cosmosarea}
\end{figure}

The COSMOS field (2 deg$^2$) boasts an exquisite photometric data set, spanning from the X-rays to the radio domain\footnote{An exhaustive overview of the COSMOS field is available at: \url{http://cosmos.astro.caltech.edu/} }. The most recent collection of multiwavelength photometry comes from the COSMOS2015 catalogue \citep{Laigle+16}, that contains 1,182,108 sources extracted from a stacked $YJHK_s$ image(blue dots in Fig.~\ref{fig:cosmosarea}). In particular, this catalogue joins optical photometry from Subaru Hyper-Suprime Cam (2 deg$^2$; \citealt{Capak+07}) and from the Canada-France-Hawaii Telescope Legacy Survey (CFHT-LS, central 1 deg$^2$; \citealt{McCracken+01}); near-infrared (NIR) bands $Y$, $J$, $H$, and $K_s$ from UltraVISTA DR2 (down to $K_s$<24.5 in the central 1.5 deg$^2$, of which 0.6~deg$^2$ are covered by ultra-deep stripes with limiting $K_s$<25.2; \citealt{McCracken+12}) and from CFHT $H$ and $K_s$ observations obtained with the WIRCam ($K_s$<23.9 outside the UltraVISTA area; \citealt{McCracken+01}). Over the full 2 deg$^2$ area, mid-infrared (MIR) photometry was obtained from the \textit{Spitzer} Large Area Survey with Hyper-Suprime-Cam (SPLASH; \citealt{Steinhardt+14}; P.~Capak et al. in prep.) using 3.6--8$\mu$m data from the Infrared Array Camera (IRAC). We refer the reader to \citet{Laigle+16} for more details.
 
In order to obtain a homogeneous galaxy selection function, we limited our study to the inner UltraVISTA DR2 area, also excluding stars and masked regions in the COSMOS2015 catalogue with less accurate photometry, which reduces the initial sample to 45\% of its size (524,061 sources). Following \citet{Jin+18}, we partly fill up these blank regions by adding 22,838 unmasked $K_s$--selected sources from the UltraVISTA catalogue of \citet{Muzzin+13} (3$\sigma$ limit of $K_s$<24.35 with 2'' aperture). This ensures a more complete coverage within the UltraVISTA area, with fluctuations in prior source density of only 2.5\%. This builds our $K_s$ prior sample of 546,899 galaxies. Given the similar selection, we confirm that excluding the slightly shallower $\sim$4\% subsample from \citet{Muzzin+13} leaves our results unchanged, and thus we keep them throughout this work.

Photometric redshifts and M$_{\star}$ estimates were retrieved from the corresponding catalogues, by fitting the optical-MIR photometry using the stellar population synthesis models of \citet{Bruzual+03}. Both redshift and M$_{\star}$ values represent the median of the corresponding likelihood distribution. \citet{Laigle+16} report an average photometric redshift accuracy of $\left \langle |\Delta z/(1 + z)| \right \rangle =$ 0.007 at z$<$3, and 0.021 at 3$<$z$<$6. A similar accuracy of 0.013 is reached in the catalogue of \citet{Muzzin+13} at z$<$4. We further inspected a subset of 5,400 sources showing a skewed redshift probability distribution function (with $\gtrsim$5\% chance to be offset from the median by $>$0.5$\times$[1+z$_p$]). However, we verified that removing such potential redshift interlopers does not have any impact on our results. As in \citet{Jin+18}, publicly available spectroscopic redshifts were collected from the new COSMOS master spectroscopic catalog (courtesy of M. Salvato, within the COSMOS team), and were prioritized over photometric measurements if deemed reliable (z$_{s}$ quality flag $>$3 $\wedge~~ |$z$_s$-z$_p| <$ 0.1$\times$(1+z$_p$) ).

Infrared/sub-mm flux densities were de-blended and re-extracted via the prior-based fitting algorithm presented in \citet{Jin+18}, that we briefly describe in Sect.~\ref{infrared}.

 \subsection{Selecting star-forming galaxies via (NUV-r)/(r-J) colours} \label{nuvrj}

We aim to study the infrared-radio correlation within an M$_{\star}$-selected sample of star-forming galaxies. To this end, we make use of the rest-frame, dust-corrected $(NUV-r)$ and $(r-J)$ colours available in the parent catalogues (hereafter $NUVrJ$). As opposed to the widely used UVJ criterion, the (NUV--r) colour is more sensitive to recent star formation (10$^6$--10$^8$ yr scales, \citealt{Salim+05}; \citealt{Arnouts+07}; \citealt{Davidzon+17}). Therefore, this criterion enables us to better distinguish between weakly star-forming galaxies (with specific-SFR, sSFR=SFR/M$_{\star} \sim$10$^{-10}$yr$^{-1}$) and fully passive systems (sSFR$<$10$^{-11}$yr$^{-1}$). 
 
We further selected galaxies with redshift 0.1$<$z$<$4.5 and 10$^{8}<$M$_{\star}$/M$_{\odot}<$10$^{12}$. This leaves us with a final sample of 413,678 star-forming galaxies (red dots in Fig.~\ref{fig:cosmosarea}), out of which 22,238 (5.4\%) are spectroscopically confirmed. The fraction of catastrophic failures ($|$z$_s$-z$_p| >$ 0.15$\times$(1+z$_s$)) is only 3.4\%. 

\begin{figure}
\centering
     \includegraphics[width=\linewidth]{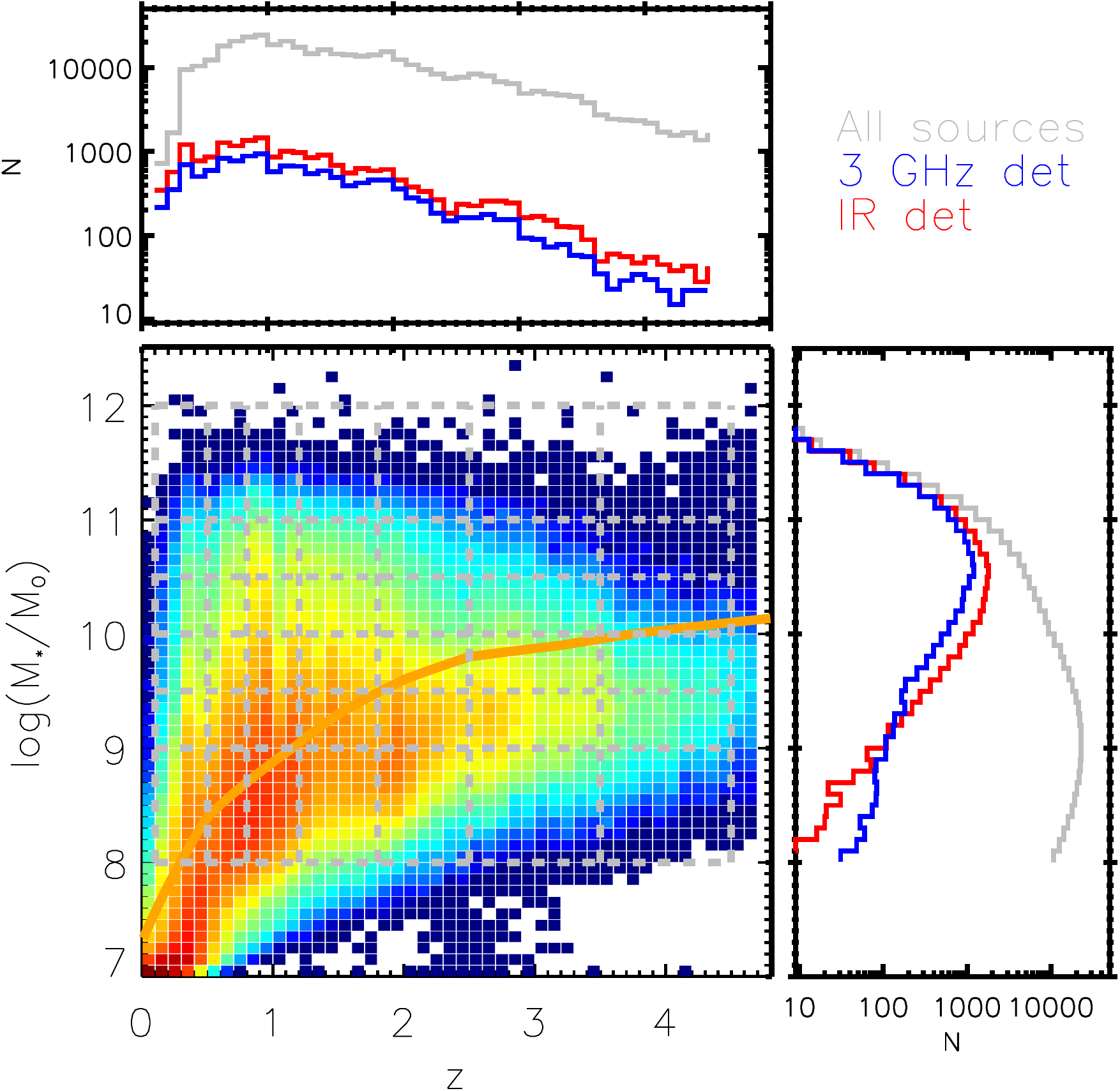}
 \caption{\small Distribution of $NUVrJ$ star-forming galaxies as a function of M$_{\star}$ and redshift. The colour scale in the central panel indicates the underlying sample size, increasing from blue to red. The grey dashed grid encloses the 42 M$_{\star}$--z bins into which we split our sample (413,678 objects). Galaxies within that grid are projected on the upper-left and bottom right histograms with redshift and M$_{\star}$, respectively (grey lines). The blue and red histograms represent to subsample with S/N$>$3 at 3~GHz and across all IR bands, respectively (see Sects.~\ref{infrared} and \ref{radio}). The orange solid line marks the 90\% M$_{\star}$ completeness limit of \citet{Laigle+16} for comparison.
 }
   \label{fig:mass_z}
\end{figure}

Such a sizable sample enables us to bin galaxies as a function of both M$_{\star}$ and redshift, while maintaining good statistical power. Fig.~\ref{fig:mass_z} shows our sample in the M$_{\star}$--redshift diagram, highlighting the chosen grid. We note that the M$_{\star}$ uncertainties taken from the parent catalogues incorporate the covariant errors on stellar population ages and dust reddening. These average M$_{\star}$ uncertainties are 0.2~dex at 10$^{8}<$M$_{\star}$/M$_{\odot}<$10$^{9}$ and 0.1~dex above, which is far smaller than the corresponding M$_{\star}$ bin width, thus not impacting our results. The 90\% M$_{\star}$ completeness limit (orange solid line, \citealt{Laigle+16}) indicates that our sample of SFGs is mostly complete down to 10$^{10}$~M$_{\odot}$ out to z$\sim$4. Although we acknowledge the increasing incompleteness towards less massive galaxies in the early Universe, we believe that including them brings a valuable addition for constraining the infrared and radio properties of galaxies down to a poorly explored regime of M$_{\star}$. This will become particularly relevant for the next generation of telescopes, such as JWST and SKA, which will routinely observe such faint sources. In addition, as we will discuss in Sect.~\ref{sed_fitting}, a very good agreement is observed between our stacked L$_{IR}$ and those extrapolated from the MS relation \citep{Schreiber+15} also at M$_{\star}<$10$^{9.5}$M$_{\odot}$, suggesting that even in this incomplete, low-M$_{\star}$ regime our galaxies are still representative of an M$_{\star}$-selected sample. We emphasize that the overall conclusions of this work are unchanged if we limit ourselves to z$<$3 and M$_{\star}>$10$^{9.5}$M$_{\odot}$, in which our sample is highly complete. Moreover, in light of our main result, i.e. q$_{IR}$ \textit{decreases} with M$_{\star}$, we anticipate that including galaxies within an incomplete M$_{\star}$ regime would at most amplify the final M$_{\star}$ dependence, thereby reinforcing our findings.

 \subsection{Infrared and sub-mm de-blended data} \label{infrared}

We complemented the existing COSMOS optical-to-IRAC photometry with cutting-edge de-blended photometry from \citet{Jin+18}, based on the de-blending algorithm developed in \citet{Liu+18} for the GOODS-North field. 

The dataset includes \textit{Spitzer}-MIPS~24$~\mu$m data (PI: D. Sanders; \citealt{LeFloch+09}); \textit{Herschel} imaging from the PACS (100-160$~\mu$m, \citealt{Poglitsch+10}) and the SPIRE (250, 350, and 500$~\mu$m, \citealt{Griffin+10}) instruments, as part of the PEP (\citealt{Lutz+11}) and HerMES (\citealt{Oliver+12}) programs, respectively. In addition, JCMT/SCUBA2 (850~$\mu$m) images are taken from the S2CLS program (\citealt{Cowie+17}; \citealt{Geach+17}), the ASTE/AzTEC (1.1~mm) data are nested maps from \citet{Aretxaga+11} over a sub-area of 0.72 deg$^2$. Finally, \citet{Jin+18} also included MAMBO data \citep{Bertoldi+07} at 1.2~mm over an area of 0.11~deg$^2$.

Briefly, \citet{Jin+18} used $K_s$-selected sources from the UltraVISTA survey (Sect.~\ref{sample}) as priors to perform PSF fitting of MIPS~24$~\mu$m, VLA-3~GHz \citep{Smolcic+17} and VLA-1.4~GHz \citep{Schinnerer+10} images down to the 3$\sigma$ level in each band. Within our final sample, this procedure identifies 67,114 MIPS~24$~\mu$m+VLA priors. Nevertheless, adopting a similar approach for extracting FIR/sub-mm flux densities of \textit{all} M$_{\star}$-selected galaxies, i.e. using the full list of $K_s$ priors, would identify up to 50 sources per beam at the resolution of the FIR/sub-mm wavelengths, causing heavy confusion. Therefore, following \citet{Jin+18}, only an M$_{\star}$-complete subset of $K_s$ priors was added, which ultimately prioritizes IR brighter sources. This leads to a total of 136,584 $K_s$+MIPS~24$~\mu$m+VLA priors, that were used to de-blend and extract the \textit{Herschel}, SCUBA2 and AzTEC flux densities (see Table~\ref{tab:sample}). 
Within our final sample of 413,678 star-forming galaxies, 20,777 (5\%) have a combined S/N$>$3 over all FIR/sub-mm bands (10,285 at S/N$>$5). These are displayed as red histograms in Fig.~\ref{fig:mass_z}. The rest of the $K_s$ sources are assumed to have negligible FIR/sub-mm flux densities, consistent with the background level in those bands. This is confirmed by the Gaussian-like behaviour of the noise (centered at zero) in the residual maps, after subtracting all S/N$>$3 sources in each band \citep{Jin+18}.

Throughout the rest of this paper, we interpret individual S/N$>$3 sources as detections, while S/N$<$3 sources will be stacked, as described in Sect.~\ref{ir_stacking}.

\begin{table}
\centering
   \caption{Main numbers of priors and detections that characterize our final sample of 413,678 star-forming galaxies. Note that subsets do not add up to make the final sample. ($^{**}$): numbers reported also in Fig.~\ref{fig:bins}. }
\begin{tabular}{l c }
\hline
\hline
Definition      &       Number of sources    \\
 \hline
Final sample (this work) & 413,678 $^{**}$ \\
- MIPS 24~$\mu$m + VLA priors         & 67,114 \\
- MIPS 24~$\mu$m + VLA + $K_s$ priors & 136,584 \\
- S/N$_{IR}>$3  & 20,777 $^{**}$ \\
- S/N$_{3GHz}>$3  & 13,808 $^{**}$  \\
\hline
\end{tabular}
\label{tab:sample}
\end{table}

\begin{figure}
\centering
     \includegraphics[width=\linewidth]{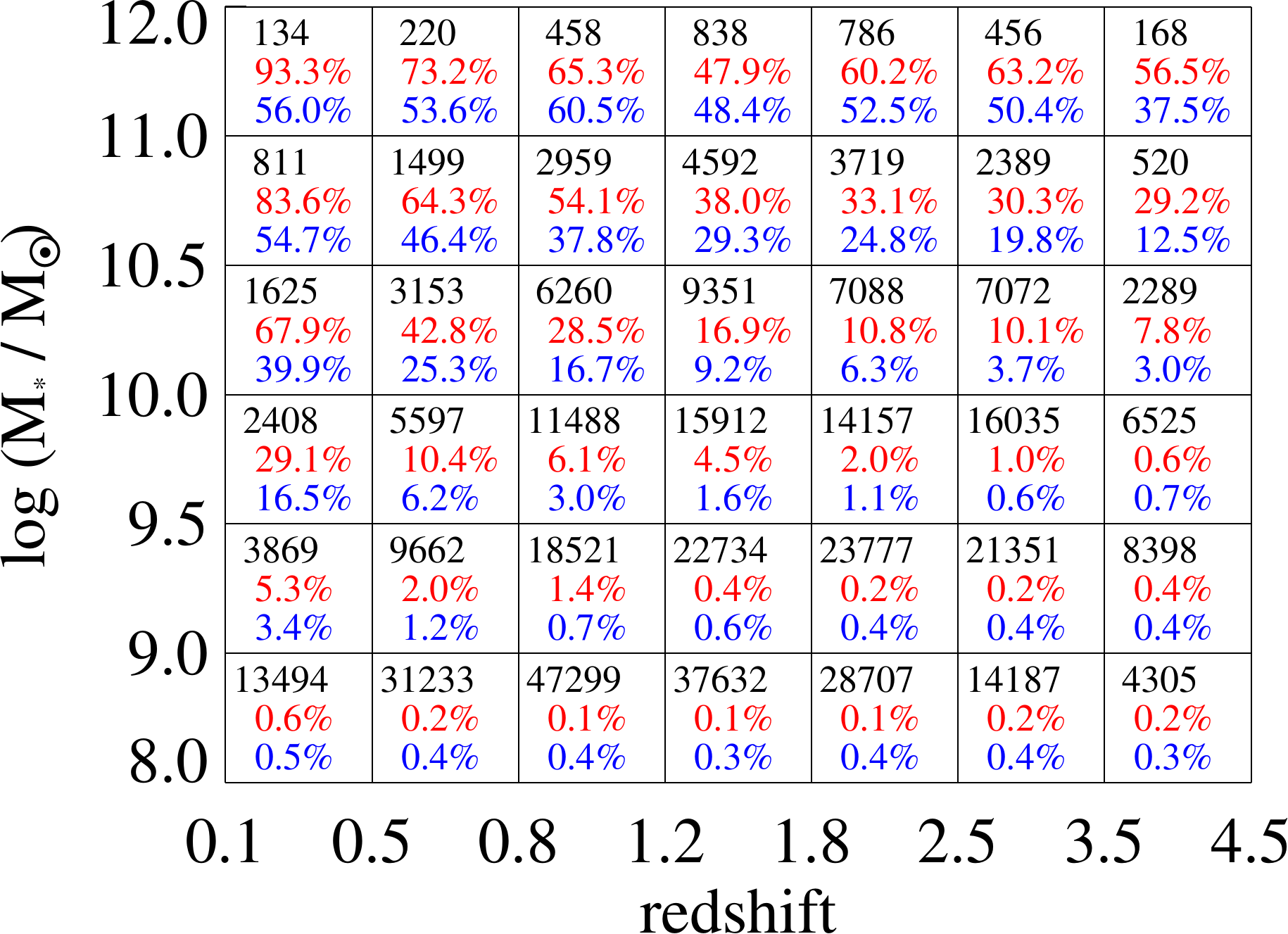}
 \caption{\small Number of $NUVrJ$--based star-forming galaxies analyzed in this work, as a function of M$_{\star}$ and redshift (black). For convenience, in each bin we report the fraction of sources with combined S/N$_{IR}>$3 over all FIR/sub-mm bands (red) and with S/N$_{3~GHz}>$3 (blue).  
 }
   \label{fig:bins}
\end{figure}

\subsection{Radio data in the COSMOS field} \label{radio}

For our analysis we exploited data from the VLA-COSMOS 3 GHz Large Project \citep{Smolcic+17}, that is one of the largest and deepest radio survey ever conducted over a medium sky area like COSMOS. With 384h of observations, the final mosaic reaches a median rms=2.3~$\mu$Jy~beam$^{-1}$ over 2.6 deg$^2$ at an angular resolution of 0.75'', the highest among radio surveys in COSMOS. A total of 10,830 sources were blindly extracted down to S/N$>$5. 

In addition, the COSMOS field boasts the current deepest radio-continuum data at 1.28~GHz from the MeerKAT International GHz Tiered Extragalactic Exploration (MIGHTEE, \citealt{Jarvis+16}) survey. With only 17h on-source over the central 1 deg$^2$ of COSMOS, the early science release of MIGHTEE reaches a thermal noise of 2.2~$\mu$Jy~beam$^{-1}$, though the effective noise is limited by confusion. This provides excellent sensitivity to large-scale radio emission. Nevertheless, given the relatively large beamsize (8.4''$\times$6.8'' FWHM), MIGHTEE flux densities were de-blended using the same list of $K_s$+MIPS~24$~\mu$m+VLA priors applied on FIR/sub-mm images (Sect.~\ref{infrared}). Similarly, VLA radio flux densities were re-extracted based on the same PSF-fitting technique down to S/N$>$3. However, for consistency with publicly available VLA catalogues, we take the 1.4 and 3~GHz radio flux densities of S/N$>$5 sources from \citet{Schinnerer+10} and \citet{Smolcic+17}, respectively. As a sanity check, we show in Appendix~\ref{Appendix_radio} that our procedure leads to fully consistent total 3~GHz flux densities.

Given its unparalleled depth and resolution over the full COSMOS area, we primarily use the VLA~3~GHz dataset for our radio analysis, which counts 13,808 S/N$>$3 sources out of 413,678 M$_{\star}$-selected star-forming galaxies ($\sim$3\%). These radio detections are shown as blue histograms in Fig.~\ref{fig:mass_z}. Fainter sources will be accounted for via stacking, as described in Sect.~\ref{radio_stacking}. Nevertheless, repeating the same stacking analysis with ancillary radio datasets at 1.3~GHz (MIGHTEE) and 1.4~GHz (VLA) is essential to validate our procedure against potential variations of radio spectral index or different angular resolutions. We refer the reader to Appendix \ref{Appendix_ancillary} for an extensive comparison between stacking at 3~GHz and with ancillary radio datasets, while through the paper we will be using radio data only at 3~GHz.

\section{Stacking analysis} \label{method}

The aim of this paper is to investigate how the IRRC evolves with M$_{\star}$ and redshift \textit{simultaneously}. Contrary to studies in which galaxies were individually detected at IR and/or radio wavelengths, leading to complex selection functions and biased samples (see discussion in \citealt{Sargent+10}), we start from a well-defined M$_{\star}$-selected sample. As a consequence, our analysis makes use of IR (Sect.~\ref{ir_stacking}) and radio (Sect.~\ref{radio_stacking}) stacking. This includes a careful treatment of some common caveats concerning IR galaxy samples, such as clustering bias (Sect.~\ref{clustering_bias}) and spectral energy distribution (SED) fitting including AGN templates (Sect.~\ref{sed_fitting}). As for stacking radio data, special care is devoted to statistically removing radio AGN from our sample (Sect.~\ref{agn_removal}). 

In addition, our notably large star-forming galaxy sample allows us to bin as a function of \textit{both} M$_{\star}$ and redshift, as shown in Fig. \ref{fig:bins}. For each bin, we also report the total number of M$_{\star}$-selected SFGs (black), as well as the corresponding fractions having combined S/N$_{IR}>$3 (red) and S/N$_{3~GHz}>$3 (blue). As can be seen, both fractions are a strong function of both M$_{\star}$ and redshift. Therefore, binning along both parameters enables us to account for the fact that galaxies of distinct M$_{\star}$ are detectable at IR and radio wavelengths over different redshift ranges. These aspects will be extensively discussed when comparing our results with previous literature (Appendix~\ref{Appendix_comparison}).

\begin{figure}
\centering
     \includegraphics[width=\linewidth]{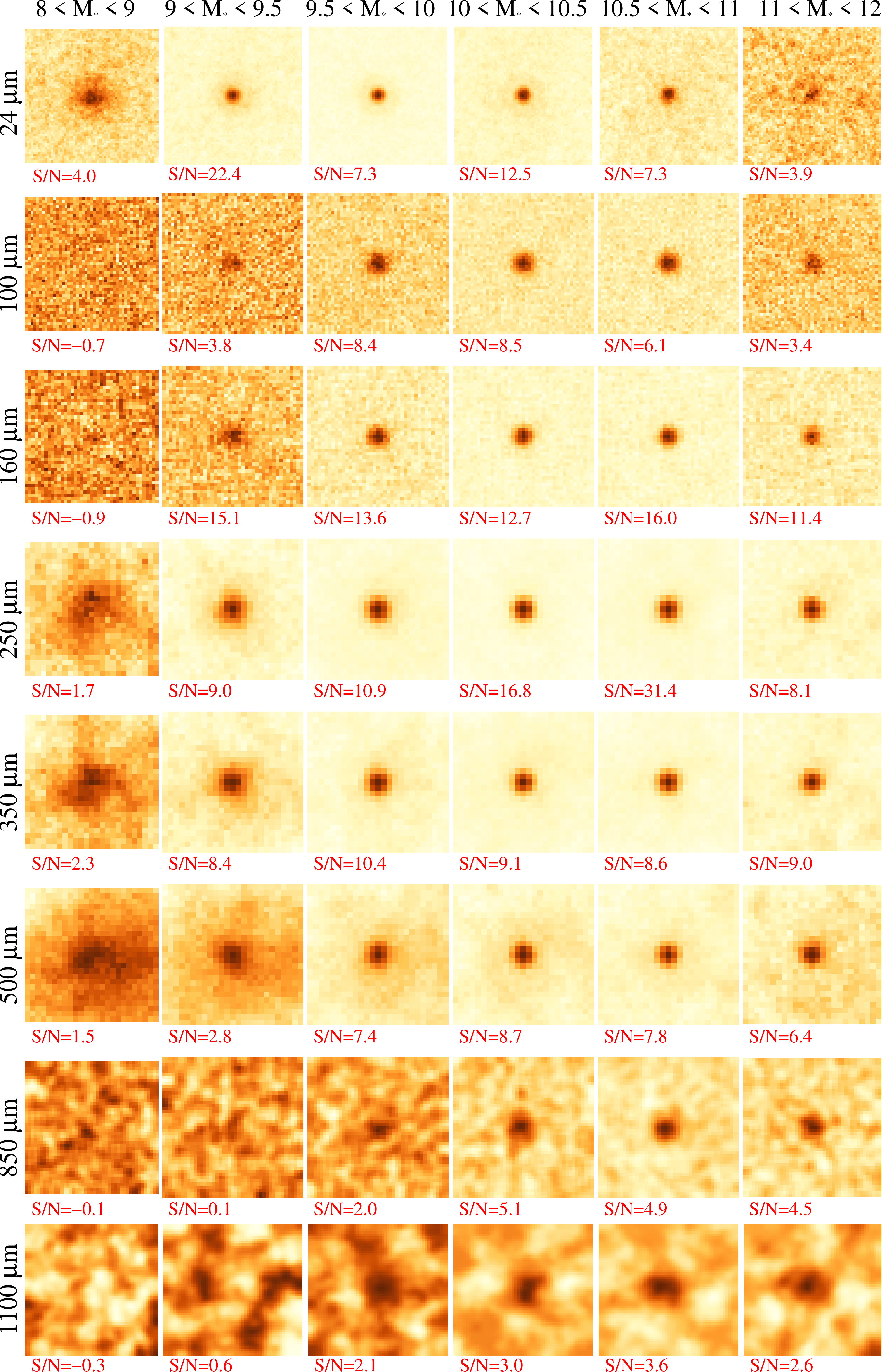}
 \caption{\small Stacked cutouts of $NUVrJ$--based SFGs at 0.8$<$z$<$1.2, as a function of M$_{\star}$ (left to right, expressed in $\log$~M$_{\odot}$). Within each bin we stacked only sources with S/N$<$3 at a given band. SCUBA 850$~\mu$m and AzTEC 1100$~\mu$m images are smoothed with a Gaussian kernel to ease the visualization. Each cutout size is 8$\times$FWHM of the PSF, while for \textit{Spitzer}-MIPS we choose 13$\times$FWHM. Below each cutout we report the corresponding S/N ratio.
 }
   \label{fig:ir_stacks}
\end{figure}

\subsection{Infrared and sub-mm stacking} \label{ir_stacking}

In this Section we estimate the average flux densities across eight infrared and sub-mm bands, namely MIPS~24$~{\mu}$m, PACS 100-160$~{\mu}$m, SPIRE 250-350-500$~{\mu}$m, SCUBA 850$~{\mu}$m and AzTEC 1100$~{\mu}$m. Similarly to other studies, we perform \textit{median} stacking on the residual maps from \citet{Jin+18}, i.e. after subtracting all detected sources with S/N$>$3 in each band (see also \citealt{Magnelli+09}). Individual S/N$>$3 detections will be added to stacked flux densities a-posteriori through a weighted average (Eq.~\ref{eq:weighted}).
Median stacking strongly mitigates contamination from bright neighbors and catastrophic outliers, and thus reduces the confusion noise for the faint sources. 
We stress that our procedure yields very consistent results with either median or mean stacking of detections and non-detections combined (e.g. \citealt{Magnelli+15}; \citealt{Schreiber+15}), as shown in Sect.~\ref{sed_fitting}.

To produce stacked and rms images in each band, we used the publicly available IAS stacking library\footnote{\url{https://www.ias.u-psud.fr/irgalaxies/downloads.php}} (\citealt{Bavouzet+08}; \citealt{Bethermin+10}). For each band, M$_{\star}$ bin and redshift bin, we stack N$\times$N pixel cutouts from the residual images, each centred on the NIR position of the M$_{\star}$-selected priors (Sect.~\ref{sample}). We choose the cutout size to be 8 times the full-width at half maximum (FWHM) of the PSF, while for \textit{Spitzer}-MIPS we choose 13$\times$FWHM, since a substantial fraction of the 24$~\mu$m flux is located in the first Airy ring. Since the AzTEC map covers only a central sub-area of 0.72 deg$^2$, at 1.1~mm we only stack within that region. We emphasize that the M$_{\star}$, z, and SFR distribution of the SFG population within the AzTEC region is fully consistent with that derived in the rest of the COSMOS field, thus not biasing the resulting stacked flux densities. To measure total flux densities, we followed different techniques depending on the input map. For MIPS and PACS images, we used a PSF fitting technique (e.g. \citealt{Magnelli+14}). A correction of 12\% is further applied to account for flux losses from the high-pass-filtering processing of PACS images (e.g. \citealt{Popesso+12}; \citealt{Magnelli+13}). For SPIRE images, the photometric uncertainties are not dominated by instrumental noise but by the confusion noise caused by neighboring sources (\citealt{Dole+03}; \citealt{Nguyen+10}). Since SPIRE, as well as SCUBA and AzTEC images are already brightness maps (in units of Jy beam$^{-1}$ or equivalent), we read the total flux from the peak brightness pixel. The total flux density was taken as the median of the input cube at the central pixel.

The uncertainties on the stacked flux densities are measured using a bootstrap technique (e.g. \citealt{Bethermin+15}). Within each M$_{\star}$--z bin, we run our stacking procedure 100 times, in all bands. For $m$ non-detections at a given band, in each random realization we re-shuffle the input sample, preserving the same total $m$ by allowing source duplication. We take the median of the resulting flux distribution as our formal stacked flux. The 1$\sigma$ dispersion around this value is interpreted as the flux error. We propagate this uncertainty in quadrature with the standard deviation of the stacked map across 100 random positions within the cutout (after masking the central PSF). Though the latter component is typically sub-dominant relative to a bootstrapping dispersion, this conservative approach accounts for the strong fluctuations seen in low S/N stacked images, especially at low M$_{\star}$.

As an example, Fig.~\ref{fig:ir_stacks} shows stacked cutouts in all IR/sub-mm bands at 0.8$<$z$<$1.2 (i.e. close to the median redshift of our sample) as a function of M$_{\star}$. As expected from the tight MS relation that links M$_{\star}$ and SFR in star-forming galaxies, stacks at low M$_{\star}$ display lower S/N, despite the larger numbers of input sources.

\subsubsection{Correcting for clustering bias} \label{clustering_bias}

The stacked flux densities calculated above can be biased high if the input sources are strongly clustered or very faint. This bias is caused by the greater probability of finding a source close to another one in the stacked sample compared to a random position. This generates an additional signal, as extensively discussed in the literature (e.g. \citealt{Bavouzet+08}; \citealt{Bethermin+10}, \citeyear{Bethermin+12}; \citealt{Kurczynski+10}; \citealt{Bourne+12}; \citealt{Viero+13}; \citealt{Schreiber+15}; \citealt{Bethermin+15}). Given the large number of stacked sources in each bin, the S/N is typically good enough to be able to correct for this effect, that becomes more prominent with increasing beam size (e.g. up to 50\% for SPIRE images, see \citealt{Bethermin+15}). Here we briefly describe our approach, referring the reader to Appendix A.2 of \citet{Bethermin+15} for a detailed explanation. 

We model the signal from stacking as the sum of three components: a central point source with the median flux of the underlying population, a clustering component convolved with the PSF, and a residual background term (Eq.~\ref{eq:clustering}). Following \citet{Bethermin+15}, we attempt at separating these components via a simultaneous fit in the stacked images (\citealt{Bethermin+12}; \citealt{Heinis+13}, \citeyear{Heinis+14}; \citealt{Welikala+16}). 
  \begin{equation}
  S(x, y) = \varphi \times PSF(x, y) + \psi \times (PSF \otimes w)(x, y) + \varepsilon
  \label{eq:clustering} 
  \end{equation}
where $S(x,y)$ is the stacked image, $PSF$ the point spread function, and $w$ the auto-correlation function. The symbol $\otimes$ represents the convolution. The parameters $\varphi$, $\psi$, and $\varepsilon$ are free normalizations of the source flux, clustering signal and background term, respectively.

We parametrize the ``clustering bias'' as bias=$\psi$/($\varphi$+$\psi$), once we have verified that residuals (i.e. $\varepsilon$) are always consistent with zero within the uncertainties. We do not see any obvious M$_{\star}$ or redshift dependence of the clustering bias. However, at fixed wavelength, this can fluctuate significantly depending on the S/N of the input stacked image. For these reasons, we prefer to use an \textit{average} clustering correction $\langle 1-bias \rangle$ for each band (see Table~\ref{tab:clustering}), drawn only from stacks with S/N$>$3. For those images, we multiply the stacked flux by $\langle 1-bias \rangle$ at the corresponding wavelength. Only MIPS~24~$\mu$m data are not shown, since their flux densities will not be used for SED fitting in Sect.~\ref{sed_fitting}. Uncertainties on the clustering corrections were propagated quadratically with the stacked flux errors obtained in Sect. \ref{ir_stacking}. 

\begin{figure}
\centering
     \includegraphics[width=\linewidth]{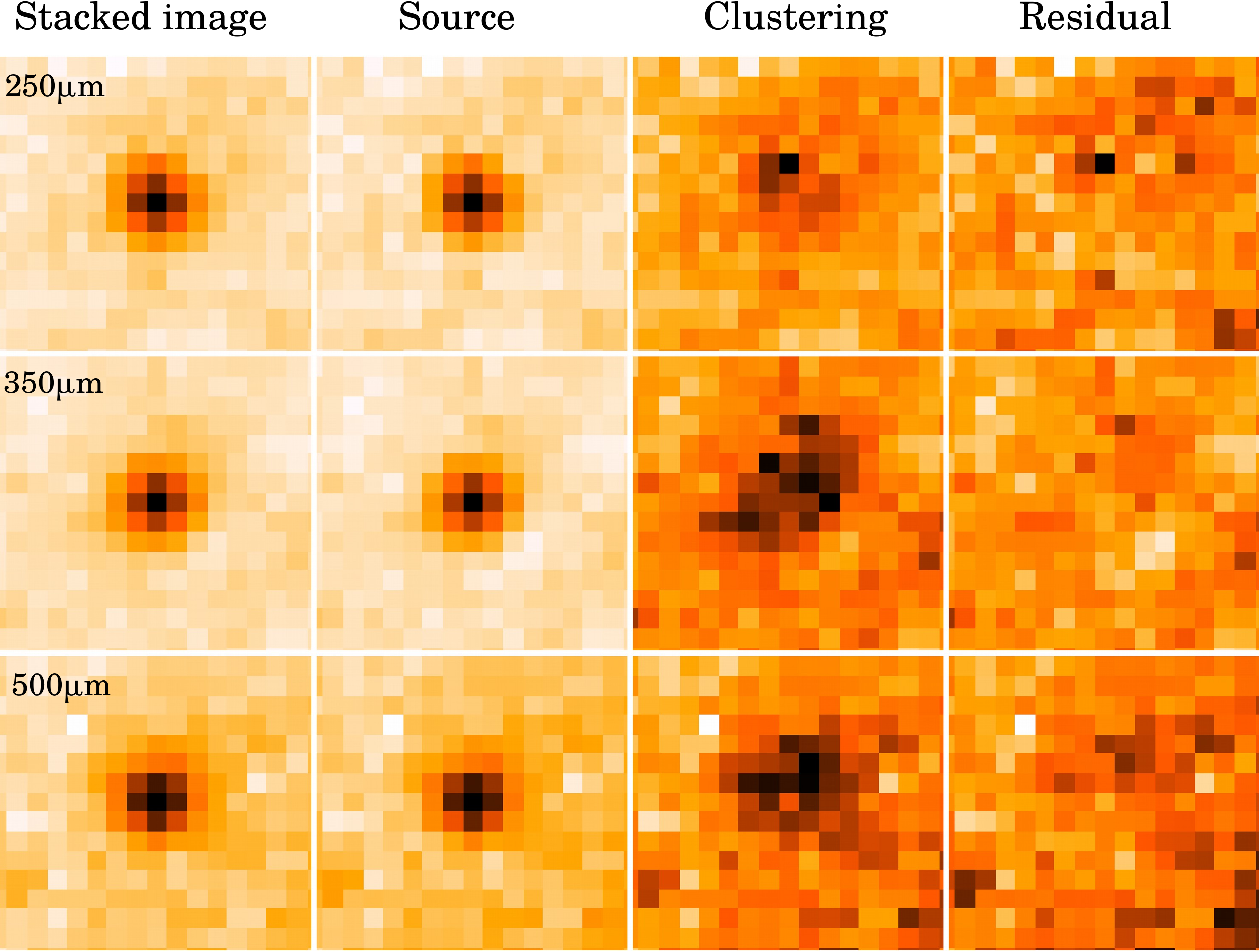}
 \caption{\small Image decomposition of median stacks at 250, 350 and 500$~\mu$m, for a specific bin at 0.8$<$z$<$1.2 and 11$<$$\log$(M$_{\star}$/M$_{\odot}$)$<$12. From left to right, the stacked image is separated among a point source PSF, the clustering signal and a residual background term, respectively. The colour scale is normalized to the maximum in each cutout for visual purposes. See Sect.~\ref{clustering_bias} for details.
 }
   \label{fig:clustering}
\end{figure}

We stress that this method is suitable if the intrinsic source size is negligible compared to the PSF. This is especially true in SPIRE images, of which we show an example in Fig.~\ref{fig:clustering}. This refers to a specific bin at 0.8$<$z$<$1.2 and 11$<$$\log$(M$_{\star}$/M$_{\odot}$)$<$12, for which all stacks give good enough S/N. Particularly for SPIRE images, the clustering bias can make up to 40\% of the total flux, and it can be recognized as a more extended and diffuse emission. However, for consistency we extend this analysis to the full set of FIR/sub-mm data.

\begin{table}
\centering
   \caption{Average fraction of clustering signal at each FIR/sub-mm band. Uncertainties indicate the 1$\sigma$ dispersion among all S/N$>$3 stacks at a given band.}
\begin{tabular}{l c }
\hline
\hline
Wavelength      &        \% Clustering signal    \\
 \hline
PACS 100~$\mu$m	     &    11.3$\pm$7.4     \\
PACS 160~$\mu$m	     &    10.2$\pm$16.5     \\
SPIRE 250~$\mu$m	 &    25.9$\pm$18.9     \\
SPIRE 350~$\mu$m	 &    31.3$\pm$20.8     \\
SPIRE 500~$\mu$m	 &    42.7$\pm$24.2     \\
SCUBA 850~$\mu$m	 &    19.2$\pm$10.7     \\
AzTEC 1100~$\mu$m    &    20.1$\pm$12.9     \\
\hline
\end{tabular}
\label{tab:clustering}
\end{table}

%
%

\begin{figure*}
\centering
     \includegraphics[width=\linewidth]{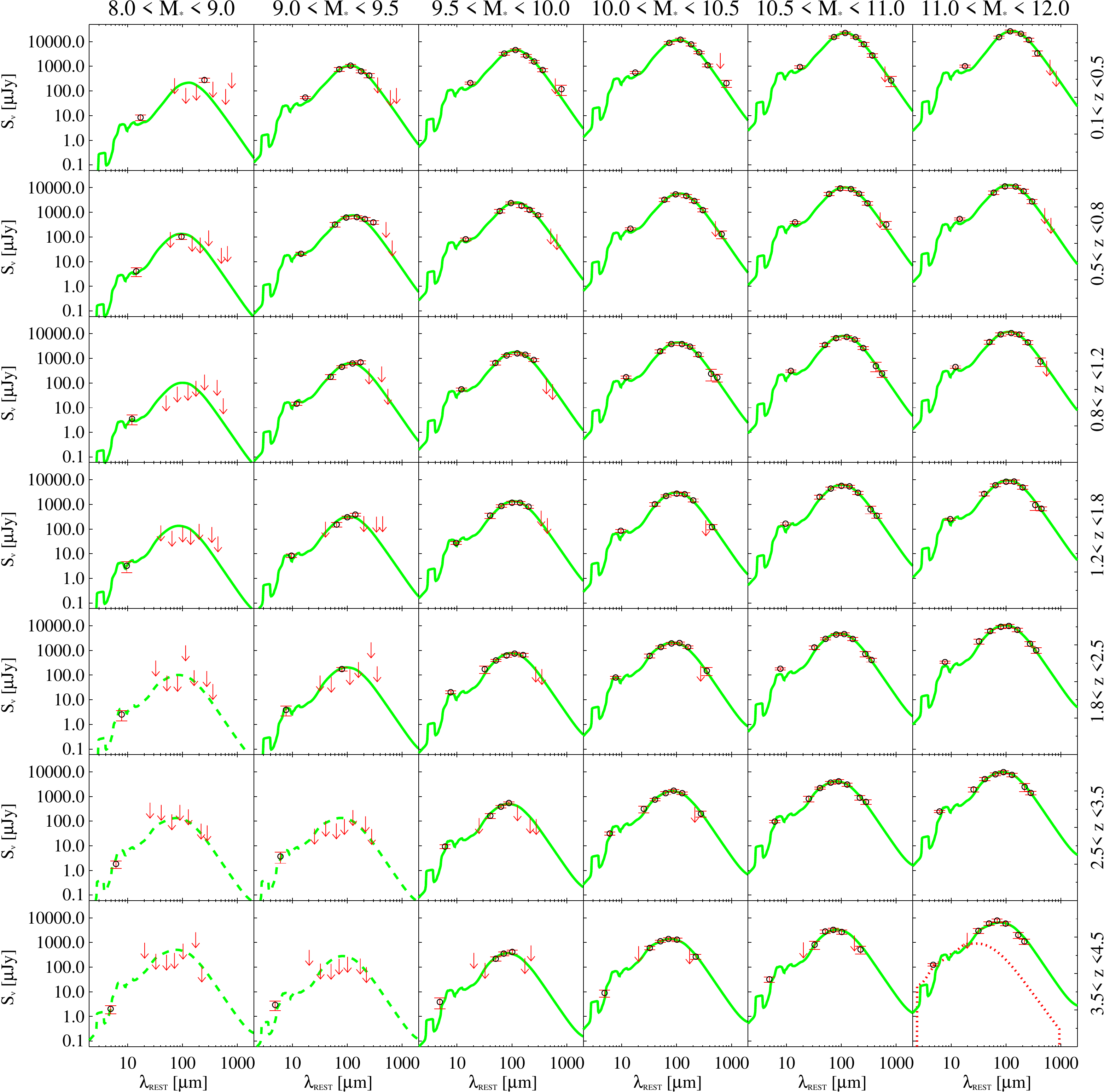}
 \caption{\small Best-fit template obtained from SED-fitting decomposition (green lines), as a function of M$_{\star}$ (left to right, expressed in $\log$~M$_{\odot}$) and redshift (top to bottom). Red circles indicate the IR/sub-mm photometry (MIPS~24$~\mu$m, PACS 100-160~$\mu$m, SPIRE 250-350-500~$\mu$m, SCUBA 850~$\mu$m and AzTEC 1.1~mm), while downward arrows mark the corresponding 3$\sigma$ upper limits. The red dotted line is the best-fit AGN template, shown in the only bin where its significance is above 3$\sigma$. Green dashed lines represent SEDs without FIR measurements and at z$\gtrsim$1.5, for which the integrated L$_{IR}$ is interpreted as 3$\sigma$ upper limit (5/42 bins). MIPS~24$~\mu$m flux densities are not used in the fitting.}
   \label{fig:sed}
\end{figure*}
%

Lastly, the clustering-corrected source flux densities ($S_{stack}$) are combined with those of individual detections ($S_{i}$) with S/N$>$3 in each band. If ($m$, $n$) is the number of stacked and detected sources, respectively, the weighted-average flux $S_{bin}$ in a given bin is derived as follows:
\begin{equation}
 S_{bin} = \frac{m \times S_{stack} +  \sum^{n}_{i=1} S_{i}}{n + m}
 \label{eq:weighted}
\end{equation}
Flux uncertainties were propagated in quadrature. For stacks with S/N$<$3 in which we could not constrain the clustering correction, $S_{stack}$ was set to the noise level of the stacked map (i.e. equal to its uncertainty). This way the weighted-average flux $S_{bin}$ and its error are mainly driven by individual detections, for which the flux could be measured more accurately \citep{Jin+18}. If the combined flux $S_{bin}$ has S/N$<$3, then it was set to 3$\times$ the noise and interpreted as 3$\sigma$ upper limit.

\subsection{Conversion to L$_{IR}$ and SFR} \label{sed_fitting}

This Section illustrates how we fit the observed FIR/sub-mm SEDs to determine the total (8-1000$~\mu$m rest-frame) IR luminosity within each M$_{\star}$--z bin. To this end, we use the two-component SED-fitting code developed by \citet{Jin+18} (see also \citealt{Liu+18}). Briefly, this includes: 3 mid-infrared AGN torus templates from \citet{Mullaney+11}; 15 dust continuum emission models by \citet{Magdis+12}, that were extracted from \citet{Draine+07} to best reproduce the average SEDs of MS (14) or SB (1) galaxies at various redshifts. While \citeauthor{Draine+07} models were based on a number of physical parameters, the library of \citet{Magdis+12} depends exclusively on the mean radiation field $\langle U \rangle$=L$_{IR}$ per unit dust mass (M$_{d}$), and on whether the galaxy is on or above the MS. However, on the MS the average dust temperature strongly evolves with redshift (e.g. \citealt{Magnelli+14}) and directly enters M$_d$. Therefore, $\langle U \rangle$ and the SED shape both vary as a function of redshift, for which \citet{Magdis+12} empirically found as $\langle U \rangle$$\propto$(1+z)$^{1.15}$ up to z$\sim$2. More recently, \citet{Bethermin+15} revised the evolution of $\langle U \rangle$ with redshift out to z$\sim$4, using IR/sub-mm data in the COSMOS field, retrieving $\langle U \rangle$$\propto$(1+z)$^{1.8}$. Here we adopt the set of 14 MS templates from \citet{Magdis+12}, fit them to our data, and we compare the $\langle U \rangle$--z trend with \citet{Bethermin+15}.

The SED-fitting routine performs a simultaneous fitting using AGN and dust emission models, looking for the best-fit solution via $\chi^2$ minimization. In order to account for the typical photo-z uncertainty of the underlying galaxy population (at fixed  M$_{\star}$,z), each template is fitted to the data across a range of $\pm$0.05$\times$(1+$\langle z \rangle$) around the median redshift $\langle z \rangle$. The code keeps track of each SED solution and corresponding normalization, generating likelihood distributions and uncertainties on e.g. L$_{IR}$, $\langle U \rangle$ and AGN luminosity, if any. 
We note that only FIR and sub-mm photometry (i.e. ignoring the MIPS~24~$\mu$m data-point) were used in the fitting procedure. This is to avoid internal variations of the MIR dust features that cannot be captured by our limited set of templates (e.g., IR to rest-frame 8$~\mu$m ratio, IR8, \citealt{Elbaz+11}), which might affect the global FIR/sub-mm SED fitting. This optimization clearly prioritizes the FIR/sub-mm part of the SED, while not impacting the final L$_{IR}$ estimates (e.g. \citealt{Liu+18}).

Fig.~\ref{fig:sed} shows the best-fit star-forming galaxy template from the \citet{Magdis+12} library (green lines), as a function of M$_{\star}$ (left to right, expressed in $\log$M$_{\odot}$) and redshift (top to bottom). Red circles indicate the IR/sub-mm photometry, while downward arrows mark 3$\sigma$ upper limits. The red dotted line is the best-fit AGN template from \citet{Mullaney+11}, shown if significant above 3$\sigma$. This is only found in the highest M$_{\star}$ and redshift bin. Green dashed lines represent SEDs without FIR measurements and at z$\gtrsim$1.5, for which the integrated L$_{IR}$ is interpreted as 3$\sigma$ upper limit (5/42 bins). Even though 24$~\mu$m has long been used as a proxy for L$_{IR}$, this is only accurate at z$\lesssim$1.5 (e.g. \citealt{Elbaz+11}; \citealt{Lutz14}). For this reason we still interpret as measurements the L$_{IR}$ obtained from SEDs without FIR data, but only at z$\lesssim$1.5. That is the case for a few bins at the lowest M$_{\star}$, in which the SED reproduces a-posteriori the 24$~\mu$m data-point. Globally, our stacking analysis yields robust L$_{IR}$ estimates in 37/42 bins.

We find $\langle U \rangle$$\propto$(1+z)$^{1.74\pm0.18}$, which is fully consistent with the revised $\langle U \rangle$--z trend of \citet{Bethermin+15}: $\langle U \rangle$$\propto$(1+z)$^{1.8\pm0.4}$. This test is reassuring, since it confirms that one single z-dependent (or $\langle U \rangle$--dependent) MS galaxy template is fully able to reproduce the observed SED across a wide M$_{\star}$ interval.

\begin{figure}
\centering
     \includegraphics[width=\linewidth]{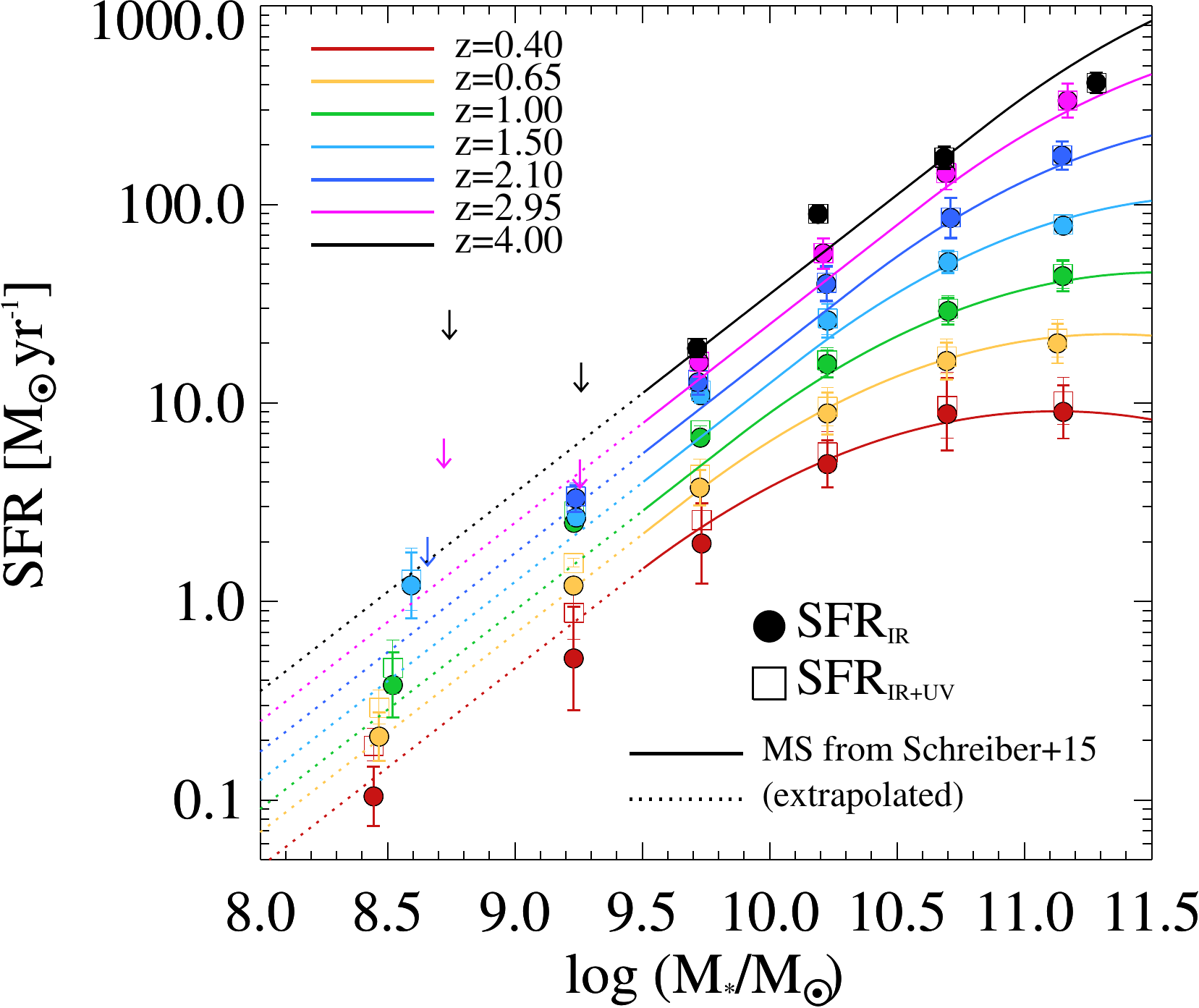}
 \caption{\small SFR--M$_{\star}$ relation of the $NUVrJ$ star-forming galaxies selected in this work, colour-coded by redshift over 0.1$<$z$<$4.5. At fixed M$_{\star}$ and redshift, SFR$_{IR}$ measurements are converted from the L$_{IR}$ obtained from IR/sub-mm SED-fitting (circles). Downward arrows indicate 3$\sigma$ upper limits. For completeness, we also show the SFR$_{IR+UV}$ estimates by combining SFR$_{IR}$ with UV-uncorrected SFRs (open squares). Solid lines mark the evolving MS relation between SFR$_{IR+UV}$ and M$_{\star}$ at different redshifts \citep{Schreiber+15}. We observe an excellent agreement with the MS, even below M$_{\star}$$\sim$10$^{9.5}$M$_{\odot}$ that relies upon a linear extrapolation from \citet{Schreiber+15} not constrained by previous data.}
   \label{fig:ms}
\end{figure}
%
%
Given the tight correlation between L$_{IR}$ and SFR, IR data have been extensively used as proxy for SFR, assuming that most of galaxy star formation is obscured by dust (\citealt{Kennicutt98}; \citealt{Kennicutt+12}). This is probably true inside the most massive star-forming galaxies (see \citealt{Madau+14} for a review). However, at decreasing M$_{\star}$, galaxies become metal poorer (e.g. \citealt{Mannucci+10}), thus less dusty and obscured. In these systems the ultraviolet (UV) domain provides a key complementary view on the unobscured star formation (\citealt{Buat+12}; \citealt{Cucciati+12};  \citealt{Burgarella+13}). 

On this basis, the comprehensive work by \citet{Schreiber+15} exploited IR-based SFRs (i.e. SFR$_{IR}$) and UV-uncorrected SFRs, in the deepest CANDELS fields, to calibrate the star-forming MS over an unprecedented M$_{\star}$ (down to M$_{\star}$=10$^{9.5}$M$_{\odot}$) and redshift range (z$\lesssim$4). Since we carried out a similar analysis, it is worth checking whether the SFR estimates based on our IR stacking reproduce or not the MS of \citet{Schreiber+15}.

For consistency, we need to collect the UV-uncorrected SFRs for our input sample. Hence, for each source we take the rest-frame NUV luminosity L$_{NUV}$ (2800\AA) given in the corresponding parent catalogue (\citealt{Laigle+16} or \citealt{Muzzin+13}) to estimate the UV-uncorrected SFR following \citet{Kennicutt98}: SFR$_{UV}$~[M$_{\odot}$yr$^{-1}$] = 3.04$\times$10$^{-10}$~L$_{2800}$/L$_{\odot}$, already scaled to a \citet{Chabrier03} IMF. We verified that this conversion agrees well with more recent SFR$_{UV}$ prescriptions (e.g. \citealt{Hao+11}; \citealt{Murphy+11}, see also \citealt{Kennicutt+12}).

Within each M$_{\star}$--z bin, we simply take the median SFR$_{UV}$, and we add it to the SFR$_{IR}$ corresponding to the stacked L$_{IR}$, calculated as SFR$_{IR}$=10$^{-10}$~L$_{IR}$/L$_{\odot}$ (\citealt{Kennicutt98}, scaled to a \citealt{Chabrier03} IMF). Fig.~\ref{fig:ms} displays our data in the SFR--M$_{\star}$ plane, colour-coded by redshift over 0.1$<$z$<$4.5. At fixed M$_{\star}$ and redshift, we show SFR$_{IR}$ (circles) and the total SFR$_{IR+UV}$ (open squares) for comparison. Downward arrows are 3$\sigma$ upper limits scaled from L$_{IR}$. 
As can be seen, our data are in excellent agreement with the evolving MS relation at all redshifts (solid lines, \citealt{Schreiber+15}). While SFR$_{UV}$ appears generally negligible compared to the total SFR, it becomes as high as SFR$_{IR}$ towards low M$_{\star}$ and low redshift (e.g. \citealt{Whitaker+12}, \citeyear{Whitaker+17}). Our median values agree with \citet{Schreiber+15} even below M$_{\star}$$\sim$10$^{9.5}$M$_{\odot}$, at which we extrapolate the MS relation due to lack of previous data (dashed lines).
This test compellingly demonstrates that our L$_{IR}$ can be deemed robust over the full M$_{\star}$ and redshift interval explored in this work.

\subsection{Radio stacking at 3~GHz} \label{radio_stacking}

In this Section we describe the equivalent stacking analysis done with radio 3~GHz \citep{Smolcic+17} data, in order to derive average rest-frame 1.4~GHz spectral luminosities (L$_{1.4~GHz}$) in each M$_{\star}$--z bin. 

As done for IR stacking (Sect.~\ref{ir_stacking}), we treat detections and non-detections separately. Total flux densities of radio sources with 3$<$S/N$<$5 were taken from \citet{Jin+18} (see Sect.~\ref{radio}), while for brighter sources we matched their flux densities to those of the corresponding catalogues. The purpose of this approach is twofold: using the same published flux densities for S/N$>$5 detections for consistency and avoiding to deal with the effect of side-lobes from bright sources in stacked images, that might complicate total flux measurements (see Appendix A of \citealt{Leslie+20} for a discussion). In addition, radio detections might contain a substantial fraction of AGN, that is expected to increase at higher M$_{\star}$ (e.g. \citealt{Heckman+14}). We will carefully deal with this issue in Sect.~\ref{agn_removal}. At relatively faint flux densities ($<$100~$\mu$Jy), most of radio emission is thought to arise from star formation (\citealt{Bonzini+15}; \citealt{Padovani+15}; \citealt{Novak+17}; \citealt{Smolcic+17}), though some AGN-related radio emission might still be contributing (e.g. \citealt{White+15}; \citealt{Jarvis+16}). For this reason, median stacking of both detections and non-detections (e.g. \citealt{Karim+11}; \citealt{Magnelli+15}) in deep VLA-COSMOS 1.4~GHz images should result in minimal radio AGN contamination. This alternative approach will be tested in Appendix~\ref{Appendix_comparison}. 

Within the UltraVISTA area analyzed in this work, the 3~GHz rms (2.3~$\mu$Jy~beam$^{-1}$) fluctuates by less than 2\% \citep{Smolcic+17b}. Indeed, we anticipate that no difference between median or rms-weighted mean stacking of non-detections is observed (see Appendix~\ref{Appendix_radio} and \citealt{Leslie+20}), as detailed below. For these reasons, we choose to perform median stacking of non-detections. Individual detections will be added a-posteriori via a simple mean weighted average, as done in Eq.~\ref{eq:weighted}. 

Our stacking routine generates cutouts with size of 8$\times$FWHM$_{3~GHz}$ (i.e. 6'' at 3 GHz), centered on the NIR position of each input galaxy. We acknowledge that an average offset of 0.1'' was found between 3~GHz \citep{Smolcic+17b} and UltraVISTA positions \citep{Laigle+16}, which is half the size of a pixel. To account for this systematic offset, our routine performs sub-pixel interpolation and searches for the peak flux ($S_{peak}$) within $\pm$1 pixel from the center of the stacked image. The peak flux uncertainty is estimated via bootstrapping 100 times, as done in Sect.~\ref{ir_stacking}. We take the median of the resulting flux distribution as our formal peak brightness. The 1$\sigma$ dispersion around this value is interpreted as the corresponding error. We also measured the standard deviation across 100 random positions in the stack (masking the central beam of 0.75''). This gives less conservative errors compared to a bootstrap, but it is used to derive the formal rms of the stacked map.

Total flux densities ($S_{tot}$) are calculated by fitting a 2D elliptical Gaussian function to the median stacked image, using the IDL routine {\sc mpfit2dfun}\footnote{\url{https://pages.physics.wisc.edu/~craigm/idl/fitqa.html}}. Given the typically high S/N ($\sim$10 on average) reached in the central pixel, we leave size, position angle and normalization of the 2D Gaussian as free parameters. We verified that adopting a circular Gaussian or forcing the normalization to the peak flux does not significantly affect any of our stacks. The total flux was calculated by integrating over the 2D Gaussian area $A_{gauss}$. The integrated flux error was computed by multiplying the peak flux error by $\sqrt{A_{gauss}/A_{beam}}$, where $A_{beam}$ is the known beam area, and adding a known 5\% flux calibration error in quadrature (\citealt{Smolcic+17}). We remind the reader that the peak flux error already incorporates the variance of the stacked sample via bootstrapping.
  
In order to assess whether our sources are clearly resolved, we follow the same criterion applied to VLA 3~GHz detections \citep{Smolcic+17} to identify resolved sources:
\begin{equation}
 \frac{S_{tot}}{S_{peak}} > 1 + (6\times {S/N}_{peak})^{-1.44}
\label{eq:res}
\end{equation}
where ${S/N}_{peak}$ is simply the peak flux divided by the rms of the image. This expression was obtained empirically to define an envelope containing 95\% of unresolved sources, below such threshold. We find that 31 stacks out of 42 are resolved, according to Eq.~\ref{eq:res}. For these, total flux densities are on average 1.8$\times$ higher than peak flux densities. Similarly, \citet{Bondi+18} found 77\% of VLA~3~GHz detected SFGs are resolved, and this fraction does not change significantly with M$_{\star}$ \citep{JimenezAndrade+19}. Of the 11 bins with unresolved emission, 3 have ${S/N}_{peak} <$3. These are all among the 5 bins without L$_{IR}$ estimates from IR stacking (Sect.~\ref{sed_fitting}). Analogously to our treatment of the IR measurements, we discard all those 5 bins from the rest of our analysis.

For the stacks with resolved emission, we prefer to use their integrated flux from 2D Gaussian fitting as the most accurate estimate. Instead, for unresolved stacks we use the peak flux, consistent with the treatment of 3~GHz detections \citep{Smolcic+17}. Fitting residuals are on average 3\% of the total flux, and always consistent with zero within the uncertainties. We validate this approach by reproducing the total flux densities of 3~GHz detections presented in \citet{Smolcic+17} at S/N$>$5 and in \citet{Jin+18} at 3$<$S/N$<$5, respectively (Appendix \ref{Appendix_radio}).

Finally, we combined the radio stacked flux densities within each M$_{\star}$--z bin together with individual detections, following Eq.~\ref{eq:weighted}. The combined 3~GHz flux densities were first scaled to 1.4~GHz assuming of $S_{\nu} \propto \nu^{\alpha}$, with spectral index $\alpha$=--0.75$\pm$0.1 (e.g. \citealt{Condon92}; \citealt{Ibar+09},\citeyear{Ibar+10}). This assumption is discussed in Appendix.~\ref{radio_slope}. Lastly, 1.4~GHz flux densities were converted to rest-frame 1.4~GHz spectral luminosities (L$_{1.4~GHz}$), again assuming $\alpha$=--0.75. Formal L$_{1.4~GHz}$ errors were calculated by propagating the uncertainties on both combined flux and spectral index.

The various checks described in Appendix~\ref{Appendix_ancillary} prove our L$_{1.4~GHz}$ robust across the full range of M$_{\star}$ and redshift analyzed in this work. We note that our L$_{1.4~GHz}$ estimates do not necessarily trace radio emission from star formation. Indeed, radio AGN are not yet removed at this stage, and they might be potentially boosting the L$_{1.4~GHz}$. This issue will be addressed in Sect.~\ref{agn_removal}.

\section{The IRRC and the contribution of radio AGN} \label{results}

Using the median L$_{IR}$ and L$_{1.4~GHz}$ obtained from stacking, we study the evolution of the IRRC as a function of M$_{\star}$ and redshift. Logarithmic uncertainties on each luminosity were propagated quadratically to get $q_{IR}$ errors. Among the 42 M$_{\star}$--z bins analyzed in this work, 37 yield robust estimates of L$_{IR}$ and L$_{1.4~GHz}$, while the remainder are discarded from the following analysis. Unsurprisingly, these latter 5 bins (three at 10$^{8}<$M$_{\star}$/M$_{\odot}<$10$^{9}$ and 1.8$<$z$<$4.5; two at 10$^{9}<$M$_{\star}$/M$_{\odot}<$10$^{9.5}$ and 2.5$<$z$<$4.5) are among the least complete in M$_{\star}$, as highlighted in Figs.~\ref{fig:mass_z} and \ref{fig:sed}. Therefore their exclusion partly mitigates the M$_{\star}$ incompleteness of the remaining sample.

\subsection{q$_{IR}$ before removing radio AGN} \label{q0}

\begin{figure}
     \includegraphics[width=\linewidth]{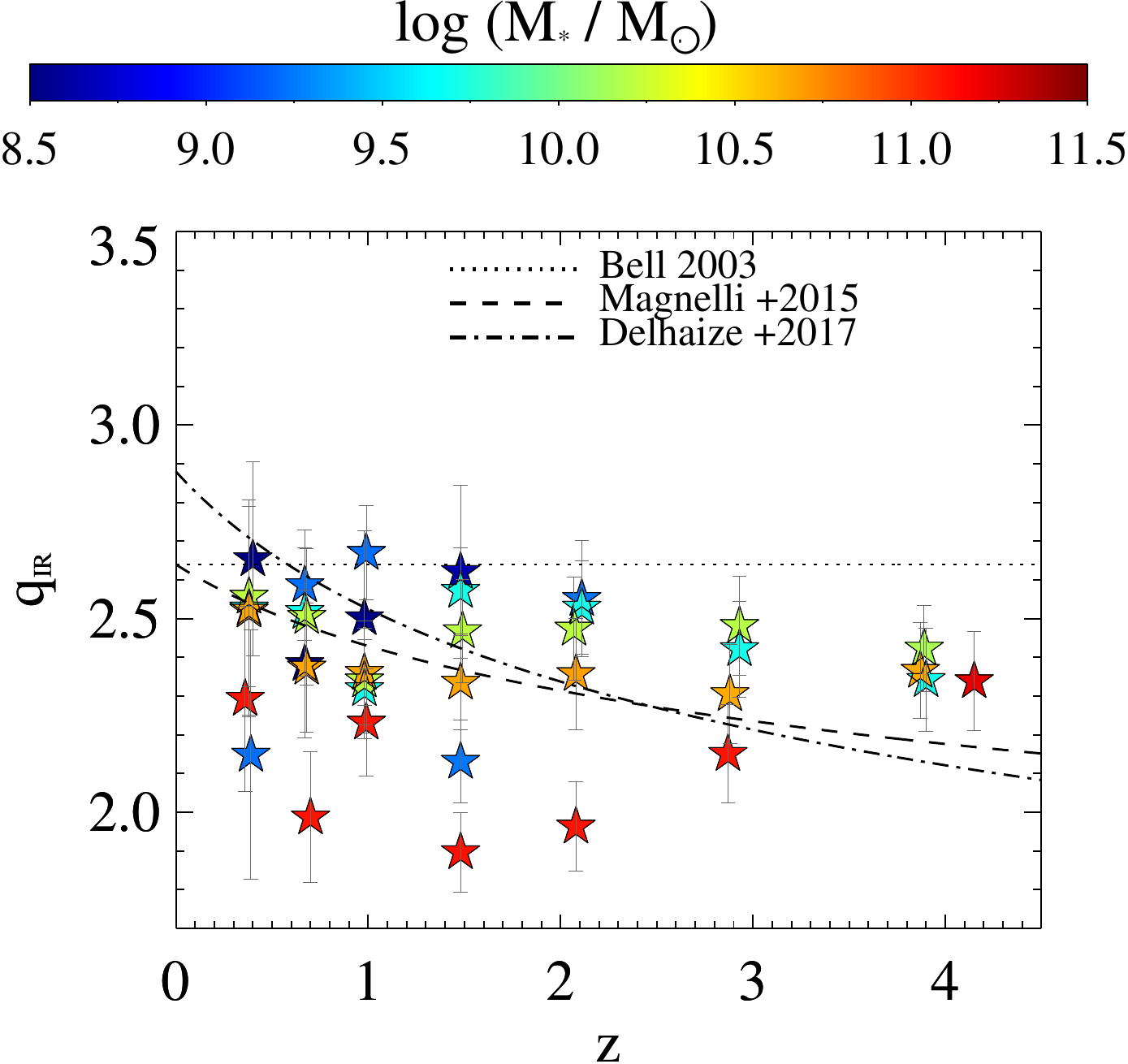}
 \caption{\small AGN-uncorrected q$_{IR}$ evolution as a function of redshift (x-axis) and M$_{\star}$ (colour bar). Errors on q$_{IR}$ represent the 1$\sigma$ scatter around the median value, estimated via bootstrapping over L$_{IR}$ and L$_{1.4~GHz}$. For comparison, other IRRC trends with redshift are taken from the literature (black lines): \citeauthor{Bell03} (\citeyear{Bell03}, dotted); \citeauthor{Magnelli+15} (\citeyear{Magnelli+15}, dashed); \citeauthor{Delhaize+17} (\citeyear{Delhaize+17}, dot-dashed). Our q$_{IR}$ values still include the contribution of radio AGN. See Sect.~\ref{q0} for details.
 }
   \label{fig:qplot0}
\end{figure}

Fig.~\ref{fig:qplot0} shows the average $q_{IR}$ as a function of redshift, colour-coded in M$_{\star}$ (stars). For comparison, other prescriptions of the evolution of the IRRC are overplotted (black lines). \citet{Bell03} inferred the average IRRC in local SFGs, finding $q_{IR}$=2.64$\pm$0.02 (dotted line), with a scatter of 0.26~dex. \citet{Magnelli+15} studied an M$_{\star}$-selected sample at z$\lesssim$2, and constrained the evolution of the far-infrared radio correlation (FIRC, parametrized via $q_{FIR}$\footnote{The far-infrared luminosity used to compute $q_{FIR}$ was integrated between 42 and 122$~\mu$m rest-frame. This is quantified to be 1.91$\times$ smaller than the total L$_{IR}$ \citep{Magnelli+15}.}) across the SFR--M$_{\star}$ plane at M$_{\star}>$10$^{10}$~M$_{\odot}$. From stacking IR and radio images, they parametrized the evolution with redshift of the FIRC as: $q_{FIR}$=(2.35$\pm$0.08)$\times$(1+z)$^{-0.12\pm0.04}$, where the normalization is scaled to 2.63 in the q$_{IR}$ space. More recently, \citet{Delhaize+17} exploited a jointly-selected sample of IR (from \textit{Herschel} PACS/SPIRE) or radio (from the VLA-COSMOS 3~GHz Large Project, \citealt{Smolcic+17}) detected sources (at $\geq$5$\sigma$) in the COSMOS field. Through a survival analysis that accounts for non-detections in either IR or radio, they inferred the evolution of the IRRC with redshift out to z$\sim$4 as: $q_{IR}$=(2.88$\pm$0.03)$\times$(1+z)$^{-0.19\pm0.01}$. While this trend appears somewhat steeper than that of \citeauthor{Magnelli+15}, we note that \citet{Delhaize+17} did not formally remove objects with significant radio excess, while \citet{Magnelli+15} performed median radio stacking to mitigate the impact of potential outliers such as radio AGN. Nevertheless, \citet{Delhaize+17} argue that the IRRC trend with redshift would flatten if applying a 3$\sigma$-clipping: $q_{IR}$=(2.83$\pm$0.02)$\times$(1+z)$^{-0.15\pm0.01}$, which becomes fully consistent with that of \citet{Magnelli+15}.

When compared to the above literature, it is evident that our q$_{IR}$ values lie systematically below other studies at M$_{\star}>$10$^{11}$~M$_{\odot}$, while lower M$_{\star}$ galaxies lie closer or slightly above them. In other words, our q$_{IR}$ estimates seem to display a clear M$_{\star}$ stratification, with the most massive galaxies having typically lower $q_{IR}$ than less massive counterparts. As mentioned before, we remind the reader that our sample, at this point, contains some fraction of radio AGN, which might be boosting the L$_{1.4~GHz}$, particularly at high M$_{\star}$ (see e.g. \citealt{Best+12}) where radio AGN feedback is known to be prevalent. Our L$_{IR}$ estimates are, instead, corrected for a potential IR-AGN contribution (Sect.~\ref{sed_fitting}). Therefore, the net effect caused by including AGN is \textit{lowering} the intrinsic $q_{IR}$. Selecting typical SFGs on the MS should, however, reduce the incidence of powerful radio AGN expected in massive hosts, since most radio AGN at z$<$1 are found to reside in quiescent galaxies (e.g. \citealt{Hickox+09}; \citealt{Goulding+14}). 

For these reasons, we caution that Fig.~\ref{fig:qplot0} should be taken as the AGN-uncorrected $q_{IR}$. However, it is worth showing it to quantify how much $q_{IR}$ will change after removing radio AGN.

\subsection{Searching for radio AGN candidates} \label{agn_removal}

In this Section we carry out a detailed study aimed at identifying potential radio AGN, removing them and ultimately deriving the intrinsic q$_{IR}$ trend purely driven by star formation. 

In our radio analysis we have combined individual radio-detections (above S/N$>$3) with undetected sources via a weighted average (Eq.~\ref{eq:weighted}). Contrary to stacking detections and non-detections together, this formalism enables us to characterize the nature of individual radio detections, i.e. whether they show excess radio emission relative to star formation.
    
We make the underlying assumption that radio-undetected AGN do not significantly affect any of our radio stacks. This is supported by the excellent agreement between mean and median stacked L$_{1.4~GHz}$ of non-detections (Fig.~\ref{fig:radio_median}, bottom panel). Indeed, if the contribution of radio-undetected AGN were substantial, the corresponding mean L$_{1.4~GHz}$ would be significantly higher than the median L$_{1.4~GHz}$ from stacking. This assumption is further supported by the fact that the fraction of identified radio AGN is a strong function of radio flux density, and the sources we stack are by construction faint in the radio. \citet{Algera+20a} argue that below 20 $\mu$Jy (at 3 GHz), the fraction of radio-excess AGN is $<$10\% (see also \citealt{Smolcic+17b}, \citealt{Novak+18}). We acknowledge that our assumption does not allow us to collect a \textit{complete} sample of radio AGN, especially at high redshift where the fraction of radio detections notably drops (Fig.~\ref{fig:bins}). Nevertheless, we will show that any residual AGN contribution does not change our conclusions.  

We briefly summarize our next steps as follows. In Section \ref{qir_detections} we explore the q$_{IR}$ distribution traced by individual 3~GHz detections as a function of M$_{\star}$ and redshift. First, we identify a subset of radio detections at M$_{\star}>$10$^{10.5}$~M$_{\odot}$ that is representative of an M$_{\star}$-selected sample. Then we decompose their q$_{IR}$ distribution between AGN and star formation components (Sect.~\ref{ragn_highmass}). This enables us to subtract potential radio AGN candidates, and calibrate the intrinsic best-fit IRRC with redshift at M$_{\star}>$10$^{10.5}$~M$_{\odot}$ (Sect.~\ref{ragn_highmass_recal}). Later we extrapolate this calibration towards lower M$_{\star}$ bins (Sect.~\ref{agn_extrapolation}), where a similar in-depth analysis was not possible due to radio-detections being strongly incomplete in this M$_{\star}$ regime. Finally, the intrinsic (i.e. AGN-corrected) IRRC as a function of M$_{\star}$ and redshift is presented in Sect.~\ref{q_mass_z}.

\begin{figure*}
\centering
     \includegraphics[width=\linewidth]{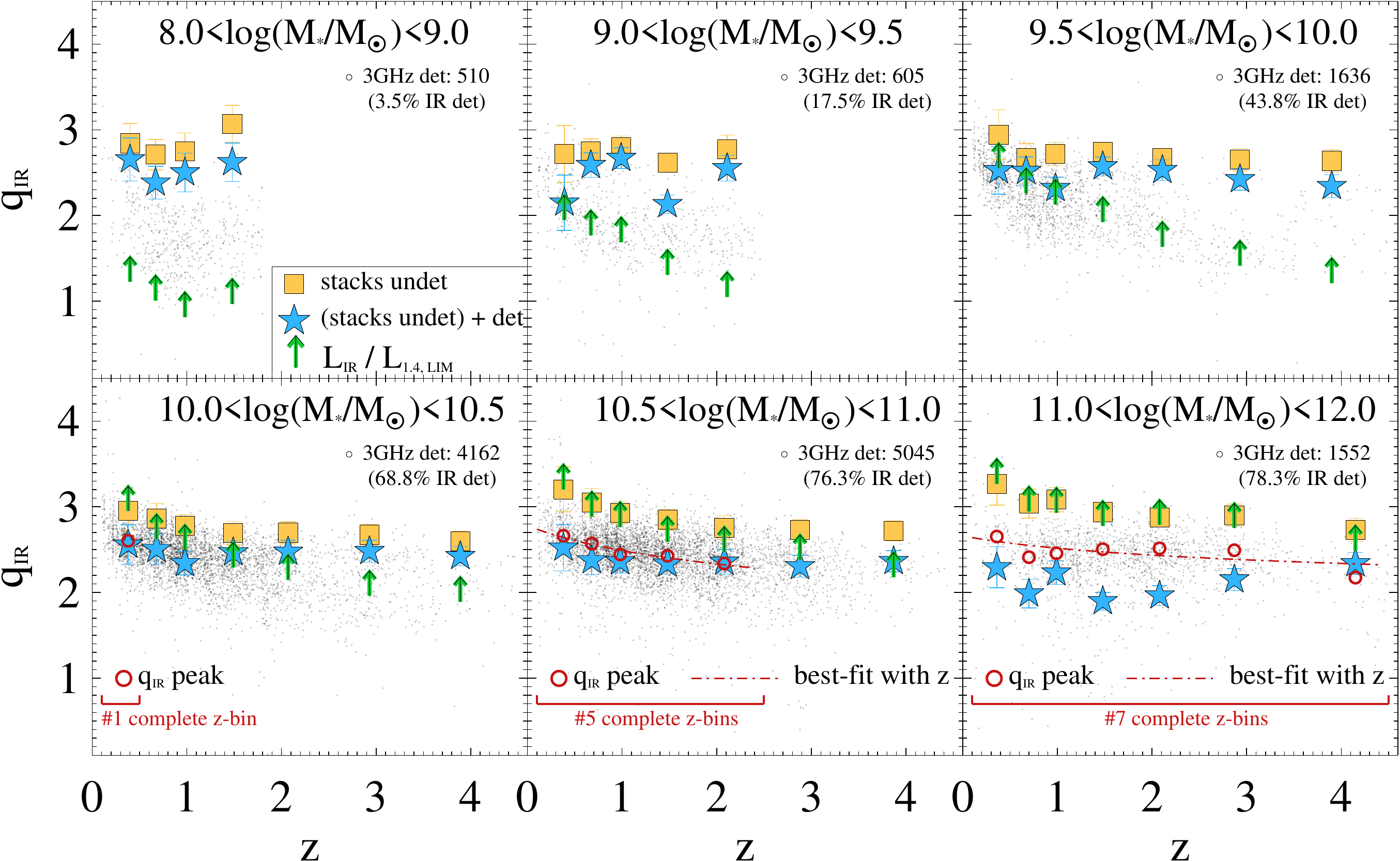}
 \caption{\small Distribution of q$_{IR}$ as a function of redshift, split across increasing M$_{\star}$ bins. In each panel, we compare the q$_{IR}$ estimates of individual radio detections (black dots) with the median stacked values of non-detections (yellow squares) and with the weighted-average q$_{IR}$ of detections and non-detections together (from Eq.~\ref{eq:weighted}, blue stars). Green upward arrows indicate the corresponding threshold q$_{IR, lim}$ above which radio detections become inaccessible from an M$_{\star}$-selected sample. We select relatively \textit{complete} M$_{\star}$--z bins, in which at least 70\% of radio detections have q$_{IR}$ below the corresponding threshold. This criterion identifies 13 bins (in red square brackets). Within these bins, the peak of q$_{IR}$ distribution (q$_{IR, peak}$) is indicated with red open circles. In the two highest M$_{\star}$ bins, the best fitting trends with redshift are shown by the red dashed lines. Each panel reports the number of individual 3~GHz sources and their fraction with S/N$_{IR}>$3. See Sect.~\ref{qir_detections} for details.
 }
   \label{fig:qplot_binned0}
\end{figure*}

\subsubsection{The q$_{IR}$ distribution of radio detections} \label{qir_detections}

In order to study the q$_{IR}$ distribution of 3~GHz detections, we need to calculate their average L$_{IR}$ as a function of M$_{\star}$ and redshift. For convenience, we refer the reader back to Fig.~\ref{fig:mass_z} (blue histograms) for visualizing the distribution of 3~GHz detections in the M$_{\star}$--z space. Out of 13,510 radio detections among our 37 bins, 8762 (65\%) have a combined S/N$_{IR}>$3, therefore reliable L$_{IR}$ measurements from SED-fitting of FIR/sub-mm de-blended photometry \citep{Jin+18}. For the remainder of the sample, we stack again their IR/sub-mm images in all bands in each M$_{\star}$--z bin. Stacked IR flux densities are corrected for clustering bias and converted to L$_{IR}$ following the same procedure adopted for the prior M$_{\star}$ sample (Sect.~\ref{ir_stacking}). Median stacked L$_{IR}$ are retrieved for the same 37/42 bins of the full parent sample, since a stacked S/N$>$3 flux was obtained in at least one FIR/sub-mm band. Then, for each source we re-scaled its median stacked L$_{IR}$ to the redshift and M$_{\star}$ of that source (assuming the MS relation), in order to reduce the variance of the underlying sample within each M$_{\star}$--z bin. We verified that our stacked L$_{IR}$ are always systematically below the 3$\sigma$ L$_{IR}$ upper limits inferred from FIR/sub-mm SED-fitting \citep{Jin+18}. This ensures that our stacking analysis provides more stringent constraints on the L$_{IR}$ of individual non-detections.

From this analysis, we are well placed to explore the full q$_{IR}$ distribution of 3~GHz detections at different M$_{\star}$ and redshifts. Fig.~\ref{fig:qplot_binned0} shows q$_{IR}$ as a function of redshift, split in six M$_{\star}$ bins. Black dots mark individual 3~GHz detections, blue stars represent the q$_{IR}$ obtained by combining detections and non-detections (same as in Fig.~\ref{fig:qplot0}), while yellow squares are the stacks of non-detections only. In each panel we report the number of 3~GHz detected sources, and the fraction of them with combined S/N$_{IR}>$3. This fraction strongly increases with M$_{\star}$, from 3.5\% at 10$^{8}<$M$_{\star}$/M$_{\odot}<$10$^{9}$ to 78.3\% at 10$^{11}<$M$_{\star}$/M$_{\odot}<$10$^{12}$, which implies that at the lowest M$_{\star}$ nearly all q$_{IR}$ estimates of radio detections rely upon IR stacking. This is because the 3~GHz detection limit sets a rough threshold in SFR (if radio emission primarily arises from star formation), therefore is biased towards high-M$_{\star}$ galaxies because of the MS relation. Because of these potential biases, it is essential to identify the bins in which radio detections give us access to a representative sample of M$_{\star}$-selected galaxies.

Indeed, our purpose is to use single radio detections to calibrate a threshold that best distinguishes radio AGN from radio SFGs, as a function of M$_{\star}$ and redshift. In order to extend this calibration to our full M$_{\star}$-selected sample, we need to make sure that our derived trends are not affected by selection biases, i.e. that the radio-detected sources we rely upon are fully representative of M$_{\star}$-selected galaxies at a given redshift. To this end, within each bin we compare the q$_{IR}$ of single radio detections against a specific threshold (q$_{IR, lim}$), corresponding to the 3~GHz survey limit at a given M$_{\star}$ and redshift (green upward arrows in Fig.~\ref{fig:qplot_binned0}). This threshold is proportional to the median stacked L$_{IR}$ of the full SFGs sample, divided by the 3$\sigma$ luminosity limit at 1.4~GHz, scaled from 3~GHz by assuming a fixed spectral index $\alpha$=-0.75 (Appendix.~\ref{radio_slope}). Specifically, q$_{IR, lim}$ indicates the limiting q$_{IR}$ at which a typical MS galaxy of a given M$_{\star}$, z and L$_{IR}$ drops below the 3~GHz detection limit, which translates into a \textit{lower} q$_{IR}$ limit. In other words, sources with q$_{IR}>$q$_{IR, lim}$ lie within an M$_{\star}$ range that is virtually inaccessible by our 3~GHz survey. Therefore, any measurement above that threshold is not representative of an M$_{\star}$-selected sample. Conversely, radio detections below that threshold would be seen also in an M$_{\star}$-selected sample of SFGs.

In this framework, we consider as \textit{complete} only those M$_{\star}$--z bins in which at least 70\% of radio detections are below q$_{IR, lim}$. This cutoff enables us to narrow down the position of the mode of the q$_{IR}$ distribution, leaving us with a total of 13 \textit{complete} bins (at $>$70\% level) across the full sample. Unsurprisingly, they are preferentially located in high-M$_{\star}$ galaxies, and/or at low redshift. These are delimited by a red segment in Fig.~\ref{fig:qplot_binned0}. 

It is quite evident that the locus populated by radio detections tends to decline with redshift, at each M$_{\star}$. However, this behaviour is far more pronounced at low M$_{\star}$, and likely driven by selection effects. In fact, by definition the L$_{1.4~GHz}$ of radio detections increases with redshift at all M$_{\star}$, because 3~GHz sources are drawn from a flux-limited sample. On the contrary, the L$_{IR}$ of radio detections behaves differently with M$_{\star}$: at higher M$_{\star}$ it is mostly based on IR-detected sources, while at lower M$_{\star}$ it comes predominantly from IR stacking. At higher M$_{\star}$, L$_{IR}$ increases with redshift similarly to L$_{1.4~GHz}$, giving rise to a nearly flat q$_{IR}$ locus. At lower M$_{\star}$, instead, L$_{IR}$ stands typically below the IR detection limit, thus not bound to a monotonic redshift increase. This effect causes an apparent \textit{decrease} of q$_{IR}$ with redshift, that is mostly driven by the radio detection limit. Indeed, a similar trend can be seen in the green arrows, that move down in redshift at each M$_{\star}$.

\begin{figure}
\centering
     \includegraphics[width=\linewidth]{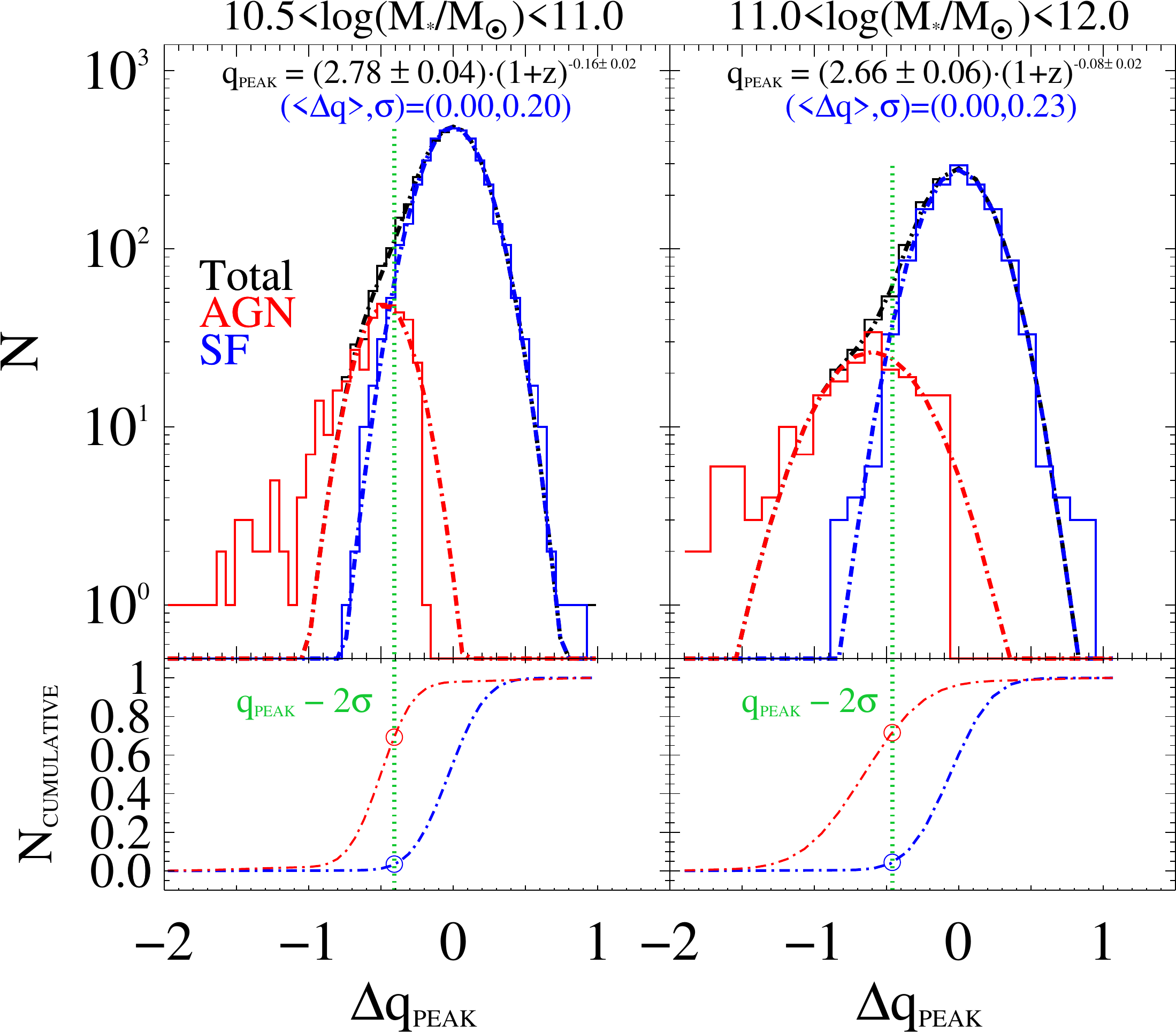}
 \caption{\small q$_{IR}$ distribution of 3~GHz detections in the two highest M$_{\star}$ bins (left and right panels), after correcting for the internal q$_{IR}$--redshift dependence. Only sources within complete z-bins were considered at each M$_{\star}$. The total distribution (black histogram) was dissected among SF-dominated (blue) and AGN-dominated (red) radio sources, and fitted with two separate Gaussian functions (dot-dashed lines). While the SF population was fitted first, the AGN part was fitted in a second step from the total--SF residuals. The 1$\sigma$ dispersion attributed to SF equals 0.20 and 0.23~dex at 10$^{10.5}<$M$_{\star}$/M$_{\odot}<$10$^{11}$ and 10$^{11}<$M$_{\star}$/M$_{\odot}<$10$^{12}$, respectively. The bottom panels display the corresponding cumulative Gaussian fits, both normalized to unity. The vertical green dotted line marks the threshold that best separates between the SF and AGN populations (see Table~\ref{tab:qthres}), which rejects about 70\% AGN and only 3--4\% SFGs (open circles).
 }
   \label{fig:histo0}
\end{figure}

Since the single complete z-bin found at 10$^{10}<$M$_{\star}$/M$_{\odot}<$10$^{10.5}$ is insufficient for us to constrain a redshift trend, we only consider the remaining 12 complete bins placed at M$_{\star}>$10$^{10.5}$. For each of them, we identify the peak of the corresponding q$_{IR}$ distribution of radio detections, namely q$_{IR, peak}$ (see red open circles in Fig.~\ref{fig:qplot_binned0}). We note that q$_{IR, peak}$ represents the \textit{mode} of radio detections, rather than the \textit{average} that is, instead, potentially affected by underlying radio AGN (Sect.~\ref{ragn_highmass}). Then we fitted a power law trend of q$_{IR, peak}$ with redshift using the IDL routine {\sc mpfit2dfun}, obtaining the best-fit expressions shown in Fig.~\ref{fig:histo0}.

  %

\subsubsection{Identifying radio AGN at high M$_{\star}$} \label{ragn_highmass}

After fitting the q$_{IR, peak}$ trend with redshift for the two M$_{\star}$ bins, we need to account for such redshift dependence while exploring the q$_{IR}$ distributions of radio detections. To this end, we align the position of q$_{IR, peak}$ in each redshift bin to match the best-fit redshift trend. This allows us to marginalize over the internal redshift trend, and merge all radio detection homogeneously within the same M$_{\star}$ bin. The resulting redshift-corrected q$_{IR}$ distribution is displayed in Fig.~\ref{fig:histo0} for the two highest M$_{\star}$ bins (left and right panel). Each total histogram (black) includes the contribution from both galaxy- and AGN-dominated radio sources. We proceed to dissecting into the two components as follows , leaving the discussion on how radio AGN affect the redshift trend in Sect.~\ref{ragn_highmass_recal}.

Assuming that the peak of the distribution is populated by radio-detected SFGs, and that the intrinsic q$_{IR}$ distribution of SFGs is symmetric around the peak, we mirror the right-hand side of the observed q$_{IR}$ distribution to the left side. This symmetric function is interpreted as the intrinsic q$_{IR}$ distribution of SFGs (blue histogram). We fitted it with a Gaussian function, leaving normalization and dispersion free to vary. The Gaussian fit yields a dispersion of 0.20 and 0.23~dex at 10$^{10.5}<$M$_{\star}$/M$_{\odot}<$10$^{11}$ and 10$^{11}<$M$_{\star}$/M$_{\odot}<$10$^{12}$, respectively (blue dot-dashed lines). The residual histogram (total-SF) is then fitted with a second Gaussian function (red dot-dashed lines), that parametrizes the additional radio-excess population ascribed to AGN. We attempted to fit the AGN population with other non-Gaussian functions, since the lowest q$_{IR}$ tail is not perfectly reproduced with a Gaussian shape. However, we stress that our purpose is separating the two populations statistically and prioritize a clean identification of SFGs, while a proper characterization of the \textit{shape} of the AGN population is beyond the scope of this paper. 

We note that our fitting approach relies on the assumption that q$_{IR, peak}$ is entirely attributed to SF. Therefore, by mirroring and fitting the SF Gaussian first, it is possible that we might be underestimating the intrinsic relative fraction of radio AGN. We discuss this potential issue in Appendix~\ref{Appendix_AGN}, though we anticipate that our main findings could only be reinforced if addressing this effect.

Another possible caveat of our approach lies in the assumption that IR-undetected sources are represented by a single stacked L$_{IR}$, though rescaled to the M$_{\star}$ and redshift of each object based on the MS relation. However, we checked that the distribution of radio detections that are also IR detected displays an average scatter of 0.22~dex, as for the full radio-detected sample shown in Fig.~\ref{fig:histo0}. This is because the vast majority of radio sources at M$_{\star}>$10$^{10.5}$~M$_{\odot}$ is also individually detected at IR wavelengths (see Fig.~\ref{fig:bins}). Therefore, taking a single stacked L$_{IR}$ in each bin does not strongly impact the calibration of the SF locus.

Choosing the best dividing line between AGN and SF-dominated radio sources is a challenging, and somewhat arbitrary task. Moving the threshold to higher q$_{IR}$ increases the purity of SFGs to the detriment of completeness, and vice-versa for a lower threshold. Here we attempt to reach low levels of cross-contamination between the SF and AGN populations, while keeping a high completeness of the SF population. For this reason, we checked the cumulative q$_{IR}$ distribution drawn by the two Gaussian fits (AGN in red, SF in blue), as shown at the bottom of Fig.~\ref{fig:histo0}, each normalized to unity.

Four different thresholds (q$_{thres}$) were examined: (1) q$_{thres}$=q$_{peak}$-1$\sigma$; (2) q$_{thres}$=q$_{peak}$-2$\sigma$; (3) q$_{thres}$=q$_{peak}$-3$\sigma$; (4) q$_{thres}$=q$_{cross, AGN=SF}$. In this formalism, q$_{peak}$ is still the peak of the SF population (blue Gaussian fit in Fig.~\ref{fig:histo0}), and $\sigma$ its dispersion, while q$_{cross, AGN=SF}$ represents the cross-over value at which the numbers of radio AGN and radio SFGs match each other. For each threshold, in Table~\ref{tab:qthres} we report the cumulative fractions of SF and AGN populations lying below it. Qualitatively speaking, an ideal compromise consists of a low fraction of SF galaxies and a high fraction of AGN below the threshold. 

This comparison highlights that the best trade-off between cross-contamination and completeness is given by the threshold q$_{thres}$=q$_{peak}$-2$\sigma$, in both M$_{\star}$ bins. This method rejects about 70\% of potential radio-excess AGN, and only 3--4\% of SFGs, that we believe is quite acceptable. The offset from the corresponding q$_{peak}$ value is on average 0.43~dex (that we deem robust in Sect.~\ref{ragn_highmass_recal}), which implies that our radio-excess AGN should have statistically at least 63\% of their total radio emission arising from AGN activity.

  %
\begin{table}
\centering
   \caption{Comparison between fractions of radio-SFGs and radio-excess AGN below some threshold q$_{thres}$, for the two highest M$_{\star}$ bins. Four different thresholds are examined. The best trade-off between cross-contamination and completeness is given by q$_{thres}$ = q$_{peak}$ -2$\sigma$ (green dotted line in Fig.~\ref{fig:histo0}), which we apply in the following analysis. See Sect.~\ref{ragn_highmass} for details.}
\begin{tabular}{ll cc }
\hline
\hline
  M$_{\star}$ (M$_{\odot}$) bin    &   q$_{thres}$  &      \% SF   &      \% AGN    \\
      &    &        (q$<$q$_{thres}$)  &   (q$<$q$_{thres}$)      \\
 \hline
 10$^{10.5}$--10$^{11}$~M$_{\odot}$ & q$_{peak}$--1$\sigma$	     &       20.1\%    &  93.6\%    \\
                                     & q$_{peak}$--2$\sigma$	     &       3.5\%     &  69.2\%    \\
                                     & q$_{peak}$--3$\sigma$	     &       0.4\%     &  27.3\%    \\
                                     & q$_{cross, AGN=SF}$        &       2.6\%     &  64.7\%    \\
  &   &   & \\
10$^{11}$--10$^{12}$~M$_{\odot}$    & q$_{peak}$--1$\sigma$	     &      23.0\%    &  89.2\%    \\
                                     & q$_{peak}$--2$\sigma$	     &      4.5\%     &  71.5\%    \\
                                     & q$_{peak}$--3$\sigma$	     &      0.7\%     &  45.2\%    \\
                                     & q$_{cross, AGN=SF}$        &      3.7\%     &  70.2\%    \\
\hline
\end{tabular}
\label{tab:qthres}
\end{table}
%

\subsubsection{Re-calibrating the radio AGN threshold}  \label{ragn_highmass_recal}

\begin{figure}
\centering
     \includegraphics[width=\linewidth]{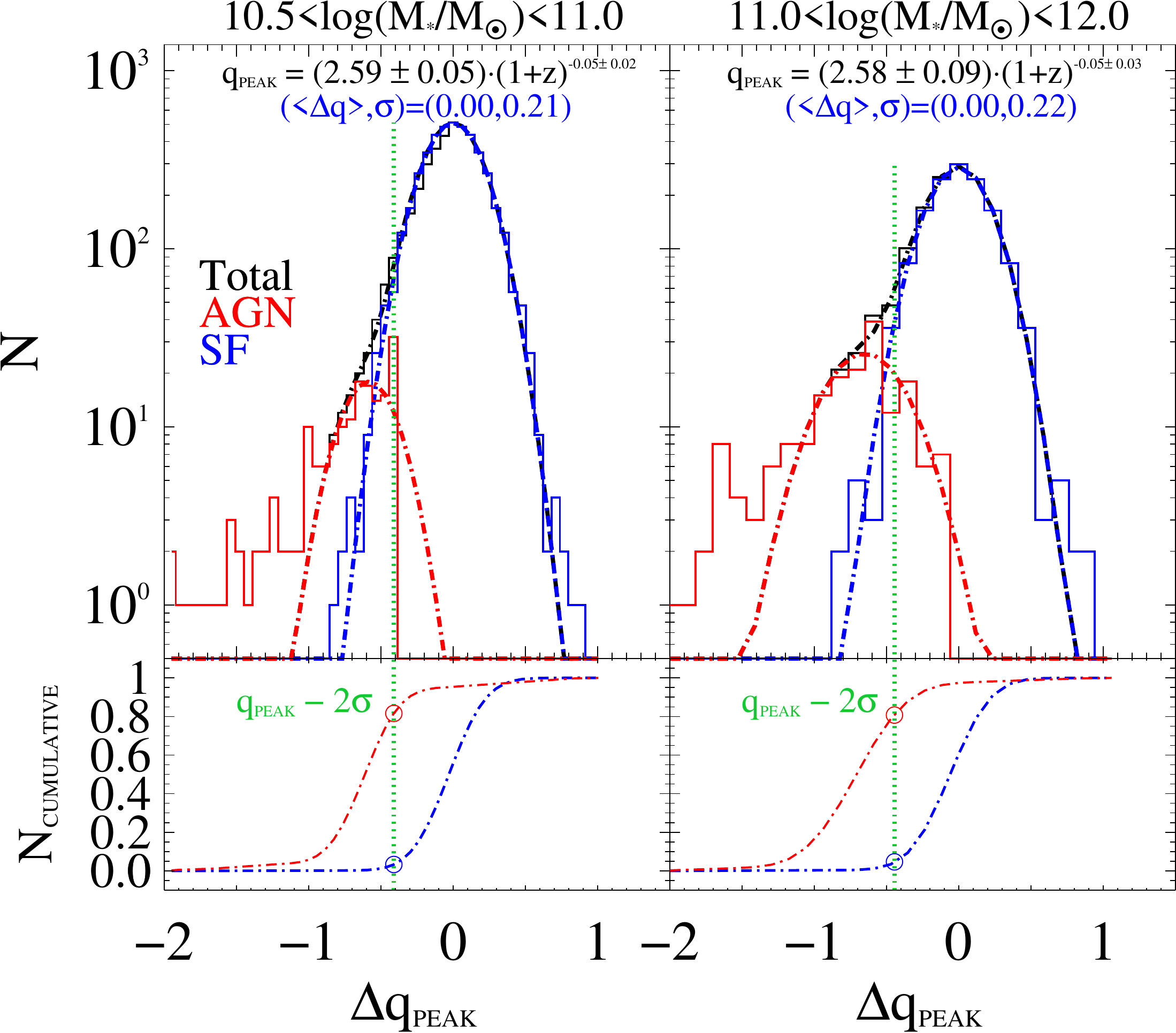}
 \caption{\small Same as Fig.~\ref{fig:histo0}, but normalizing the peak to the flatter q$_{IR}$--z trend calibrated after removing AGN (Sect.~\ref{ragn_highmass_recal}). This two-fold approach slightly improves the q$_{IR}$ decomposition, as highlighted by the larger cumulative fraction of radio AGN that are rejected below the q$_{IR}$ threshold (red open circles, 81\% against the previous 70\%).
 }
   \label{fig:histo0_corr}
\end{figure}

According to the threshold defined above, we removed radio-excess AGN from our 12 complete z-bins at M$_{\star}>$10$^{10.5}$~M$_{\star}$. Then we combined the remaining radio-detected SFGs with stacks of non-detections to compute the new L$_{1.4~GHz}$ in those bins, which should be free from AGN contamination. We verified that the new L$_{1.4~GHz}$ shifts the previously determined q$_{IR}$ (blue stars in Fig.~\ref{fig:qplot_binned0}) upward by a certain amount. In those complete bins, we fitted the AGN-corrected q$_{IR}$ with redshift, obtaining a significantly flatter relation than before, as shown in Fig.~\ref{fig:histo0_corr}. This suggests that the steeper redshift trend seen before (Sect.~\ref{qir_detections}) might be driven by radio AGN contamination, while the intrinsic redshift trend is significantly flatter, and possibly M$_{\star}$ invariant. 

To test the robustness of the newly derived q$_{IR}$--z trend, we again shift the q$_{IR}$ measurements of individual detections by the offset from such a trend at each z-bin, and perform a second q$_{IR}$ decomposition, as shown in Fig.~\ref{fig:histo0_corr}. The Gaussian fit that parametrizes star formation is nearly unchanged, with a dispersion of 0.21-0.22~dex in the two highest M$_{\star}$ bins. The 2$\sigma$ threshold below the peak is also very similar: 0.42 and 0.44~dex in the two bins (therefore we use an average $\Delta q_{AGN}$=0.43~dex). Moreover, the cumulative histograms (bottom panels) underline that this latter decomposition rejects about 81\% of radio AGN below the threshold, as opposed to 70\% estimated in the first step (see red open circles in Figs.~\ref{fig:histo0} and \ref{fig:histo0_corr}), while missing a comparable 3--4\% of SFGs. This confirms the effective improvement led by our re-calibration of the SF locus in removing radio AGN. 

\begin{figure*}
\centering
     \includegraphics[width=\linewidth]{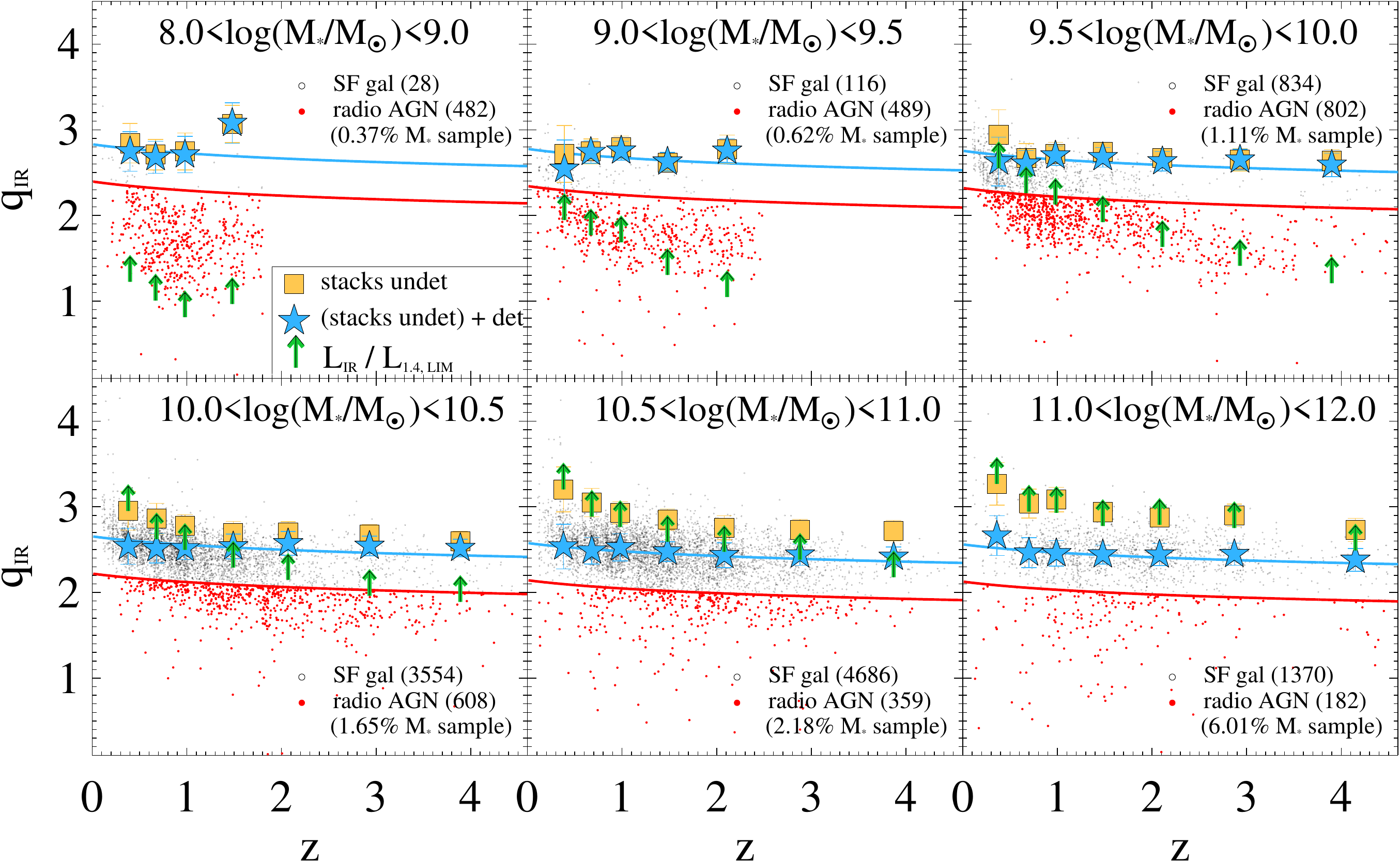}
 \caption{\small Distribution of q$_{IR}$ as a function of redshift and M$_{\star}$, after removing radio AGN (red dots). Symbols are the same of Fig.~\ref{fig:qplot_binned0}, except for the median q$_{IR}$ estimates (blue stars), which are here re-calculated after removing radio-excess AGN (red dots). Fractions of radio-detected AGN and SFGs are reported in each M$_{\star}$-bin, as well as the fraction of AGN relative to the full M$_{\star}$ sample analyzed in this work (in brackets). The blue and red solid lines denote the intrinsic IRRC of SFGs and the locus below which we classify radio sources as AGN (0.43~dex below the IRRC), respectively. 
 }
   \label{fig:qplot_binned1}
\end{figure*}
  
As shown in the updated Fig.~\ref{fig:qplot_binned1} (at M$_{\star}>$10$^{10.5}$~M$_{\odot}$), subtracting radio AGN (red dots) based on this latter locus shifts all the median q$_{IR}$ (blue stars) exactly on the fitted q$_{IR}$--z trend (blue solid lines). This agreement suggests that no further AGN subtraction is needed in those complete bins. Therefore, we can confidently assume that the new median q$_{IR}$ coincide with the instrinsic peak of the SF population. Given the robustness of our analysis, we compute a weighted-average redshift slope among the two highest M$_{\star}$ bins, by simply weighing each slope by the inverse square of its uncertainty. This way, we obtain an average slope of -0.055$\pm$0.018, i.e. not flat at a significance of 3$\sigma$. 

While at 10$^{11}<$M$_{\star}$/M$_{\odot}<$10$^{12}$ all z-bins (0.1$<$z$<$4.5) were used to constrain this trend, at 10$^{10.5}<$M$_{\star}$/M$_{\odot}<$10$^{11}$ we only used the first 5/7 z-bins (0.1$<$z$<$2.5). We now extrapolate the same relation also at 2.5$<$z$<$4.5, finding a good agreement with the median q$_{IR}$ estimates.

The resulting fractions of radio AGN identified in the two highest M$_{\star}$ bins should be quite representative of the overall incidence of radio AGN in these galaxies. This is suggested by the tightness of the SF Gaussian fit ($\sigma$$\sim$0.21--0.22~dex), that we interpret as the instrinsic scatter of the IRRC in these galaxies. Therefore, radio-undetected AGN that are not captured in our analysis, if any, are expected to be mostly composite (AGN+SF) radio sources whose total emission is predominantly arising from star formation processes.

It is worth noting that about 20\% radio AGN still lie within our clean sample of SFGs, as shown in Fig.~\ref{fig:histo0_corr}. As highlighted in \citet{Molnar+20}, while this high-q tail of AGN is SF-dominated in the radio, it could add to the intrinsic scatter of the underlying pure SFG sample. Therefore, our inferred scatter of 0.21--0.22~dex could be slightly overestimated (see e.g. 0.16~dex in \citealt{Molnar+20} for local SFGs), also due to larger uncertainties on L$_{1.4~GHz}$ and L$_{IR}$ than in the local Universe.

The fact that in both M$_{\star}$ bins the subtraction of radio-excess AGN leads to a \textit{flattening} of the q$_{IR}$--z trend might be also induced by a larger relative fraction of radio AGN with increasing redshift. As a sanity check, in both M$_{\star}$ bins we split and decomposed the q$_{IR}$ distribution of Fig.~\ref{fig:histo0_corr} separately at z$<$1.2 and z$>$1.2, examining the evolution of the relative fraction of radio AGN. Though we do find that radio AGN are slightly more prevalent at higher redshifts (i.e. on average from 12\% at z$<$1.2 to 18\% at z$>$1.2), we confirm that the dispersion of the SF population is redshift-invariant ($\sim$0.20~dex), both before and after removing radio AGN. This implies that the relative offset between the AGN and SF loci is unchanged, therefore our sample of $>$2$\sigma$ radio-excess AGN is globally preserved.

After removing those AGN, the flatter, yet declining q$_{IR}$ evolutionary trend could be explained by residual radio AGN activity \textit{within} the SF locus. We estimate the overall fraction of ``pure'' SFGs to be 95\% at z$<$1.2 and 90\% at z$>$1.2. Such minimal AGN contamination is probably more important at higher redshifts because SFGs are intrinsically IR brighter, so the radio-excess contrast (at fixed L$_{1.4~GHz}$) is less evident. Therefore, we argue that any further correction for mis-classified radio AGN would induce an even flatter q$_{IR}$ trend with redshift.

Finally, our approach leads to the following fractions of radio-excess AGN. At 10$^{10.5}<$M$_{\star}$/M$_{\odot}<$10$^{11}$, radio AGN are 7.1\% of all radio-detections and 2.2\% of the full M$_{\star}$ sample of SFGs. At 10$^{11}<$M$_{\star}$/M$_{\odot}<$10$^{12}$, radio AGN are 11.7\% of all radio-detections and 6.0\% of the full M$_{\star}$ sample of SFGs (see Table~\ref{tab:fagn}). These numbers are consistent with the known prevalence of radio AGN in the most massive galaxies (e.g. \citealt{Heckman+14}; \citealt{Hardcastle+20}). An increasing incidence of (X-ray) AGN activity with M$_{\star}$ has also been reported in recent studies (\citealt{Aird+19}; \citealt{Delvecchio+20}; \citealt{Carraro+20}), and possibly driven by the ability of the dark matter halo mass to regulate the amount of cold gas that trickles to the central black hole \citep{Delvecchio+19}.
  
Our empirical threshold identifies as radio-excess AGN sources with at least 63\% of the total radio emission arising from AGN activity. Therefore, radio sources with lower, yet substantial AGN contribution could still be mis-classified as radio-SFGs (e.g. \citealt{White+15}, \citealt{Wong+16}; \citealt{White+17}). We attempt at quantifying this fraction by comparing our classification against ancillary VLBA data in the COSMOS field (\citealt{HerreraRuiz+17}, \citeyear{HerreraRuiz+18}). This excellent dataset contains 468 VLBA sources detected at $>$5$\sigma$, targeted from a pre-selected sample of VLA-COSMOS 1.4~GHz sources at S/N$_{1.4}>$5.5 (\citealt{Schinnerer+10}, 2,864~sources). Since the brightness temperature reached by VLBA observations at about 0.01'' resolution exceeds 10$^6$~K, detections are most likely to be radio AGN \citep{HerreraRuiz+17}. Therefore, this sample provides an unambiguous method to test our source classification, though for a very tiny fraction of our sample with 1.4~GHz flux S$_{1.4}>$55~$\mu$Jy, typically hosted in massive galaxies (M$_{\star}>$10$^{10}$~M$_{\odot}$). Out of 13,510 3~GHz radio detections among our 37 bins, we found only 189 VLBA counterparts within 0.5'' search radius. A fraction as high as 90\% (170/189) were identified as ``radio-excess AGN'' based on our recursive approach. The remaining 10\% AGN mis-classified as SFGs from our approach are all IR-detected sources with typically high SFRs, which clearly reduces the apparent contrast between AGN- and SF-driven radio emission at arcsec scales. Although limited to a relatively bright and highly incomplete subsample, the comparison with VLBA data further demonstrates the reliability of our radio AGN identification method.

\begin{table*}
\centering
   \caption{Table summarizing the numbers and fractions of radio AGN and SFGs in different M$_{\star}$ bins, after fitting the AGN-corrected q$_{IR}$ with redshift (Sect.~\ref{agn_extrapolation} and Fig.~\ref{fig:qplot_binned1}). Columns are sorted as follows> (1) M$_{\star}$ range; (2) best-fit normalization of the q$_{IR}$--z trend, in the form q$_{IR}\propto$(1+z)$^{\gamma}$, by imposing $\gamma$=-0.055$\pm$0.018 as found in the two highest M$_{\star}$ bins (Sect.~\ref{ragn_highmass_recal}); (3,4) number of identified radio AGN and radio SFGs, respectively. In brackets we report their fractions relative to the radio-detected sample, and relative to the full M$_{\star}$ sample; (5) Number of M$_{\star}$-selected SFGs analyzed in this work. $^{(*)}$: calculated over four redshift bins (0.1$<$z$<$1.8). $^{(**)}$: calculated over five redshift bins (0.1$<$z$<$2.5).}
\begin{tabular}{lcccc }
\hline
\hline
  M$_{\star}$ (M$_{\odot}$) bin      &   q$_{IR}$--z fit  &     $\#$ radio AGN                                   &    $\#$ radio SFGs                                      &    $\#$ M$_{\star}$ sample  \\
                                     &     (normalization) &    (\%  radio-det, \% M$_{\star}$ sample)  &  (\% radio-det, \% M$_{\star}$ sample)    &    \\
        (1)       &     (2) &    (3)  &  (4)    &  (5)  \\
 \hline
 10$^{8}$--10$^{9}$~M$_{\odot}$      &     2.83$\pm$0.10$^{(*)}$     &  482 (94.5\%, 0.4\%)$^{(*)}$      &  28 (5.5\%, 0.0\%)$^{(*)}$    &   129,658$^{(*)}$    \\
 10$^{9}$--10$^{9.5}$~M$_{\odot}$    &     2.78$\pm$0.03$^{(**)}$    &  489 (80.8\%, 0.6\%)$^{(**)}$     &  116 (19.2\%, 0.1\%)$^{(**)}$  &   78,563$^{(**)}$     \\
 10$^{9.5}$--10$^{10}$~M$_{\odot}$   &     2.75$\pm$0.02             &  802 (51.0\%, 1.1\%)             &  834 (49.0\%, 1.2\%)     &   72,122     \\
 10$^{10}$--10$^{10.5}$~M$_{\odot}$  &     2.65$\pm$0.03             &  608 (14.6\%, 1.7\%)             &  3,554 (85.4\%, 9.6\%)   &   36,838     \\
 10$^{10.5}$--10$^{11}$~M$_{\odot}$  &     2.58$\pm$0.01             &  359 (7.1\%, 2.2\%)         &  4,686 (92.9\%, 28.4\%)       &   16,489     \\
 10$^{11}$--10$^{12}$~M$_{\odot}$    &     2.56$\pm$0.02             &  182 (11.7\%, 6.0\%)        &  1,370 (88.3\%, 44.8\%)       &   3,060      \\
\hline
\end{tabular}
\label{tab:fagn}
\end{table*}
%

\subsubsection{Extrapolating the SF-vs-AGN loci at low M$_{\star}$} \label{agn_extrapolation}

We extrapolate the q$_{IR}$--z trend of non-AGN galaxies calibrated in the previous Section towards less massive counterparts. As mentioned in Sect.~\ref{qir_detections}, 3~GHz detections placed at M$_{\star}<$10$^{10.5}$~M$_{\odot}$ are not representative of an M$_{\star}$-selected sample. In particular, a galaxy of a given M$_{\star}$ and redshift, with infrared luminosity L$_{IR}$ of a typical MS galaxy would likely fall below the 3~GHz detection limit, as indicated by the green arrows in Fig.~\ref{fig:qplot_binned0}. Radio detections at these masses are therefore quite peculiar relative to the overall galaxy population. 

This is further suggested in Fig.~\ref{fig:qplot_binned0} by the q$_{IR}$ offset between median measurements (blue stars) and individual radio detections (black dots). The latter lie systematically below the median q$_{IR}$, deviating more and more at lower M$_{\star}$. For these reasons, we refrain from calibrating the IRRC directly on those radio detections. We prefer to use the median q$_{IR}$ values as benchmark, since they should be sensitive to a more representative sample of galaxies of that M$_{\star}$. 

We proceed as follows. Within each M$_{\star}$ bin, the redshift trend of q$_{IR}$ is extrapolated from that calibrated at higher M$_{\star}$, in the form q$_{IR}\propto$(1+z)$^{-0.055\pm0.018}$ (Sect.~\ref{ragn_highmass_recal}). Only the normalization is left free to vary, in order to best fit the median q$_{IR}$. In other words, at M$_{\star}<$10$^{10.5}$~M$_{\odot}$, we \textit{assume} a constant q$_{IR}$--z slope. This approach is preferable to leaving also the slope as a free parameter, since the small number of bins is insufficient for us to constrain the redshift trend as accurately as previously done with single detections.  However, we stress that if we leave the slope free when fitting the q$_{IR}$ in each M$_{\star}$ bin, we always obtain slopes consistent between zero and -0.055, within 1$\sigma$ uncertainties.

Following the iterative approach already tested at higher M$_{\star}$, the best-fitting trend of q$_{IR}$ with redshift enables us to identify radio AGN as sources lying $>$0.43~dex below the best-fit SF locus. After subtracting those radio AGN, we re-calculate the weighted-average q$_{IR}$ and search again for the best normalization that fits the new AGN-corrected q$_{IR}$ measurements with redshift. We repeat this procedure twice, i.e. until all median q$_{IR}$ are unchanged within the uncertainties, at each M$_{\star}$. This condition sets the end of our recursion. 

The final, AGN-corrected q$_{IR}$ are shown in Fig.~\ref{fig:qplot_binned1} for all M$_{\star}$ bins (blue stars). This plot highlights the sample of radio-detected AGN that was removed (red dots) and the final SF locus (blue solid lines) that we eventually inferred after subtracting those AGN. The numbers of radio-detected AGN and SFGs are reported in each panel for convenience.

In most bins at M$_{\star}<$10$^{10.5}$~M$_{\odot}$, the AGN-corrected q$_{IR}$ measurements nearly coincide with those obtained from stacking non-detections alone (yellow squares). These latter values delimit the highest q$_{IR}$ that could be reached if removing, by definition, all radio detections. The result of similarity between the two sets of q$_{IR}$ measurements is due to a heavy subtraction of radio AGN from the sample of radio detections. Within the sample of radio detections, the fraction of radio AGN identified at M$_{\star}<$10$^{10.5}$~M$_{\odot}$ \textit{increases} with \textit{decreasing} M$_{\star}$. From the first to the fourth M$_{\star}$ bin, these fractions are: 94.5\%, 80.8\%, 51.0\% and 14.6\%, respectively. However, when compared to the size of our full M$_{\star}$ sample in each bin, they drop to (in the same order): 0.4\%, 0.6\%, 1.1\% and 1.7\%, respectively (see Table~\ref{tab:fagn}). These latter numbers are consistent with a decreasing incidence of radio AGN towards lower M$_{\star}$ systems, following the trend constrained at M$_{\star}>$10$^{10.5}$~M$_{\odot}$ (Sect.~\ref{ragn_highmass_recal}). Nevertheless, according to our analysis the vast majority of radio-detected dwarf galaxies (M$_{\star}<$10$^{9.5}$~M$_{\odot}$, e.g. \citealt{Mezcua17}) in COSMOS are expected to be radio AGN.

Bearing this in mind, we note that the weighted average q$_{IR}$ (blue stars) are yet mostly driven by non-detections (yellow squares), which outnumber individual detections (dots) by a factor of $>$100 at M$_{\star}<$10$^{9.5}$~M$_{\odot}$. However, those few radio detections (mostly radio AGN, red dots) stand out from the stacks of non-detections (yellow squares) typically by over a factor of ten, up to one-hundred. As a consequence, after removing radio AGN at M$_{\star}<$10$^{9.5}$~M$_{\odot}$, the new average q$_{IR}$ still move upward by 0.2--0.3~dex.

\begin{figure}
\centering
     \includegraphics[width=\linewidth]{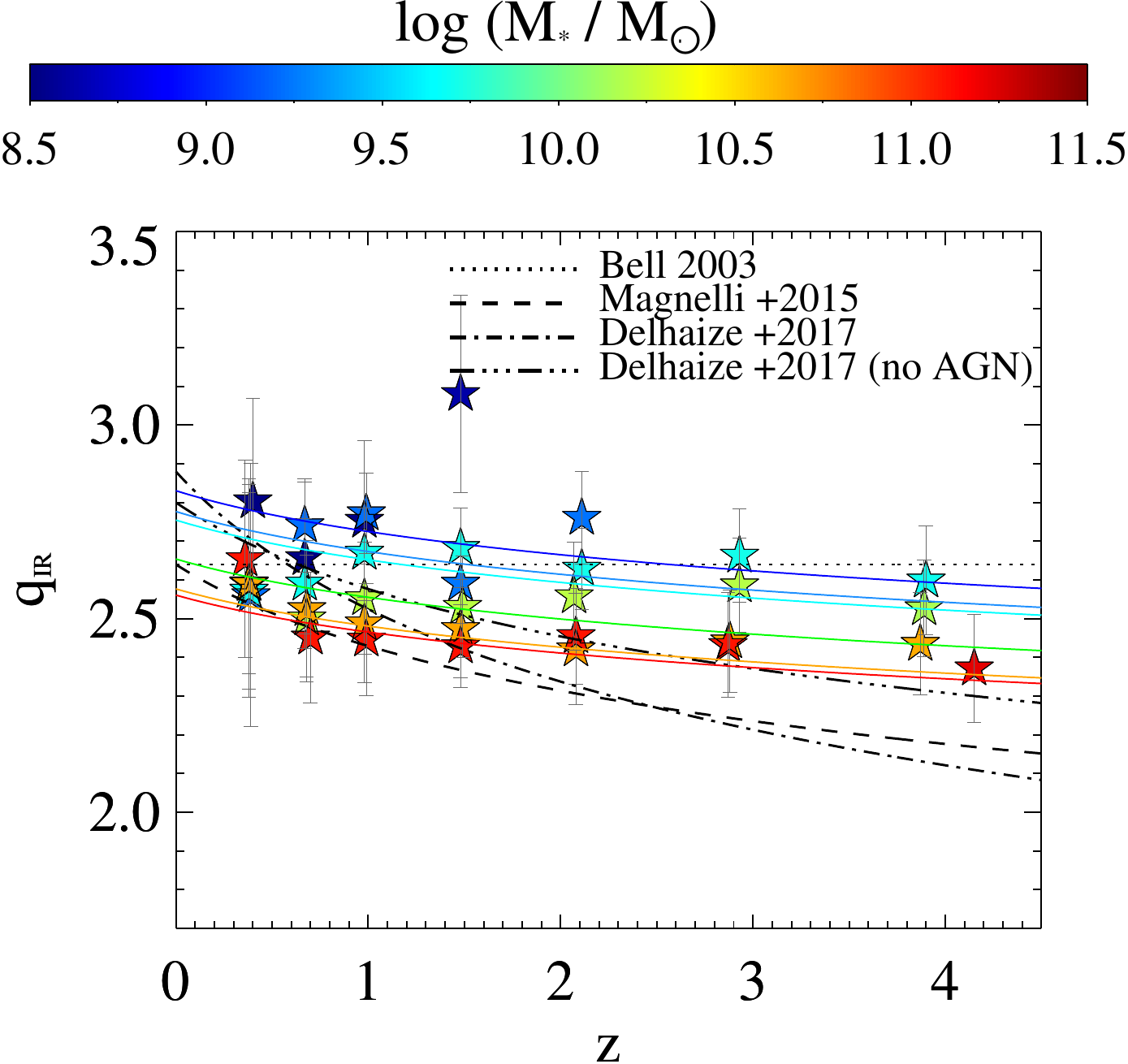}
 \caption{\small Intrinsic (i.e. AGN-corrected) q$_{IR}$ evolution as a function of redshift (x-axis) and M$_{\star}$ (colour bar). The L$_{IR}$ estimates are the same reported in Fig.~\ref{fig:qplot0}, while L$_{1.4~GHz}$ measurements have been re-calculated after excluding radio-detected AGN (Sect.~\ref{agn_removal}). For comparison, other IRRC trends with redshift are taken from the literature (black lines): \citeauthor{Bell03} (\citeyear{Bell03}, dotted); \citeauthor{Magnelli+15} (\citeyear{Magnelli+15}, dashed); \citeauthor{Delhaize+17} (\citeyear{Delhaize+17}, dot-dashed) and their AGN-corrected version after removing 2$\sigma$ outliers (triple dot-dashed lines).  
 }
   \label{fig:qplot1}
\end{figure}

\subsection{The intrinsic IRRC evolves primarily with M$_{\star}$} \label{q_mass_z}

After correcting our combined L$_{1.4~GHz}$ measurements for radio AGN contamination, we are finally able to examine the evolution of the intrinsic IRRC as a function of M$_{\star}$ and redshift, as presented in Fig.~\ref{fig:qplot1}. For each M$_{\star}$ bin, we show the best-fit power-law trend, whose slope was directly inferred in Sect.~\ref{ragn_highmass_recal} in the two highest M$_{\star}$ bins (i.e. -0.055$\pm$0.018). We verified that our median L$_{IR}$ estimates are, instead, totally unchanged after removing radio-excess AGN, as expected given their minimal fraction relative to the parent M$_{\star}$-selected SFG sample.

The colour bar highlights a clear stratification of q$_{IR}$ with M$_{\star}$, with more massive galaxies showing systematically lower q$_{IR}$ values. This behaviour was already seen in Fig.~\ref{fig:qplot0} before removing radio AGN, but here it suggests that some additional mechanisms unrelated to AGN activity might be boosting (reducing) radio emission in more (less) massive systems, relative to the IR.

For comparison, other IRRC trends with redshift are reported from \citeauthor{Bell03} (\citeyear{Bell03}, dotted line), \citeauthor{Magnelli+15} (\citeyear{Magnelli+15}, dashed line) and \citeauthor{Delhaize+17} (\citeyear{Delhaize+17}, dot-dashed line). Since \citet{Delhaize+17} did not remove radio-excess AGN, we also show their AGN-corrected relation by removing 2$\sigma$ outliers (as reported in \citealt{Delvecchio+18}): q$_{IR}$$\propto$(1+z)$^{-0.12\pm0.01}$ (triple dot-dashed line). This trend is flatter than the previous one, more consistent with that of \citet{Magnelli+15} and more appropriate for a comparison with our approach.

In the following, we examine the significance of the M$_{\star}$ dependence at fixed redshift, and we provide a multi-parametric fit as a function of both parameters.
%
%
\begin{figure}
\centering
     \includegraphics[width=\linewidth]{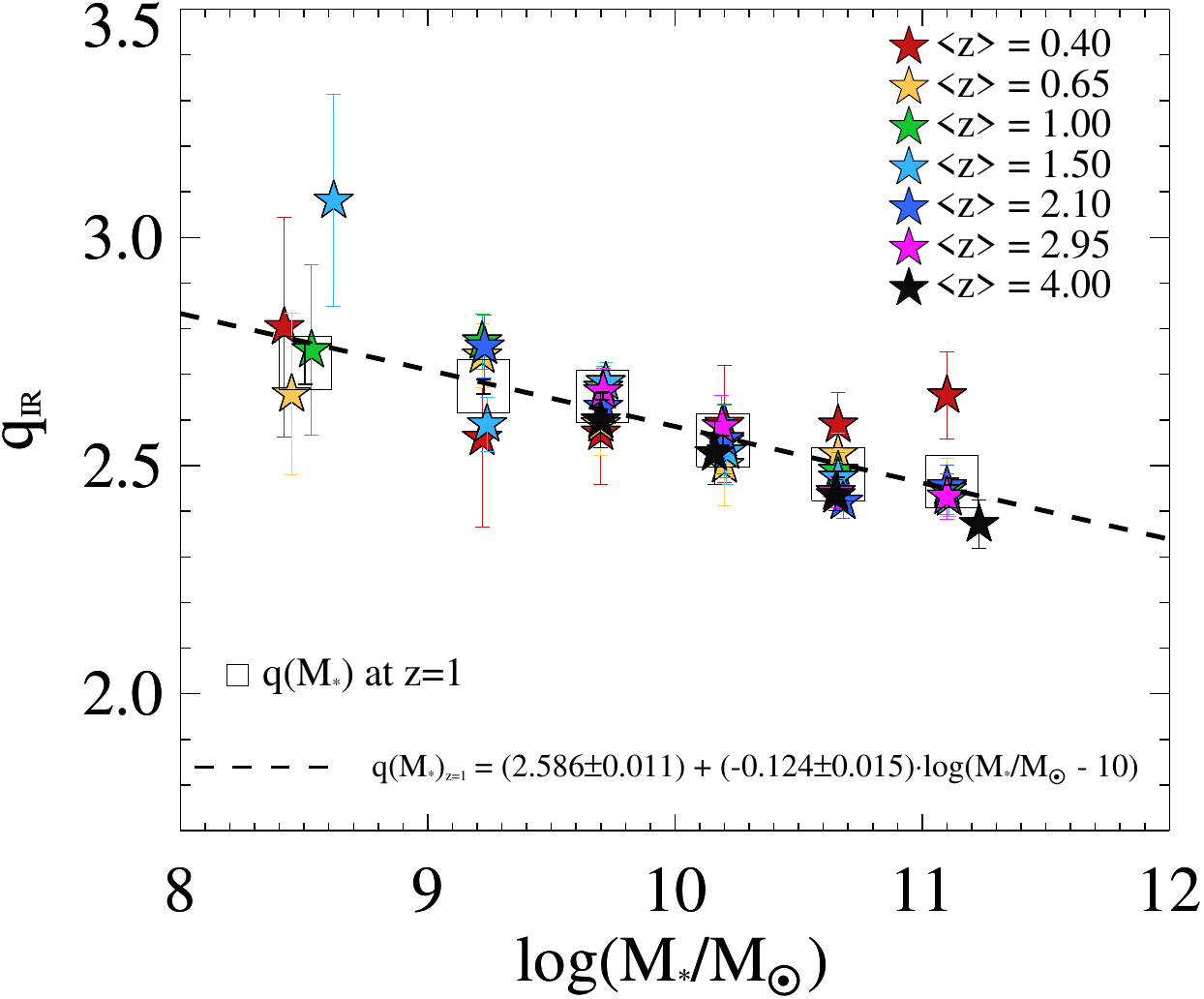}
 \caption{\small Distribution of AGN-corrected q$_{IR}$ as a function M$_{\star}$, colour-coded by redshift (stars). At each M$_{\star}$, open squares indicate the median q$_{IR}$ values at z=1, obtained after propagating the uncertainties of slope and normalization of the corresponding q$_{IR}$--z fit and interpolating each at z=1. These values were fitted with a linear function in $\log$--$\log$ space (black dashed line).
 }
   \label{fig:qmass}
\end{figure}

Fig.~\ref{fig:qmass} shows the equivalent of Fig.~\ref{fig:qplot1} but projected on M$_{\star}$, with redshift bins in different colours. Error bars on each q$_{IR}$ were re-scaled by a factor of $\sqrt \chi^{2}_{red}$ in each M$_{\star}$ bin, to bring the reduced $\chi^2$ of the corresponding q$_{IR}$--z fit to unity. It is quite evident that an M$_{\star}$ dependence reduces substantially the scatter of the average q$_{IR}$ around a single trend. To better quantify this, first we bootstrapped over the uncertainties of slope and normalization obtained from each q$_{IR}$--z trend (see Table~\ref{tab:fagn}). Then, at each M$_{\star}$, we interpolated the full range of bootstrapped IRRCs at z=1, in correspondence of the 16$^{th}$, 50$^{th}$ and 84$^{th}$ percentiles. Interpolating at z=1, besides being at roughly the median redshift of our sample, reduces the increasing divergence of each q$_{IR}$--z fit at lower or higher redshifts. This leaves us with the interpolated median q$_{IR}$(z=1) as a function of M$_{\star}$ (black open squares). Error bars indicate the uncertainty on the median value. The black dashed line marks the corresponding linear best-fit trend: q(M$_{\star}$)$_{z=1} \propto$ (-0.124$\pm$0.015)$\times$$\log$(M$_{\star}$/M$_{\odot}$-10). This function yields a $\chi^{2}_{red}$=0.87, with an M$_{\star}$ slope close to that commonly found when fitting q$_{IR}$ as a function of redshift (e.g. \citealt{Magnelli+15}), and significant at over 8$\sigma$. Though the interpolated fit at z=1 is purely indicative, this check suggests that M$_{\star}$ might be the primary driver of the evolution of the IRRC across redshift.

Moreover, in order to incorporate the dependence of the IRRC on both M$_{\star}$ and redshift \textit{simultaneously}, we performed a multi-parametric fit in the 3-dimensional q$_{IR}$--M$_{\star}$--z space. This yields the following best-fit expression: 
\begin{equation}
q_{IR} (M_{\star},z) = (2.646\pm0.024) \times A^{(-0.023 \pm  0.008)} - B\times(0.148\pm0.013)
   \label{eq:bestq}
\end{equation}
where A=(1+z) and B=$(\log M_{\star}$/M$_{\odot}-10)$. The corresponding $\chi^{2}_{red}$=0.90. The best-fit slopes with redshift and M$_{\star}$ are significant at 2.9$\sigma$ and 11$\sigma$ levels, respectively. This further strengthens the need for a primary M$_{\star}$ dependence, followed by a weaker and less significant redshift dependence. These numbers and confidence levels refer to the median trend. However, we acknowledge that, if assuming a constant IRRC scatter of 0.21--0.22~dex across all M$_{\star}$ galaxies, the weak co-dependence on redshift could be easily washed out. This dilution might also hide a mildly increasing redshift trend, which could be expected by Inverse Compton cooling of cosmic ray electrons \citep{Murphy09}. Nevertheless, the main argument of our analysis is to demonstrate how previously reported \textit{best-fitting} IRRC trends with redshift are likely a red herring, whereas the M$_{\star}$ (or a related proxy) is a better predictor of the average IRRC in SFGs. 

The need for an additional M$_{\star}$-dependence of the IRRC (in the form of L$_{radio}$--SFR) has been also highlighted in previous low-z studies (\citealt{Gurkan+18}; \citealt{Read+18}) and recently extended out to z$\sim$1 \citep{Smith+20} using deep LOFAR-150~MHz data. A similar conclusion was indepedently reached in \citet{Molnar+20}, when considering the dependence of the IRRC on galaxy spectral radio luminosity. To mitigate selection effects, they exploited a depth-matched sample of SFGs at z$<$0.2. After performing a radio decomposition analysis in different bins of L$_{1.4~GHz}$, \citet{Molnar+20} report that q$_{IR}$ decreases with increasing L$_{1.4~GHz}$. Assuming that radio emission comes predominantly from star formation, this is in line with our inferred M$_{\star}$ dependence, since more massive SFGs are also brighter in radio \citep{Leslie+20}. This further corroborates the idea that the IRRC varies across different types of galaxies, at fixed redshift (but see e.g. \citealt{Pannella+15} for an alternative interpretation). Therefore, we conclude that our results are in qualitative agreement with \citet{Molnar+20}, who also demonstrate the implications of such a non-linearity for decreasing q$_{IR}$ vs. z trends in the literature.

\section{Discussion} \label{discussion}

The main result of this work is the finding that the IRRC primarily evolves with M$_{\star}$, and only weakly with redshift (Eq.~\ref{eq:bestq}). While the M$_{\star}$ dependence has not been explored in detail so far, except in the local Universe (e.g. \citealt{Gurkan+18}, see Sect.~\ref{origin}), for several years much effort has been devoted in understanding the mild, but significant decline of the IRRC with redshift from both an observational (e.g. \citealt{Murphy09}; \citealt{Ivison+10a}; \citealt{Sargent+10}; \citealt{Magnelli+15}; \citealt{Delhaize+17}; \citealt{CalistroRivera+17}; \citealt{Molnar+18}) and a theoretical (\citealt{Lacki+10b}; \citealt{Schleicher+13}; \citealt{Schober+16}; \citealt{Bonaldi+19}) perspective. In Appendix~\ref{Appendix_comparison} we expand on the role played by various assumptions in deriving different IRRC trends presented in the literature. In this Section, instead, we interpret our results and discuss the many implications of our findings in the context of the origin and evolution of the IRRC. In particular, we split the discussion in several sections, each focusing on a specific issue. First, we explore some physical interpretations of the origin of an M$_{\star}$ and redshift-dependent IRRC (Sect.~\ref{origin}). We further investigate the possible evolution of the IRRC above the MS (Sect.~\ref{q_sb}). A discussion on the incidence of AGN activity is also presented (Sect.~\ref{dwarves}). Finally, we comment on the use of radio emission as a SFR tracer in the light of our results (Sect.~\ref{qsfr}).

\subsection{What drives the primary M$_{\star}$ dependence?} \label{origin}

Our main finding is that the IRRC decreases primarily with M$_{\star}$, and only weakly with redshift. In particular, within the range 10$^{9}<$M$_{\star}$/M$_{\odot}<$10$^{12}$, the median q$_{IR}$ decreases by 0.25~dex (a factor of 1.8), at fixed redshift, and with high significance ($\sim$10$\sigma$, see Eq.~\ref{eq:bestq}). This suggests that the dependence L$_{1.4~GHz}$--M$_{\star}$ is steeper than the dependence L$_{IR}$--M$_{\star}$ (i.e. the MS). To translate this result into the corresponding IR-radio slope, we take our best q$_{IR}$--M$_{\star}$ relation (Eq.~\ref{eq:bestq}) at fixed redshift, and assume for simplicity a linear MS between M$_{\star}$ and SFR (i.e. L$_{IR}$). This yields L$_{IR}\propto$L$_{1.4~GHz}^{0.90}$. In the past years, the deviation from a linear trend has been gaining increasing momentum, due to several studies finding a similar sub-linear behaviour in the local Universe (L$_{IR}\propto$L$_{1.4~GHz}^{0.75-0.90}$, \citealt{Bell03}; \citealt{Hodge+08}; \citealt{Davies+17}; \citealt{Brown+17}; \citealt{Gurkan+18}; \citealt{Molnar+20}). This might challenge the idea of calibrating radio emission as a universal SFR tracer, as we discuss later in Sect.~\ref{qsfr}.

Here we explore some physical parameters behind this non-linearity, that might induce an M$_{\star}$-evolving q$_{IR}$ similar to our findings. First, we discuss the possible role of a top-heavy IMF. Later, we test some radio synchrotron models (e.g. \citealt{Lacki+10b}) by studying the relation between q$_{IR}$ and SFR surface density.

\subsubsection{The role of the IMF}

We quantify whether a deviation from a canonical IMF slope (e.g. \citealt{Chabrier03}; n(M)$\propto$M$^{-2.35}$ at 0.8$<$M$<$100~M$_{\odot}$) could justify an M$_{\star}$-decreasing q$_{IR}$. In particular, we note that reprocessed IR light comes predominantly from stars with M$>$5M$_{\odot}$, while radio synchrotron emission comes from more massive stars with M$>$8M$_{\odot}$. We check whether a systematically flatter IMF in more massive galaxies could explain the observed decreasing q$_{IR}$. 

A top-heavy IMF has been directly constrained only in massive early-type galaxies at z$\sim$0 \citep{Cappellari+12} from the comparison between dynamical masses and optical light, but only proposed or indirectly inferred otherwise (e.g. \citealt{Baugh+05}; \citealt{Hopkins+06}; \citealt{Dave+08}; \citealt{vanDokkum08}; \citealt{Dabringhausen+09}). To quantify the change of q$_{IR}$ as a function of IMF slope, we integrate the IMF over the ranges 5-100~$M_{\odot}$ and 8-100~$M_{\odot}$, with varying IMF slope. The ratio between the two integrals is somewhat proportional to L$_{IR}$/L$_{1.4~GHz}$. However, we find only 8\% variation of the integral ratio across the full range of slopes [-2.35, 0] , as compared to 80\% (i.e. 0.25~dex) q$_{IR}$ variation across all M$_{\star}$. In line with the conclusions of \citet{Murphy09}, we argue that a top-heavy IMF in the most massive galaxies proves insufficient to explain the evolving q$_{IR}$ with M$_{\star}$.

\begin{figure}
\centering
     \includegraphics[width=\linewidth]{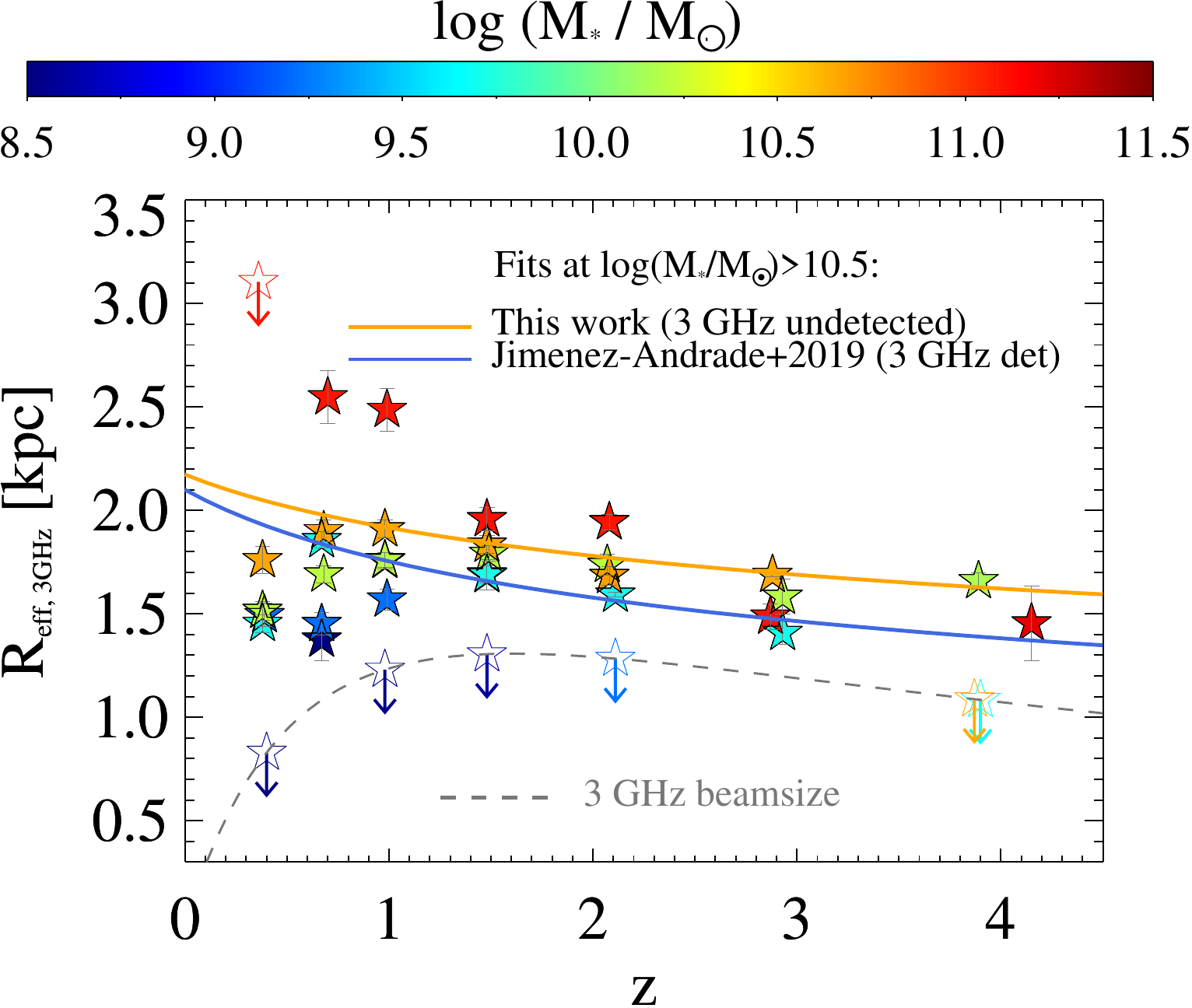}
 \caption{\small Distribution of 3~GHz effective radius (in kpc) as a function of redshift and colour-coded by M$_{\star}$ (stars). Size measurements are taken from median stacked 3~GHz images of non-detections. Upper limits are given for unresolved stacks and correspond to the angular 3~GHz beamsize (0.75'', grey dashed line). We observe a clear increase of R$_e$ with galaxy M$_{\star}$. The bins at M$_{\star}>$10$^{10.5}$~M$_{\odot}$ with resolved emission are fitted with a power-law redshift trend, which yields R$_e \propto$(1+z)$^{-0.18\pm0.07}$ (orange solid line). A comparison study by \citet{JimenezAndrade+19} is shown (blue solid line) for 3~GHz detected SFGs at similar M$_{\star}$ in COSMOS, obtaining R$_e \propto$(1+z)$^{-0.26\pm0.08}$. 
 }
   \label{fig:sizes}
\end{figure}

\subsubsection{The role of the SFR surface density ($\Sigma_{SFR}$)} \label{q_sigma_sfr}

The model proposed by \citet{Schleicher+13} postulates that the magnetic field strength scales with SFR, boosting radio synchrotron emission during shocks or galaxy interactions (e.g. \citealt{Donevski+15}; \citealt{Tabatabaei+17}). Because of the MS relation, this model implicitly predicts a net enhancement of radio emission with increasing M$_{\star}$, as well as an increase with redshift due to higher gas density in galaxies. Related to this, the single-zone galaxy model of \citet{Lacki+10b}, which includes a detailed CR description, suggests a variation of q$_{IR}$ as a function of SFR surface density ($\Sigma_{SFR}$), from ``normal galaxies'' ($\Sigma_{SFR}$$\lesssim$0.06~M$_{\odot}$~yr$^{-1}$~kpc$^{-2}$) to ``starbursts'' (SBs, $\Sigma_{SFR}$$\gtrsim$2--4~M$_{\odot}$~yr$^{-1}$~kpc$^{-2}$). In particular, this model argues that q$_{IR}$ slightly declines with $\Sigma_{SFR}$ (up to $\Sigma_{SFR}$$\sim$1~M$_{\odot}$~yr$^{-1}$~kpc$^{-2}$) due to the escape of CRe, generating a radio dimming especially in lower M$_{\star}$ (or smaller size) galaxies. This effect is also expected to become more pronounced with redshift, due to IC scattering off the CMB that is expected to dominate over synchrotron cooling \citep{Murphy09}. Conversely, at $\Sigma_{SFR}$$\gtrsim$1~M$_{\odot}$~yr$^{-1}$~kpc$^{-2}$, \citet{Lacki+10b} invoke a ``conspiracy'' of ionization losses to balance spectral ageing, and additional synchrotron emission from secondary CRs, that together flatten q$_{IR}$ with $\Sigma_{SFR}$ and redshift.

Here we test the above models by relating q$_{IR}$ and average $\Sigma_{SFR}$ measured in this work. These estimates were obtained by using the total SFR$_{IR+UV}$ calculated from IR stacking and adding the dust-uncorrected UV contribution (Sect.~\ref{sed_fitting}). Galaxy sizes are drawn from median radio stacking of non-detections, carried out in Sect.~\ref{radio_stacking} at each M$_{\star}$--z bin via 2D elliptical Gaussian fitting. Though these measurements do not include the contribution of single 3~GHz detections, they come from about 97\% of all M$_{\star}$-selected galaxies in our sample, hence they should be statistically representative of their average radio properties. This approach implicitly assumes that radio emission encloses the total star formation of the host, that is quite plausible especially in high-M$_{\star}$ galaxies, where the dominant obscured SF traced by IR is also seen in the radio (e.g. \citealt{JimenezAndrade+19}). To scale angular sizes $\theta_{FWHM}$ into effective radius (R$_e$, enclosing half of the total flux density), we assume that our galaxies follow a disk-like surface brightness profile (S\'ersic index $n$=1), as found for MS galaxies (e.g., \citealt{Nelson+16}). Under this assumption, the major-axis R$_{e, maj}$ can be calculated as R$_{e, maj}$ = $\theta_{FWHM}$ / 2.43 \citep{Murphy+17}. Lastly, we take the circularized radius R$_e$ = R$_{e, maj}$ / $\sqrt{A_r}$, where $A_r$ is the axial ratio. 

\begin{figure}
\centering
     \includegraphics[width=\linewidth]{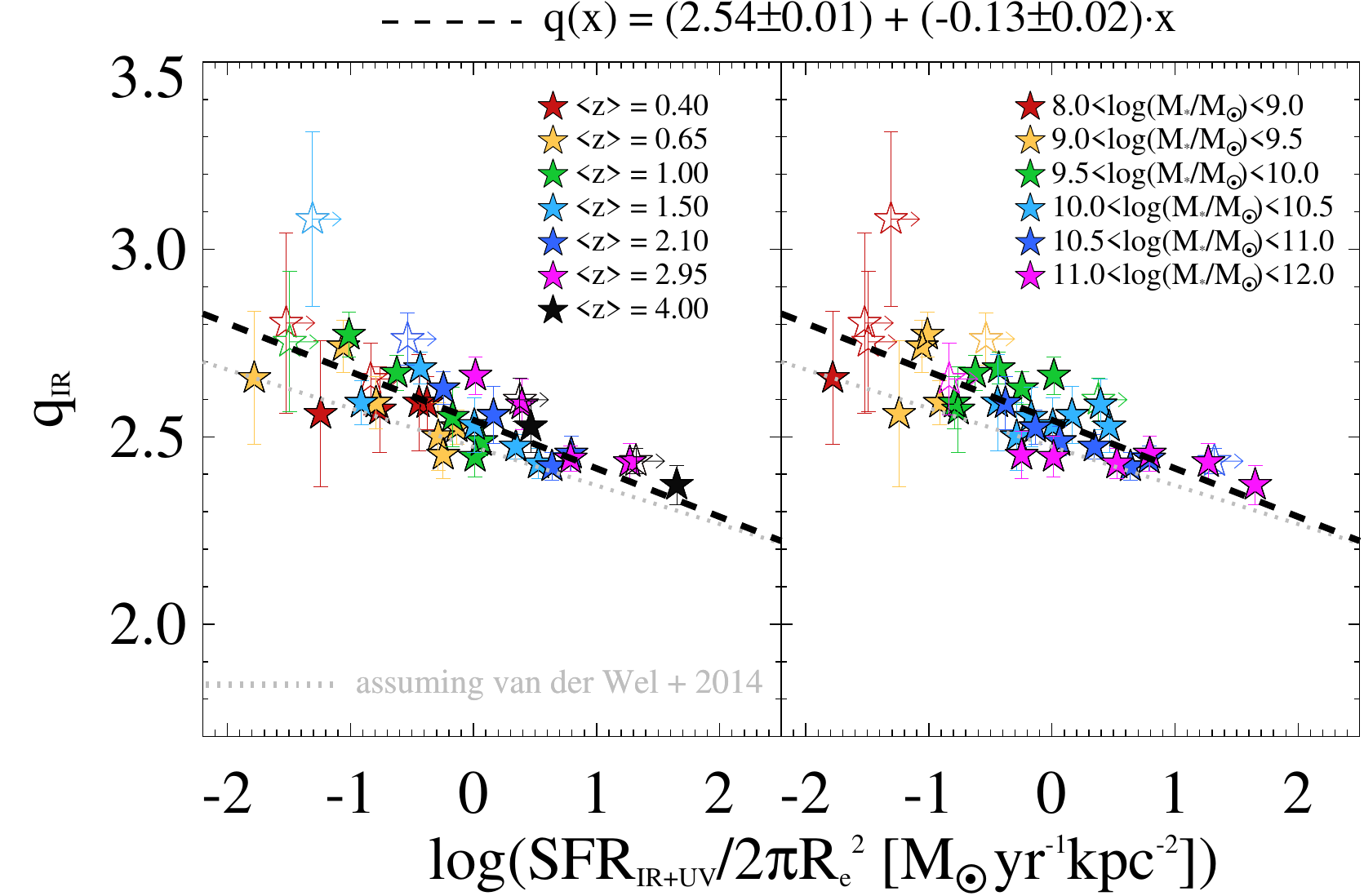}
 \caption{\small Evolution of q$_{IR}$ as a function of SFR surface density ($\Sigma_{SFR}$=SFR$_{IR+UV}$/2$\pi$R$_e^2$), colour coded by redshift (left panel) and by M$_{\star}$ (right panel). The SFR$_{IR+UV}$ estimates are taken from Sect.~\ref{sed_fitting}, while the effective radius $R_e$ is measured from stacked 3~GHz images via 2D Gaussian fitting. This plot shows a significant anti-correlation similar in slope to that observed with M$_{\star}$ (Sect.~\ref{q_mass_z}), marked by the black dashed line (q$_{IR} \propto$ (-0.13$\pm$0.02)$\times \log{\Sigma_{SFR}}$). For comparison, the best-fit trend with rest-frame optical (5000\AA) sizes estimated from \citet{vanderWel+14} scaling relation is shown (grey dotted line). 
 }
   \label{fig:qphys}
\end{figure}

Fig.~\ref{fig:sizes} displays our median stacked 3~GHz size measurements (or upper limits) as a function of redshift and M$_{\star}$. Error bars are obtained from the IDL routine {\sc mpfit2dfun}. Upper limits are shown for unresolved stacks and correspond to the angular 3~GHz beamsize (0.75'', grey dashed line), except for the highest M$_{\star}$ bin at z$<$0.5 that was convolved with a Gaussian kernel of 3''~FWHM (see Appendix~\ref{Appendix_ancillary}). Our R$_e$ measurements are well consistent with VLA 3~GHz sizes indipendently derived in the recent study of \citet{JimenezAndrade+19}. The authors used the same VLA~3~GHz COSMOS images to construct a M$_{\star}$-complete sample of 3,184 radio-detected SFGs with M$_{\star}>$10$^{10.5}$~M$_{\odot}$, most of which lie around the MS relation \citep{Schreiber+15}. The best-fitting R$_e$ trend with redshift reported by \citet{JimenezAndrade+19} for MS galaxies (blue solid line, R$_e \propto$(1+z)$^{-0.26\pm0.08}$) is broadly consistent with our evolutionary trend based on median 3~GHz stacks (orange solid line, R$_e \propto$(1+z)$^{-0.18\pm0.07}$). Our slightly larger size measurements are likely due to radio-detected SFGs \citep{JimenezAndrade+19} having a more centrally peaked surface brightness compared to our stacks \citep{Bondi+18}.

We calculate $\Sigma_{SFR}$ = SFR$_{IR+UV}$/2$\pi$R$_e^2$ (see e.g. \citealt{JimenezAndrade+19}) and show its relation with q$_{IR}$ in Fig.~\ref{fig:qphys}, colour-coded by redshift (left panel) and M$_{\star}$ (right panel). Empty symbols highlight 7/37 bins with unresolved 3~GHz stacked emission, which translates into a lower limit in $\Sigma_{SFR}$. By fitting only the 30 q$_{IR}$ and $\Sigma_{SFR}$ measurements, we obtain a significant anti-correlation similar in slope to that observed with M$_{\star}$ (Sect.~\ref{q_mass_z}), marked by the black dashed line (q$_{IR}\propto$~(-0.13$\pm$0.02)$\times \log{\Sigma_{SFR}}$). From the M$_{\star}$ and redshift of each bin, we obtain a surrogate trend with rest-frame optical (5000\AA) sizes estimated indirectly from the \citet{vanderWel+14} scaling relation for SFGs (grey dotted line). We note that the mild difference between the two latter trends originates from the 2$\times$ smaller radio sizes compared to rest-frame optical sizes \citep{Bondi+18}.

Since more massive galaxies are characterized by more compact star formation \citep{Elbaz+11}, the decreasing q$_{IR}$--$\Sigma_{SFR}$ trend is linked to that with M$_{\star}$. Nevertheless, unlike the trend with optical sizes, our $\Sigma_{SFR}$ measurements are not bound to M$_{\star}$ \textit{by construction}, but rather measured from independent tracers (IR+UV and 3~GHz data). We thus stress that our proposed q$_{IR}$--$\Sigma_{SFR}$ dependence is simply meant to be a more physical proxy for the observed M$_{\star}$ dependence. At fixed M$_{\star}$, the SFR surface density increases with redshift (left panel of Fig.~\ref{fig:qphys}), in qualitative agreement with our (weakly) decreasing q$_{IR}$ trend.

Both the slope and significance of the q$_{IR}$--$\Sigma_{SFR}$ relation are consistent with those found between q$_{IR}$ and M$_{\star}$ (Sect.~\ref{q_mass_z}). We argue that the declining q$_{IR}$--$\Sigma_{SFR}$ slope is primarily driven by the SFR, and only weakly by radio sizes. Indeed, at fixed redshift, the SFR(IR+UV) increases along the MS by a factor $>$30 from 10$^9$ to 10$^{11}$~M$_{\odot}$ (Fig.~\ref{fig:ms}), while R$_e^2$ only increases by a factor of 1.5--2.5 in the same interval. Though this is not a conclusive evidence, our analysis seems to suggest that the larger SFR per unit area in more massive (and higher-z) galaxies might be driving the sub-linear behaviour of the IRRC.

The comparison with the model of \citet{Lacki+10b} comes with a few caveats. First, we take a fixed spectral index $\alpha$=-0.75 for all galaxies (which is supported by radio ancillary data in Appendix~\ref{Appendix_ancillary}), while \citet{Lacki+10b} model a curved radio spectrum. Second, we label as ``SBs'' those galaxies that lie $>$4$\times$ above the MS (Sect.~\ref{q_sb}), while \citet{Lacki+10b} identify them as $\Sigma_{SFR}\gtrsim$2--4~M$_{\odot}$~yr$^{-1}$~kpc$^{-2}$. As we show in Fig.~\ref{fig:qphys}, the vast majority of our MS galaxies has $\Sigma_{SFR}$ below the ``SB'' threshold of \citet{Lacki+10b}.

That being said, their model predicts a decreasing q$_{IR}$ with $\Sigma_{SFR}$ (see their Fig. 1), that steepens with redshift, then followed by a flattening (or a reversal) at $\Sigma_{SFR}\gtrsim$1~M$_{\odot}~yr^{-1}~kpc^{-2}$. This behaviour is not clearly seen in our data, that instead display a smoothly declining q$_{IR}$--$\Sigma_{SFR}$ trend, and nearly redshift-invariant. Our trend is consistent with the low-q values recently inferred by \citet{Algera+20b} in compact (R$_e \sim$1~kpc) and massive (M$_{\star}>$10$^{10}$~M$_{\odot}$) sub-millimetre galaxies at 1.5$<$z$<$3.5. Indeed, their average q$_{IR}$=2.20$\pm$0.03 lies close to the extrapolation of our best-fit q$_{IR}$--$\Sigma_{SFR}$ trend at $\Sigma_{SFR}\sim$100~M$_{\odot}~yr^{-1}~kpc^{-2}$, thus further corroborating the relation between q$_{IR}$ and SFR per unit area in SFGs.

To explain low-q$_{IR}$ SBs, a further fine-tuning in the model of \citet{Lacki+10b} is to invoke ``puffy SBs'' with larger disk scale height ($h$=1~kpc, i.e. SMG-like) than canonical ``compact SBs'' ($h$=100~pc, i.e. ULIRG-like). Indeed, in puffy SBs, CRe can travel longer distances before escaping the galaxy, creating secondary hadrons that induce an extra boost of radio emission. However, this process should globally steepen the observed radio spectra ($\alpha$$\sim$-0.9:-1.0), due to bremsstrahlung and ionization losses being weak with respect to synchrotron and IC losses. This prediction is again not confirmed by our data (Fig.~\ref{fig:radio_lcomp}). 

Another speculative hypothesis could be linked to the amplification of the magnetic field strength at higher SFRs, that boosts radio emission in more massive galaxies along the MS (e.g. \citealt{Tabatabaei+17}), though a fine-tuned balance between concomitant CRe losses and secondary CRe production is also required \citep{Algera+20b}. 

In summary, our checks cannot firmly elucidate the main physical driver of the IRRC with M$_{\star}$, but they seem to support an empirical link between q$_{IR}$ and SFR surface density. Our findings do not seem to follow the q$_{IR}$ flattening or spectral index variations with $\Sigma_{SFR}$ predicted by models (e.g. \citealt{Lacki+10b}; \citealt{Schleicher+13}). Of course, our data do not have enough statistical power to discern all the underlying physical mechanisms and spectral variations that the model obviously addresses. We postulate that a more detailed data-vs-model comparison would require depth-matched observations at multiple radio frequencies of massive compact galaxies.

\subsection{Does the IRRC evolve above the MS?} \label{q_sb}

We investigate the behaviour of the average q$_{IR}$ above the MS. This is important to test whether radio emission follows a similar enhancement as L$_{IR}$ when moving above the MS, or instead q$_{IR}$ is not a good tracer of starburstiness (i.e. offset from the MS). This issue is still highly debated. For instance, \citet{Condon+91} found that the most extreme ULIRGs at z$\sim$0 have higher q$_{IR}$ and larger scatter compared to the MS population, which can be explained by flatter radio spectra due to free-free absorption (see also \citealt{Murphy+13}). On a different note, \citet{Helou+85} and \citet{Yun+01} do not report any significant deviation of q$_{IR}$ in local SB galaxies, though they also observed a larger scatter for this population. More recently, \citet{Magnelli+15} found a mild (+0.2~dex) enhancement of q$_{FIR}$ above the MS, though not significant. Such apparent tension is probably also due to different definitions of ``starburst'' galaxies and different sample selections. 

For sake of consistency with \citet{Magnelli+15}, in this Section we define ``SBs'' as galaxies with SFR$>$4$\times$SFR$_{MS}$ (e.g. \citealt{Rodighiero+11}), where SFR$_{MS}$ corresponds to the SFR predicted by the MS \citep{Schreiber+15}, at each M$_{\star}$ and redshift. Our measured SFR estimates come from IR+UV, as described in Sect.~\ref{sed_fitting}. However, following \citet{Carraro+20}, we select as SBs only \textit{individually} IR-detected galaxies (S/N$_{IR}>$3) that meet the above criterion. This is because our stacked SFR$_{IR}$ estimates are mostly dominated by MS galaxies, while the SB subsample is likely washed out in all median stacks. Especially at low M$_{\star}$ and high-redshift, this approach yields an incomplete SB sample due to galaxies being IR fainter. In order to mitigate possible selection biases, we only focus on SB galaxies with M$_{\star}>$10$^{10.5}$~M$_{\odot}$ and z$\lesssim$2.5. This interval is set to ensure that all SB galaxies (i.e. lying $>$4$\times$ above the MS) lie above the limiting L$_{IR}$ of \textit{Herschel} PACS+SPIRE data in COSMOS \citep{Bethermin+15}, and thus are IR detected. We further remove radio-excess AGN (pre-identified in Sect.~\ref{agn_removal}) from the SB subsample of radio detections, in order to consider only bona-fide SFGs and fairly compare the AGN-corrected q$_{IR}$ between the SB and MS populations. This leaves us with a sample of 554 SBs. As done for the full SFG sample, we performed median stacking at 3~GHz and combined the stacked signal with radio-detected SBs. 

\begin{figure}
\centering
     \includegraphics[width=\linewidth]{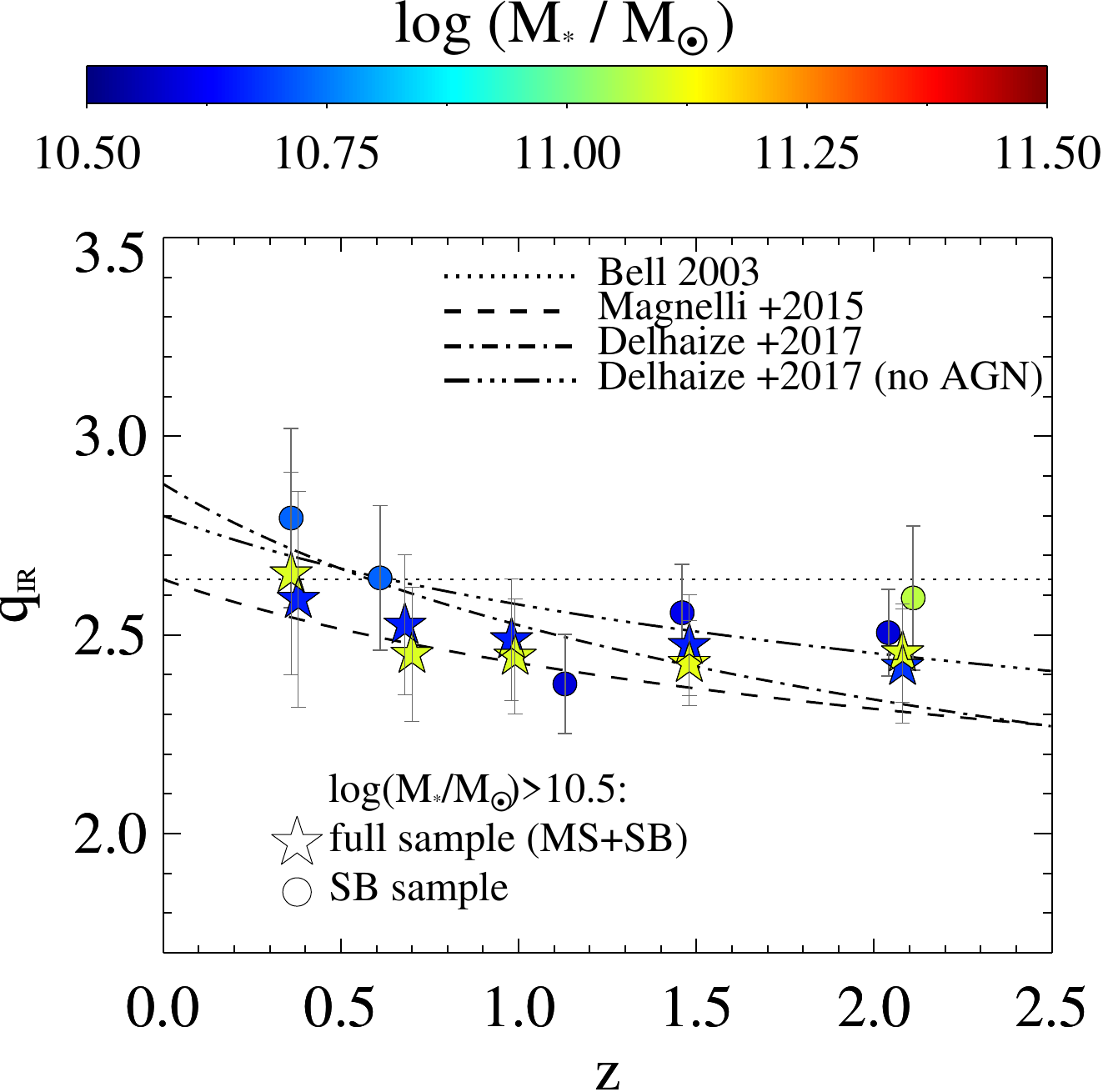}
 \caption{\small Comparison of q$_{IR}$ between our full SFG sample (MS+SB, stars) and the SB subsample (circles), as a function of redshift. To mitigate the incompleteness of an IR-based selection of SBs, we only show bins with M$_{\star}>$10$^{10.5}$~M$_{\odot}$. Black lines highlight best-fit IRRC trends from the literature for comparison. This test suggests that q$_{IR}$ evolves irrespective of whether a galaxy is on or above the MS.}
   \label{fig:q_sb}
\end{figure}

Fig.~\ref{fig:q_sb} shows the resulting q$_{IR}$ of SBs (circles) relative to the full SFG sample (MS+SB, stars) out to z$\lesssim$2.5, at M$_{\star}>$10$^{10.5}$~M$_{\odot}$. For comparison, some previous IRRC trends are reported (black lines), as in Fig.~\ref{fig:qplot1}. While some possible hints of ($\sim$0.05~dex) higher q$_{IR}$ in SBs could be present, these are consistent with MS analogues within 1$\sigma$ in all bins. Therefore, this test suggests that q$_{IR}$ evolves primarily with M$_{\star}$, irrespective of whether a galaxy is on or above the MS.

Though the (lack of) evolution of q$_{IR}$ above the MS is still debated, our decreasing q$_{IR}$--$\Sigma_{SFR}$ trend (Fig.~\ref{fig:qphys}) would predict lower q$_{IR}$ in SB than in MS galaxies, due to SBs being more compact. However, we note that our IR-detected SBs are \textit{both} $>$4$\times$ more star forming \textit{and} smaller in size (R$_e\lesssim$1~kpc at z$<$2, see \citealt{JimenezAndrade+19}) than MS analogues. Therefore, both parameters add to boost $\Sigma_{SFR}$. Since our SBs appear mostly unresolved in 3~GHz stacks, we can only place lower $\Sigma_{SFR}$ limits which prevent us from ruling out a possible flattening of q$_{IR}$ at the largest SFR surface densities. Higher resolution radio observations of these objects would be crucial to test such a behaviour.

On a side note, the sample of sub-millimetre galaxies for which \citet{Algera+20b} obtained an average q$_{IR}$=2.20 includes SFGs within a factor of three from the MS relation, thus not formally SBs. It might be possible that SB galaxies follow a different regime of q$_{IR}$, while our results predominantly reflect the behaviour of the MS population. 

As mentioned in Sect.~\ref{q_sigma_sfr}, \citet{Lacki+10b} distinguish between ``compact SBs'' ($h$=100~pc, ULIRG-like) similar to local merging galaxies, and ``puffy SBs'' ($h$=1~kpc, SMG-like) with lower q$_{IR}$ values, that are more common at high-z \citep{Genzel+08}. Nevertheless, a compact/puffy-SB transition above the MS should be reflected to a steepening of their radio spectra indices \citep{Lacki+10b}, though we are unable to discern it from our data. In this respect, \citet{Magnelli+15} did not report any significant spectral index variation above the MS. Therefore, we caution that a simple dependence of q$_{IR}$ on the SF compactness might not be suitable for unveiling the physics behind the IRRC in SBs, which might be also connected to the geometry of the SF regions or multiple mechanisms at play.

\subsection{Is there widespread AGN activity in radio-detected dwarves?}  \label{dwarves}

A noteworthy implication raised from our radio AGN subtraction is the possibly widespread AGN activity within radio-detected dwarf galaxies (M$_{\star}<$10$^{9.5}$~M$_{\odot}$). As highlighted in Sect.~\ref{agn_extrapolation} and Table~\ref{tab:fagn}, about 90\% of radio-detected dwarves are classified as radio AGN. This fraction drops down to only $\sim$0.5\% relative to the full M$_{\star}$ sample of dwarves. Such huge difference suggests that radio-detected dwarves are a quite peculiar and not representative subsample of these low-M$_{\star}$ galaxies. 

From an IR perspective, nearly all radio-detected dwarves ($>$99\%) are completely undetected (S/N$_{IR}<$3) at any IR/sub-mm band (Fig.~\ref{fig:bins}). This is likely a natural effect due to the increasing incompleteness of IR selection towards low M$_{\star}$ galaxies. From IR/sub-mm stacking, however, we obtain SFR$_{IR}>$4$\times$ higher than the MS relation, placing these sources in the SB region (e.g. \citealt{Rodighiero+11}; \citealt{Sargent+12}). This might apparently support a SF-driven origin of radio emission in dwarves.  

Nevertheless, on the radio side, these sources display on average lower L$_{1.4~GHz}$ values than more massive counterparts, but still 100$\times$ larger than those obtained from median radio stacking of non-detections. This effect fully counter-balances the high starburstiness seen in the IR, causing an overall drop of q$_{IR}$ in radio-detected dwarves by over a factor of 10, with respect to the stacked population (see black dots relative to yellow squares in Figs~\ref{fig:qplot_binned0} and \ref{fig:qplot_binned1}). These arguments let us suppose that radio-detected dwarves are consistent with being AGN-dominated in the radio.

While there is broad consensus on the prevalence of radio AGN within massive galaxies (e.g. \citealt{Heckman+14}), in which AGN-driven feedback could hamper star formation, little is known about its incidence and impact in dwarves. These systems are thought to host the pristine relics of the first black hole seeds, whose growth has been long believed to be disfavoured by SNa-driven feedback (e.g. \citealt{Reines+13}; \citealt{Dubois+15}; \citealt{Mezcua+16}; \citealt{Marleau+17}). However, there is mounting evidence that AGN feedback may also play a role at the low-mass end of the galaxy population.
 
From a theoretical perspective, cosmological simulations find that starbursting dwarf galaxies triggered by major mergers can be very frequent (\citealt{Fakhouri+10}; \citealt{Deason+14}). These events can induce widespread AGN feedback at low-M$_{\star}$ regimes, that could help solve the so-called ``too-big-to-fail'' problem, whereby simulated dwarves outnumber by several factors their observed counterparts (\citealt{GarrisonKimmel+13}; \citealt{Kaviraj+17}). This excess number cannot be suppressed via SNa feedback alone, but through additional AGN feedback (\citealt{Keller+16}; \citealt{Silk17}; \citealt{Koudmani+20}).

To search for observational AGN signatures in dwarf galaxies, spatially-resolved emission line diagnostics \citep{Mezcua+20}, deep X-ray and high angular resolution radio observations have been used (e.g. \citealt{Reines+11}, \citealt{Reines+12}; \citealt{Reines+14}; \citealt{Mezcua+19}). In the local Universe, these campaigns led to the confirmation of on-going AGN activity in starbursting dwarf galaxies \citep{Reines+12}. At higher redshifts, \citet{Mezcua+19} performed a statistical study of radio-detected dwarf galaxies at z$<$3.4 using deep VLA-COSMOS~3~GHz data \citep{Smolcic+17}. They isolated a sample of 35 bona-fide dwarf galaxies, which displayed radio jets powers and efficiencies as high as those of more massive galaxies. These studies argue that AGN feedback may be more common than previously thought, and potentially impactful for regulating galaxy star formation \citep{Kaviraj+19}. Our findings that most radio-detected dwarves stand above the MS and display excess radio emission are therefore not surprising, and in broad agreement with the above literature.

To investigate the possible AGN nature of our radio-detected dwarves, we stacked them using deep \textit{Chandra} images (\citealt{Civano+16}; \citealt{Marchesi+16}) at different redshifts, finding no X-ray detection in any of them. However, the 3$\sigma$ upper limits are about 5$\times$ higher than the level of X-ray emission predicted by star formation \citep{Lehmer+16}, which does not rule out their AGN nature.

\subsection{Is radio emission a good SFR tracer in all galaxies?}  \label{qsfr}

In this Section we discuss the link between the IRRC and SFR in galaxies. As mentioned in Sect.~\ref{sed_fitting}, the conversion from L$_{IR}$ to SFR is quite accurate in massive galaxies, while towards less massive and less obscured systems the UV may contribute as much as the IR to the global SFR. The observed correlation between L$_{IR}$ and L$_{1.4~GHz}$ is therefore not rigidly proportional to SFR.  

For this reason, we express q$_{IR}$ through a slightly different formalism that accounts for the addition of dust-uncorrected UV emission, in order to study the connection between radio emission and total SFR (=SFR$_{IR+UV}$). We thus define the parameter $q_{SFR}$ as:
\begin{equation}
q_{SFR} = \log \Bigg(\frac{L_{SFR}~[W]}{3.75\times10^{12}~Hz} \Bigg) - \log (L_{1.4~GHz}~[W Hz^{-1}])
\label{eq:qsfr}
\end{equation}
where L$_{SFR}$ is simply the SFR$_{IR+UV}$ scaled back to spectral luminosity units. This formalism enables us to keep similar units as for q$_{IR}$, while switching from luminosity to total SFR.

We repeated the analogous q$_{SFR}$ decomposition analysis at M$_{\star}>$10$^{10.5}$~M$_{\odot}$ to calibrate the AGN-vs-SF locus of radio detections (Sect.~\ref{agn_removal}). Within the two highest M$_{\star}$ bins, the best-fitting trend of q$_{SFR}$ with redshift has slope -0.057$\pm$0.002, which is strikingly similar to that inferred for q$_{IR}$ (-0.055$\pm$0.018, Sect.~\ref{ragn_highmass_recal}). Then we extrapolated such trend at lower M$_{\star}$ bins (Sect.~\ref{agn_extrapolation}) and recursively removed radio AGN to derive the AGN-corrected IRRC. 

Using the same approach as for Eq.~\ref{eq:bestq}, the multi-parametric fitting in the q$_{SFR}$--M$_{\star}$--z plane yields the following expression:
\begin{equation}
 q_{SFR} (M_{\star},z) = (2.743\pm0.034) \times A^{(-0.025 \pm  0.012)} - B\times(0.234\pm0.017)
   \label{eq:bestqsfr}
\end{equation}
where A=(1+z) and B=$(\log M_{\star}/M_{\odot}-10)$. Similarly to the fit in the q$_{IR}$ space, the redshift dependence is weaker and less significant than the M$_{\star}$ dependence, which is unsurprisingly steeper than before. This suggests that radio emission drops considerably more than SFR in low-M$_{\star}$ non-AGN galaxies. Reversing the argument, at fixed L$_{IR}$, radio emission underestimates the total SFR by a larger factor as compared to the IR light. The sub-linear trend L$_{IR}\propto$L$_{1.4~GHz}^{0.90}$ that we inferred in our analysis (see also \citealt{Bell03}; \citealt{Hodge+08}; \citealt{Davies+17}; \citealt{Brown+17}; \citealt{Gurkan+18}) becomes even steeper when adding the UV contribution to L$_{IR}$, i.e. SFR$_{IR+UV}\propto$L$_{1.4~GHz}^{0.81}$. Such a radio deficit in the dwarf-galaxy regime could be possibly linked to shorter CRe scale heights (e.g. \citealt{Helou+93}; \citealt{Lacki+10b}) or weaker magnetic fields (\citealt{Donevski+15}; \citealt{Tabatabaei+17}) that are common in less dense SF enviroments. 

Moreover, our adopted L$_{IR}$--SFR conversion does not account for the ``cirrus'' emission associated with cold dust heated by old ($>$A-type) stellar populations, which might lower the intrinsic SFR at fixed L$_{IR}$. However, this effect is expected to contribute in low-sSFR galaxies, i.e. at high M$_{\star}$ and low redshift (e.g. \citealt{Yun+01}). Hence, we expect it (if any) to further flatten the q$_{SFR}$--z trend or to amplify the M$_{\star}$ dependence of q$_{SFR}$.

In addition, we note that the lower efficiency in producing synchrotron emission in low-SFR, low-M$_{\star}$ galaxies is already factored in recent synchrotron emission models of SFGs (e.g. \citealt{Massardi+10}; \citealt{Mancuso+15}; \citealt{Bonaldi+19}) based on empirical matching between local L$_{1.4~GHz}$ and SFR functions. Therefore, our results reinforce the need for M$_{\star}$-dependent, non-linear calibrations between radio-continuum emission and SFR, in order to develop successful observing strategies for targeting low-M$_{\star}$ galaxies at radio wavelengths. These can be complemented with higher frequency observations that are more sensitive to thermal free-free emission as SFR tracer in high-redshift galaxies (e.g. \citealt{Murphy+17}, \citealt{Penney+20}, \citealt{VanderVlugt+20}; \citealt{Algera+20c}).

These considerations are relevant in the context of the forthcoming SKA. In particular, the SKA mid-frequency receivers will be equipped with five bands, of which the SKA Band2 (0.95--1.76 GHz) will be the workhorse for radio-continuum based SFR measurements. Even the faintest and least massive galaxies in our sample will be routinely observed by SKA, probing diverse populations of SFGs (and composite AGN+SF objects). Our findings highlight that a detailed understanding of the physics behind the relation between radio synchrotron emission and SFR is fundamental for fully exploiting the unique SKA capabilities in terms of depth and angular resolution.

\section{Summary and conclusions} \label{summary}

In this manuscript we calibrate the IRRC of SFGs as a function of \textit{both} M$_{\star}$ and redshift, out to z$\sim$4. Starting from an M$_{\star}$-selected sample of 413,678 galaxies SFGs selected via (NUV-r)/(r-J) colours in the COSMOS field, we leverage new de-blended IR/sub-mm data \citep{Jin+18}, as well as deep radio images from the VLA COSMOS 3~GHz Large Project \citep{Smolcic+17}. 

In each M$_{\star}$--z bin, we performed stacking of undetected sources at both IR (Sect.~\ref{ir_stacking}) and radio (Sect.~\ref{radio_stacking}) frequencies, and combined the stacked signal with individual detections a-posteriori to infer the average q$_{IR}$ as a function of M$_{\star}$ and redshift (Sect.~\ref{q0}). We develop a recursive approach for identifying and then subtracting radio-excess AGN in different M$_{\star}$ and redshift bins (Sect.~\ref{agn_removal}). This technique is calibrated on a ($>$70\%) M$_{\star}$-complete subsample of 3~GHz detections at M$_{\star}>$10$^{10.5}$~M$_{\odot}$ and extrapolated to the rest of the sample to infer the AGN-corrected IRRC (Sect.~\ref{q_mass_z}). Finally, we interpret our findings in the context of existing IRRC studies, from both models and observations. The main results of this work are listed below.

1)~The IRRC evolves primarily with M$_{\star}$, with more massive galaxies displaying systematically lower q$_{IR}$. A secondary, weaker dependence on redshift is also observed. The multi-parametric best-fitting expression is the following: q$_{IR}$(M$_{\star}$,z) = (2.646$\pm$0.024) $\times$ (1+z)$^{(-0.023 \pm  0.008)}$ - (0.148$\pm$0.013) $\times$ ($\log M_{\star}$/M$_{\odot}$ - 10). At fixed redshift, this trend translates into an IRRC of L$_{IR}\propto$L$_{1.4~GHz}^{0.90}$, which corroborates the similar sub-linear behaviour reported in the literature (e.g. \citealt{Bell03}; \citealt{Hodge+08}; \citealt{Gurkan+18}). The typical scatter of the IRRC at M$_{\star}>$10$^{10.5}$~M$_{\odot}$ is around 0.21--0.22~dex (a factor of 1.7), consistent with other studies (\citealt{Yun+01}; \citealt{Bell03}; \citealt{Molnar+20}) and roughly constant with M$_{\star}$ and z.

2) Our recursive approach for removing radio AGN enables us to statistically decompose radio-detected SFGs and AGN (Figs.~\ref{fig:histo0} and \ref{fig:histo0_corr}) as a function of M$_{\star}$ and redshift. Removing radio AGN substantially flattens the observed q$_{IR}$--z trend at M$_{\star}>$10$^{10.5}$~M$_{\odot}$ to a nearly flat slope. This correction nicely aligns the mode q$_{IR}$ of radio SFGs to the median stacked q$_{IR}$ of the full M$_{\star}$ sample of non-AGN galaxies. Therefore, we interpret the resulting AGN-corrected q$_{IR}$ measurements as robust against further AGN removal. We acknowledge that residual radio AGN activity within radio-detected SFGs (10--20\%) could be possible. Nevertheless, we expect this effect, if any, to further flatten out the evolution of q$_{IR}$ with redshift, and to induce an even steeper M$_{\star}$ dependence, thus reinforcing our main findings.

3)~The fraction of radio AGN identified within the full M$_{\star}$ sample strongly increases with M$_{\star}$, spanning from 0.4\% to 6\% across the full range (Table~\ref{tab:fagn}), in agreement with previous studies (e.g. \citealt{Heckman+14}). However, when limited to 3-GHz detected sources, about 90\% of radio-detected dwarves (M$_{\star}<$10$^{9.5}$M$_{\odot}$) are radio-excess AGN. We argue this is likely a selection effect induced by our 3~GHz-limited being biased towards the brightest radio sources in such low-M$_{\star}$ systems. We test the reliability of our radio AGN identification owing to available VLBA data of radio AGN \citep{HerreraRuiz+17}, confirming the AGN nature for 90\% of them.

4)~We examined the evolution of q$_{IR}$ as a function of SFR surface density ($\Sigma_{SFR}$), as a proxy for M$_{\star}$, finding a very similar trend both in slope and statistical significance. In agreement with recent observations of high-redshift dusty SFGs \citep{Algera+20b}, our results support a decreasing q$_{IR}$ in MS galaxies towards higher $\Sigma_{SFR}$. Nevertheless, radio synchrotron models (e.g. \citealt{Lacki+10b}; \citealt{Schleicher+13}) predict a much stronger q$_{IR}$ evolution with redshift, and $\Sigma_{SFR}$- (i.e. M$_{\star}$-) dependent radio spectral indices, neither of which are seen in our data. Another possibility links to magnetic field amplification in massive highly SFGs \citep{Tabatabaei+17}.

5) We compare the average q$_{IR}$ between MS galaxies and an M$_{\star}$-complete subsample of SBs with SFRs $>$4$\times$ above the MS (Sect.~\ref{q_sb}). Despite SBs being more compact than MS analogues \citep{JimenezAndrade+19}, we do not observe a significant difference in q$_{IR}$, apparently at odds with our expectations. According to radio synchrotron models \citep{Lacki+10b}, a ``conspiracy'' of different factors might induce a q$_{IR}$ flattening at $\Sigma_{SFR}\gtrsim$1~M$_{\odot}~yr^{-1}~kpc^{-2}$. Our findings do not seem to support this prediction. However, our current data do not allow us to discriminate between various model scenarios in this $\Sigma_{SFR}$ regime. Alternatively, we postulate that SB galaxies might follow a different q$_{IR}$ relation with $\Sigma_{SFR}$ than MS analogues, in which multiple mechanisms could play a role.

6)~We verified that adding the UV dust-uncorrected contribution to the IR, as a proxy for the total SFR, would further steepen the q$_{SFR}$--M$_{\star}$ trend, leaving the evolution with redshift unchanged. These findings imply that using radio-synchrotron emission as a SFR tracer requires M$_{\star}$-dependent conversion factors. Finally, our results can be useful to make accurate calibrations for future radio-continuum surveys as SFR machines down to dwarf galaxy regimes, especially in the upcoming SKA era.


\begin{acknowledgements}
The authors are grateful to the referee for a detailed and constructive report that greatly helped us clarify the results and implications of this work. ID is supported by the European Union's Horizon 2020 research and innovation program under the Marie Sk\l{}odowska-Curie grant agreement No 788679. ID thanks R.~Gobat for useful discussions. MJJ acknowledges support from the UK Science and Technology Facilities Council [ST/N000919/1], the Oxford Hintze Centre for Astrophysical Surveys which is funded through generous support from the Hintze Family Charitable Foundation and a visiting Professorship from SARAO. SJ acknowledges financial support from the Spanish Ministry of Science, Innovation and Universities (MICIU) under grant AYA2017-84061-P, co-financed by FEDER (European Regional Development Funds). DL acknowledges funding from the European Research Council (ERC) under the European Union's Horizon 2020 research and innovation programme (grant agreement No. 694343). IHW acknowledges support from the Oxford Hintze Centre for Astrophysical Surveys which is funded through generous support from the Hintze Family Charitable Foundation. RC acknowledges financial support from CONICYT Doctorado Nacional N$^\circ$\,21161487 and the Max-Planck Society through a Partner Group grant with MPA. JD acknowledges the financial assistance of SARAO. MN acknowledges support from the ERC Advanced Grant 740246 (Cosmic Gas). IP acknowledges financial support from the Italian Ministry of Foreign Affairs and International Cooperation (MAECI Grant Number ZA18GR02) and the South African Department of Science and Technology's National Research Foundation (DST-NRF Grant Number 113121) as part of the ISARP RADIOSKY2020 Joint Research Scheme. SMR hereby acknowledged the financial assistance of the National Research Foundation (NRF) towards this research. JS acknowledges the funding from the Swiss National Science Foundation under Grant No. 185863. Opinions expressed and conclusions arrived at, are those of the author and are not necessarily to be attributed to the NRF. The MeerKAT telescope is operated by the South African Radio Astronomy Observatory, which is a facility of the National Research Foundation, an agency of the Department of Science and Innovation. We acknowledge use of the Inter-University Institute for Data Intensive Astronomy (IDIA) data intensive research cloud for data processing. IDIA is a South African university partnership involving the University of Cape Town, the University of Pretoria and the University of the Western Cape. The authors acknowledge the Centre for High Performance Computing (CHPC), South Africa, for providing computational resources to this research project. 

\end{acknowledgements}

 \bibliographystyle{aa} 
 \bibliography{references_v2} 

\begin{appendix}

\section{Testing total radio flux densities} \label{Appendix_radio}

\begin{figure}
     \includegraphics[width=\linewidth]{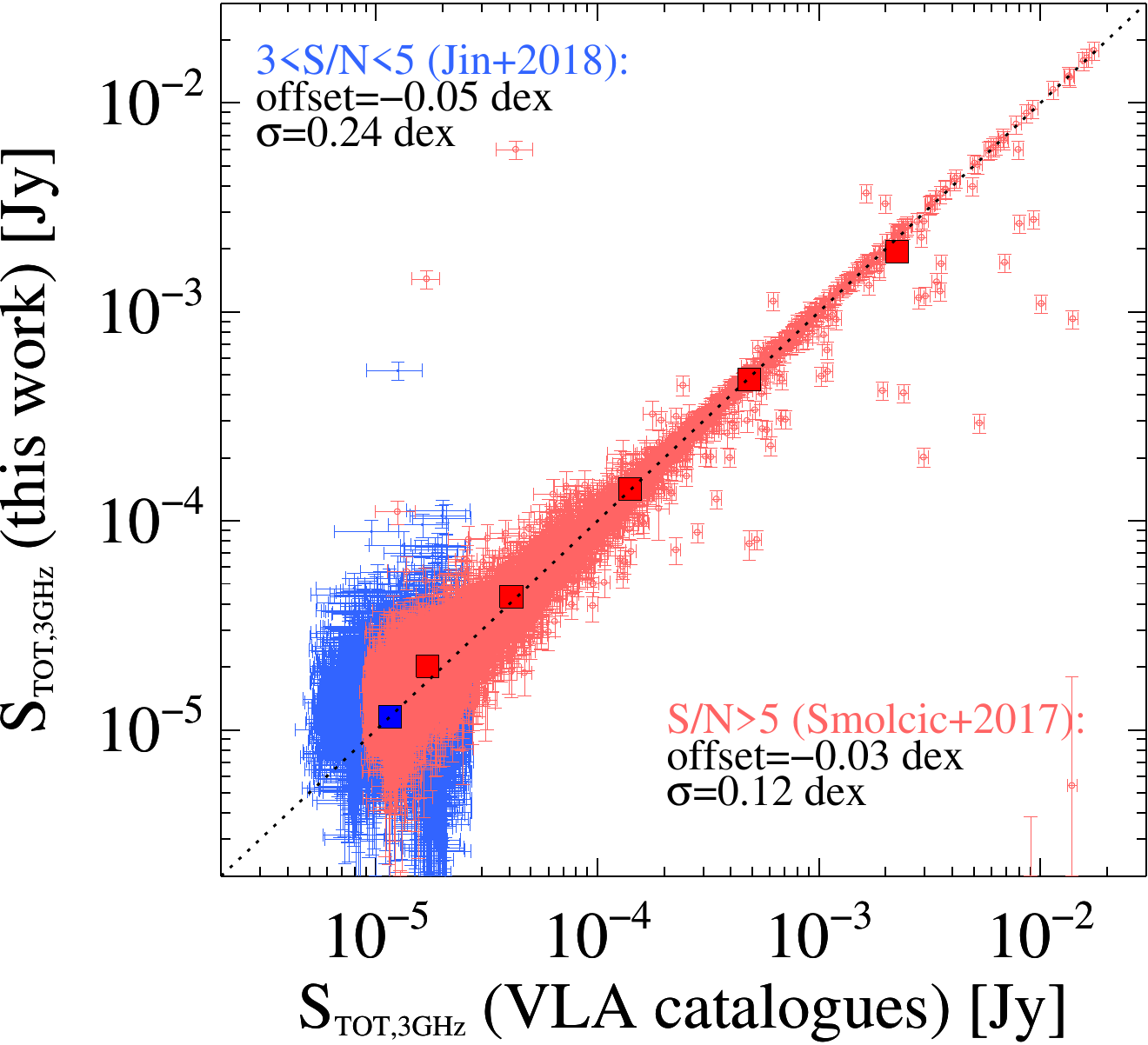}
 \caption{\small Comparison of total flux densities of 3~GHz detections between our procedure and catalogue flux densities, both at S/N$>$5 (\citeauthor{Smolcic+17}, red dots) and at 3$<$S/N$<$5 (\citealt{Jin+18}; blue dots). Squares highlight the median ratio at various intervals. The global offset and dispersion suggest a good agreement within the uncertainties down to S/N$\sim$3.
 }
   \label{fig:radio_smolcic}
\end{figure}

We validate our total flux estimation against individual detections taken from published VLA catalogues at 3~GHz. At S/N$>$5 we used the catalogue of \citet{Smolcic+17}, while total flux densities at 3$<$S/N$<$5 were taken from \citet{Jin+18}. After excluding the 67/10830 multi-component sources identified in \citet{Smolcic+17}, we calculate peak and total flux densities of each source, following the approach described in Sect.~\ref{radio_stacking}. Fig.~\ref{fig:radio_smolcic} displays the comparison between total flux densities (dots), highlighting the corresponding median ratio at various intervals (squares). It is worth noting that \citeauthor{Smolcic+17} (\citeyear{Smolcic+17}, red) used the software {\sc blobcat} (\citealt{Hales+12}; \citeyear{Hales+14}) to sum over all blobs identified in the 3~GHz image above a certain S/N cut. Therefore, it is best suited for non-Gaussian shapes. On the other hand, our approach described in Sect.~\ref{radio_stacking} assumes a 2D elliptical Gaussian, with size, angle and normalization being free to vary. Despite the different techniques, we find a good agreement at S/N$>$5, with a logarithmic offset of -0.028~dex and dispersion of 0.12~dex. At 3$<$S/N$<$5, total flux densities from \citeauthor{Jin+18} (\citeyear{Jin+18}, blue) were computed via Gaussian PSF fitting, using a circular beamsize of 0.75''. Despite the low S/N regime, we also observe a fair agreement, with an offset of -0.05~dex and dispersion of 0.24~dex. This check proves our total flux densities fully consistent with the published values of \citet{Smolcic+17} and \citet{Jin+18} for individual 3~GHz detections down to S/N$\sim$3.

\begin{figure}
     \includegraphics[width=\linewidth]{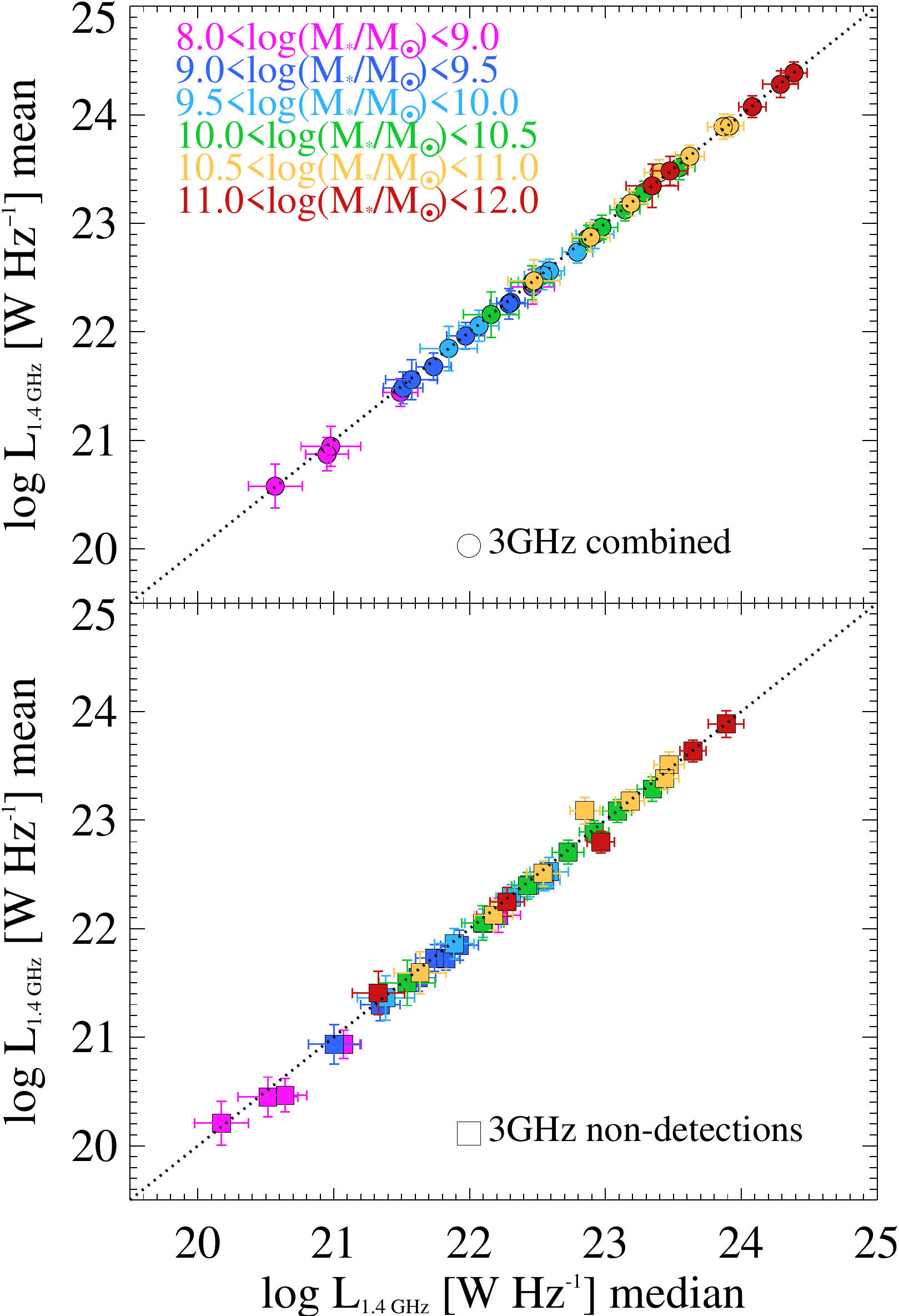}
 \caption{\small Top panel: comparison between median L$_{1.4~GHz}$ (x-axis) and rms-weighted mean L$_{1.4~GHz}$ (y-axis) for combined 3~GHz detections and non-detections, following Eq.~\ref{eq:weighted}. Colours indicate various M$_{\star}$ bins. Bottom panel: same comparison, but referred to 3~GHz undetected sources only.
 }
   \label{fig:radio_median}
\end{figure}

We further demonstrate that our choice of performing median stacking at 3~GHz, rather than rms-weighted mean stacking, does not impact our final L$_{1.4~GHz}$ estimates. A comparison between median and mean L$_{1.4~GHz}$ is presented in Fig.~\ref{fig:radio_median}. The top panel displays L$_{1.4~GHz}$ from the combined flux of detections and non-detections (see Eq.~\ref{eq:weighted}), while the bottom panel refers to the case of purely undetected sources. Colours indicate different M$_{\star}$ bins. Only stacks in which peak flux densities have S/N$>$3 are shown. No systematics is observed, at any M$_{\star}$, between mean and median stacked L$_{1.4~GHz}$. This is consistent with \citet{White+07}, who showed that, in the noise-dominated regime, the stacked median traces the population mean. Moreover, such excellent agreement confirms that the uniform 3~GHz sensitivity across the full map ensures that either stacking method can reliably recover the average flux of the underlying galaxy population. 

The fact that non-detections (bottom panel) display consistent L$_{1.4~GHz}$ between mean and median stacking suggests that, if any, radio AGN do not dominate the total radio emission in our stacks. 
The same argument cannot be implicitly extended to the combined flux densities, since these mean weighted-average flux densities could be biased towards fewer and brighter radio detections, which reduces the statistical weight of non-detections. Indeed, L$_{1.4~GHz}$ of radio detections (top panel) are always $>$3$\times$ larger than L$_{1.4~GHz}$ of non-detections (bottom panel), despite the smaller numbers. This partly smooths over the initial fluctuations between mean and median stacking, thus delivering an even tighter agreement, as we observe.

\section{Stacking ancillary VLA and MIGHTEE data} \label{Appendix_ancillary}

Here we perform a radio stacking analysis, as for 3~GHz data (Sect.~\ref{radio_stacking}), in order to check whether our 3~GHz based L$_{1.4~GHz}$ are stable against different resolutions or spectral frequencies. We exploit VLA data from the 1.4 GHz Deep Project \citep{Schinnerer+10} map. It covers 1.7 deg$^2$ with an angular resolution of 2.5'', reaching rms=12~$\mu$Jy~beam$^{-1}$ in the central 50''$\times$50''. A total of 2,864 sources were blindly extracted down to S/N$>$5. In addition, we make use of 1.3~GHz MIGHTEE (\citealt{Jarvis+16}; I.~Heywood et al. in prep.) data. MIGHTEE images formally reach 2.2$~\mu$Jy~beam$^{-1}$ at 8.4''$\times$6.8'' resolution over 1 deg$^2$ in the MIGHTEE early science data, but the effective depth is limited by confusion ($\sim$5.5$~\mu$Jy~beam$^{-1}$ in the central part).

Source flux densities in VLA 1.4~GHz and MIGHTEE 1.3~GHz maps were re-extracted, using $K_s$+MIPS~24 positional priors. While the angular resolution at VLA 1.4 GHz is high enough to yield a negligible fraction of overlapping priors within the beam, MIGHTEE data suffer from blending issues. To this end, MIGHTEE flux densities were de-blended as in \citet{Jin+18} down to 3$\sigma$ level. Then, individual S/N$>$3 detections were removed from the original image, and we used the residual map for stacking 1.3~GHz non-detections. Of course, only sources within the MIGHTEE area (central 1~deg$^2$) were stacked, containing roughly half of the sample size used for VLA stacking.

The stacking analysis follows the same reasoning and assumptions presented in Sect.~\ref{radio_stacking}. Stacked MIGHTEE flux densities are measured in the central pixel, that is assumed to trace the total flux. VLA~1.4~GHz peak flux densities were, instead, scaled to total flux densities as done at 3~GHz. Nonetheless, a different, yet empirical relation was adopted to identify resolved sources at VLA 1.4~GHz, calibrated on 1.4~GHz detections \citep{Schinnerer+10}: ${S_{tot}}$/${S_{peak}} >$ 0.35$^{-11 / ({S/N}_{peak}^{1.45})}$. Because of the larger beamsize compared to 3~GHz, we find fewer resolved stacks (17/25). Total flux densities were converted to rest-frame L$_{1.4~GHz}$ assuming $\alpha$=--0.75$\pm$0.1, that was propagated along with flux errors to deliver reliable L$_{1.4~GHz}$ uncertainties. Upper limits at 3$\sigma$ were assigned for S/N$<$3 stacks.

\begin{figure}
\centering
     \includegraphics[width=\linewidth]{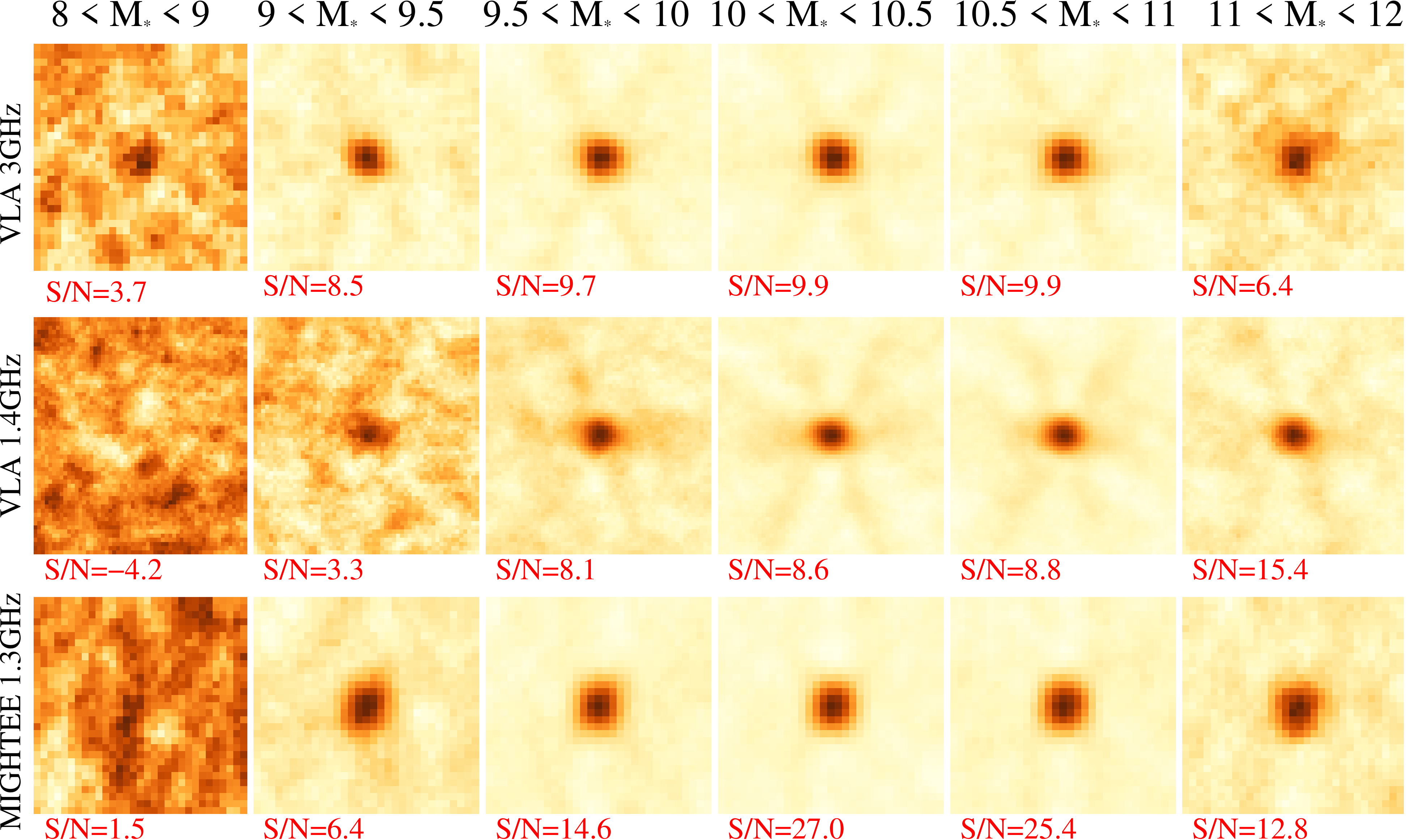}
 \caption{\small Stacked cutouts of our sample at 0.8$<$z$<$1.2, as a function of M$_{\star}$ (left to right, expressed in $\log$~M$_{\odot}$). Only individual undetected sources (S/N$<$3) are stacked. The top, middle and bottom rows show VLA 3~GHz, VLA 1.4~GHz and MIGHTEE 1.3~GHz data, respectively. Each cutout size corresponds to 8$\times$FWHM of the beam. Below each cutout we report the corresponding S/N of the total stacked flux.
 }
   \label{fig:radio_stacks}
\end{figure}

Fig.~\ref{fig:radio_stacks} shows stacked cutouts at 0.8$<$z$<$1.2 at VLA~3~GHz (top), VLA~1.4~GHz (middle) and MIGHTEE~1.3~GHz (bottom) data, as a function of M$_{\star}$ (increasing from left to right). While stacking at 3~GHz delivers S/N$>$3 in 39/42 bins, only 25/42 and 29/42 have S/N$>$3 in VLA 1.4~GHz and MIGHTEE 1.3~GHz stacked images, respectively. For VLA 1.4~GHz, the small number of bins is attributed to shallower than 3~GHz observations. For MIGHTEE, instead, this is probably induced by the confusion-limited signal in the stacks due to the larger MeerKAT primary beam at 1.3~GHz (e.g. \citealt{Mauch+20}). Nevertheless, because VLA 3~GHz data are much less sensitive than MeerKAT to large-scale radio emission, total radio flux densities might be underestimated at 3~GHz. This issue can be, however, especially relevant at low redshift (z$<$0.5) and for resolved/multi-component radio sources (e.g. \citealt{Delhaize+21}). In fact, a visual inspection of the median 3~GHz stacks of non-detections does not reveal clearly missing flux in the residual images at the scales of the MIGHTEE beam, except in the bin at z$<$0.5 and 10$^{11}<$M$_{\star}$/M$_{\odot}<$10$^{12}$. To quantify this effect, we convolved all the original 3~GHz stacked cutouts with a Gaussian kernel of 3''~FWHM, re-calculating the total flux densities and comparing them with the previous measurements. The reason why this specific beamwidth was chosen is that beyond 3'' it has already been shown that no significant missing flux is recorded at 3 GHz (see Table 2 in \citealt{Delhaize+17}). Of course, this convolution drastically reduces the global S/N of the final stacks, leaving us with S/N$>$3 in only 16/42 bins (as opposed to 39/42 before). However, only the bin at the lowest z and highest M$_{\star}$ displays on average 0.3~dex larger total flux density, while the other bins show consistent estimates within the uncertainties. Since no extra flux is visible in the new residual image, we replaced the total flux density of that single bin with the 3'' convolved value and used this value in the rest of our analysis. In any case, we stress that the final L$_{1.4~GHz}$ obtained by combining both detections and non-detections is unchanged, since the fraction of radio detections is about 56\% at z$<$0.5 and M$_{\star}>$10$^{11}$~M$_{\odot}$ (see Fig.~\ref{fig:mass_z}), thus washing out the difference in the stacked flux. As a consequence, this effect has no impact on the rest of our analysis. In addition, we emphasize that any extra missing 3~GHz flux at low redshift would further strengthen our final redshift-invariant IRRC (Sect.~\ref{q_mass_z}). This motivates our choice of using primarily VLA~3~GHz images for our analysis.

\begin{figure}
\centering
     \includegraphics[width=\linewidth]{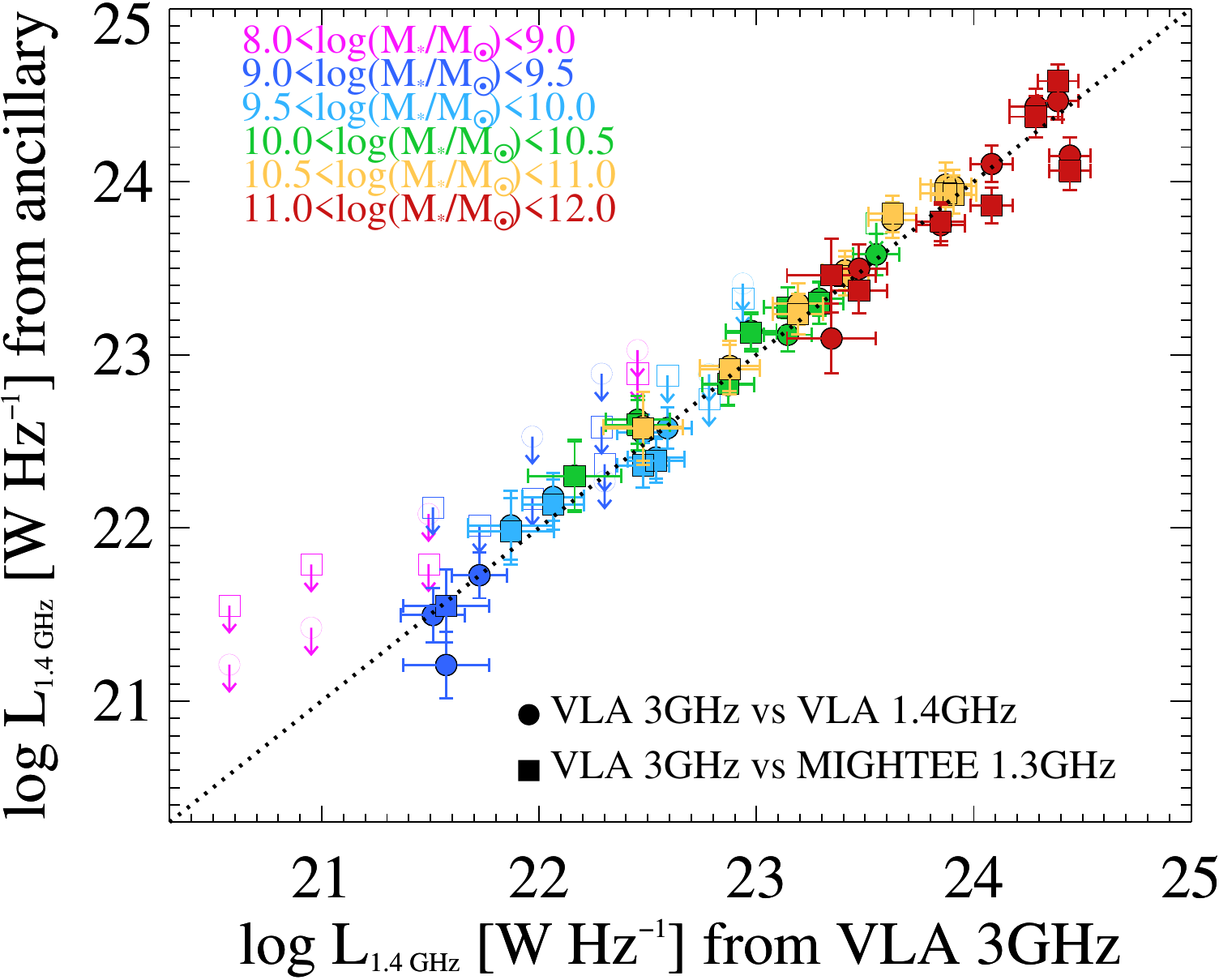}
 \caption{\small Comparison between rest-frame 1.4~GHz spectral luminosity L$_{1.4~GHz}$ obtained from 3~GHz stacks (x-axis) and ancillary radio stacks (y-axis) using VLA (1.4~GHz, circles) and MIGHTEE (1.3~GHz, squares) data. We assumed a single spectral index $\alpha$=--0.75 to scale flux densities from 3~GHz to 1.4~GHz. Colours indicate different M$_{\star}$ ranges. Downward arrows with open symbols mark 3$\sigma$ upper limits if S/N$<$3. The broad agreement between the various datasets suggests that using a single $\alpha$=--0.75 is a reasonable assumption across the full M$_{\star}$ range explored in this work.
 }
   \label{fig:radio_lcomp}
\end{figure}

\subsection{Considerations on the radio spectral index} \label{radio_slope}

We briefly discuss and test our assumption of taking a single spectral index $\alpha$=--0.75 by comparing L$_{1.4~GHz}$ estimates independently inferred from stacking the three above datasets. In Fig.~\ref{fig:radio_lcomp}, we compare L$_{1.4~GHz}$ obtained from 3~GHz stacks (x-axis) and ancillary radio stacks (y-axis), using either VLA (1.4~GHz, circles) or MIGHTEE (1.3~GHz, squares) data, colour-coded by M$_{\star}$. Downward arrows with open symbols mark 3$\sigma$ upper limits where the stacked S/N$<$3. We find a good agreement between all the various datasets, suggesting that using a single $\alpha$=--0.75 is a reasonable assumption across the full M$_{\star}$ range explored in this work. As a sanity check, the median spectral index traced by VLA 3~GHz and MIGHTEE 1.3~GHz individual detections is --0.77, in agreement with our assumption. However, we prefer to adopt a fixed $\alpha$=--0.75 in our calculation to treat both radio detections and non-detections in a self-consistent manner.

\citet{Magnelli+15} measured the average spectral index exploiting VLA 1.4~GHz and GMRT 610~MHz data for an M$_{\star}$-selected galaxy sample. They found that the observed 610~MHz--1.4~GHz slope, that probes closer to rest-frame 1.4~GHz than our 3~GHz data, does not seem to change with M$_{\star}$ or SFR, at least out to z$\sim$2. More recently, \citet{CalistroRivera+17} exploited Low Frequency Array (LOFAR) data at 150~MHz in the Bo\"{o}tes field, out to z$\sim$2.5. Interestingly, they observed a spectral flattening of the radio SED of SFGs in the observed range [150~MHz--1.4~GHz] (see also \citealt{Read+18}; \citealt{Gurkan+18}). However, they argue that this feature should not affect the k-correction for the rest-frame 1.4~GHz luminosities L$_{1.4~GHz}$. Therefore, these studies provide mounting evidence that using a single power-law spectral index $\alpha$=--0.75 at our frequency is a reasonable assumption.

\section{Impact of a different radio AGN-vs-SFG fitting approach} \label{Appendix_AGN}

We discuss a potential caveat related to our AGN-vs-SF decomposition presented in Sect.~\ref{ragn_highmass}. Specifically, our procedure relies on the assumption that the mode of the observed q$_{IR}$ distribution (q$_{IR, peak}$) of radio detections is entirely attributed to SF. Though this is supported by a number of previous studies arguing that radio AGN are a sub-dominant population in the sub-mJy regime (e.g. \citealt{Padovani+15}; \citealt{Smolcic+17}; \citealt{Novak+18}; \citealt{Ceraj+18}; \citealt{Algera+20a}), the contribution of radio-faint AGN to q$_{IR, peak}$ might not be negligible. If this is the case, by mirroring and fitting the SF Gaussian first, it is possible that we are underestimating the true fraction of radio AGN relative to SFGs. To quantify this potential issue and test how much it would affect our final M$_{\star}$-dependence of q$_{IR}$, here we follow a different approach.

\begin{figure}
\centering
     \includegraphics[width=\linewidth]{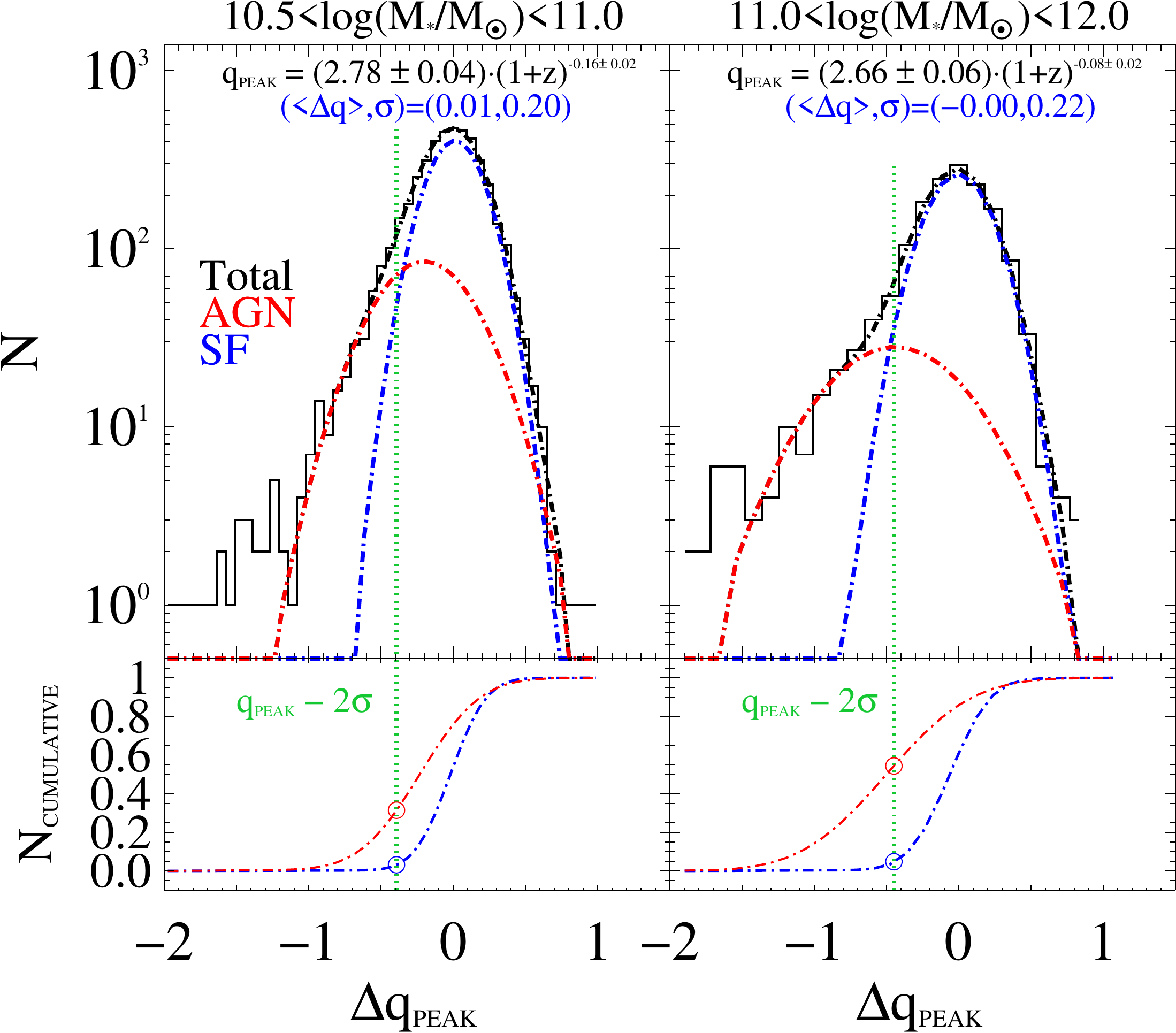}
 \caption{\small Same as Fig.~\ref{fig:histo0}, but fitting the total q$_{IR}$ distribution of 3~GHz detections (black histogram) simultaneously with a SF Gaussian (blue) and an AGN Gaussian (red dashed). The 1$\sigma$ dispersion has remained unchanged to about 0.21~dex. The bottom panels display the corresponding cumulative Gaussian fits, both normalized to unity. The vertical green dotted line marks the 2$\sigma$ threshold of 0.42~dex below which we consider a source as radio-excess AGN. As opposed to Fig.~\ref{fig:histo0}, this cutoff removes only 30-50\% of the total radio AGN population. However, we estimate this effect to be more prevalent in lower M$_{\star}$ galaxies. This implies that accounting for such mis-classified radio AGN would likely strengthen our final M$_{\star}$-dependent q$_{IR}$.
 }
   \label{fig:histo0_simult}
\end{figure}
%
The observed q$_{IR}$ distribution is fitted with two Gaussian functions \textit{simultaneously}, which parametrize the contribution of SFGs and radio-excess AGN. Contrary to Sect.~\ref{ragn_highmass}, we do not set the SF peak to q$_{IR, peak}$, but we leave it free to vary along with the dispersion and normalization for both functions. In this simultaneous fitting we give equal input weights to all bins, regardless of the number of sources in each. This approach is thus expected to return a rather conservative AGN contribution relative to SFGs. 

The results are shown in Fig.~\ref{fig:histo0_simult} for the two highest M$_{\star}$ bins. As in Fig.~\ref{fig:histo0}, the best-fit SF (blue) and AGN (red) Gaussians head up to reproduce the total distribution (black). However, we clearly notice two main differences compared to the previous approach. Firstly, the AGN distribution is far broader than the SF distribution in both M$_{\star}$ bins. Secondly, the relative fraction of radio AGN that we mis-classify as SFGs (red tail at $>$q$_{peak}$--2$\sigma$) is as high as 40--70\%, hence much higher than the 30\% obtained in Sect.~\ref{ragn_highmass} when fitting and mirroring the SF part first. This is clearly displayed by the cumulative AGN fraction in the bottom panels. Instead, the relative fractions of ``pure'' SFGs above the 2$\sigma$ threshold are about 80\% at 10$^{10.5}<$M$_{\star}$/M$_{\odot}<$10$^{11}$ and 90\% at at 10$^{11}<$M$_{\star}$/M$_{\odot}<$10$^{12}$.

Despite the lower level of purity of the SFG population, we emphasize that the main results of this paper are quite robust against the AGN-vs-SF fitting procedure. Indeed, both peak and dispersion ($\sim$0.21~dex) of the SF population are essentially unchanged, the peak being identical to q$_{IR, peak}$ and the dispersion reaching $\sim$0.21~dex. Therefore, it is reasonable to \textit{assume} that the mode of the observed q$_{IR}$ distribution is attributed to radio-detected SFGs. Related to this, the threshold q$_{peak}$--2$\sigma$ is still equal to 0.42 dex, implying that roughly the same exact sources as in Sect.~\ref{ragn_highmass} would be identified as radio-excess AGN. This agreement demonstrates that our recursive radio AGN removal would lead to the same final IRRC, regardless of the assumed shape of the AGN distribution.

If we were able to statistically remove the underlying radio AGN contribution within the SF population (though impossible with the present data), this would systematically increase q$_{IR}$ by a larger amount towards \textit{lower M$_{\star}$ galaxies}. Indeed, at 10$^{11}<$M$_{\star}$/M$_{\odot}<$10$^{12}$ the radio AGN distribution is clearly broader, but far more offset than at 10$^{10.5}<$M$_{\star}$/M$_{\odot}<$10$^{11}$, thus at higher M$_{\star}$ the two populations are more distinguishable. As a consequence, we argue that a proper correction for such an effect would further exacerbate the M$_{\star}$ stratification of q$_{IR}$ reported in this work.

\section{Differences compared to the literature} \label{Appendix_comparison}

Our best-fit relation of q$_{IR}$ as a function of M$_{\star}$ and redshift (Eq.~\ref{eq:bestq} in Sect.~\ref{q_mass_z}) is fully consistent with the average q$_{IR}$ value measured in local SFGs (i.e. 2.64 in \citealt{Bell03}) for a typical galaxy with M$_{\star}$$\sim$10$^{10}$~M$_{\odot}$. At higher redshifts, instead, our average q$_{IR}$ measurements follow flatter evolutionary trends compared to previous studies (Fig.~\ref{fig:qplot1}), while the best-fit normalization appears broadly consistent with the literature only at M$_{\star}>$10$^{10.5}$~M$_{\odot}$. In order to interpret these differences in a quantitative fashion, we identify three key points that combined differentiate our approach from that adopted in the previous literature: (i) removing radio AGN via a recursive approach in each M$_{\star}$ and redshift bin; (ii) exploiting an M$_{\star}$-selected sample of SFGs; (iii) binning the derived q$_{IR}$ as a function of both M$_{\star}$ and redshift. To test our results against different techniques, we expand on each of these aspects below.

\subsection{Radio AGN subtraction} \label{test_magnelli}

In Sect.~\ref{agn_removal}, we performed a recursive subtraction of radio AGN as a function of M$_{\star}$ and redshift, carefully calibrated on high-M$_{\star}$ galaxies, and then extrapolated to lower M$_{\star}$ analogues. However, other studies followed alternative approaches to discard radio AGN when deriving the intrinsic IRRC. For instance, \citet{Magnelli+15} performed median stacking of both radio detections and non-detections out to z$\sim$2. This method strongly reduces the contribution of a few bright outliers, assuming that the bulk radio population is made of SFGs. This assumption is quite reasonable, since \citet{Magnelli+15} started from an M$_{\star}$-selected sample, of which radio detections make a negligible fraction.

\begin{figure}
\centering
     \includegraphics[width=\linewidth]{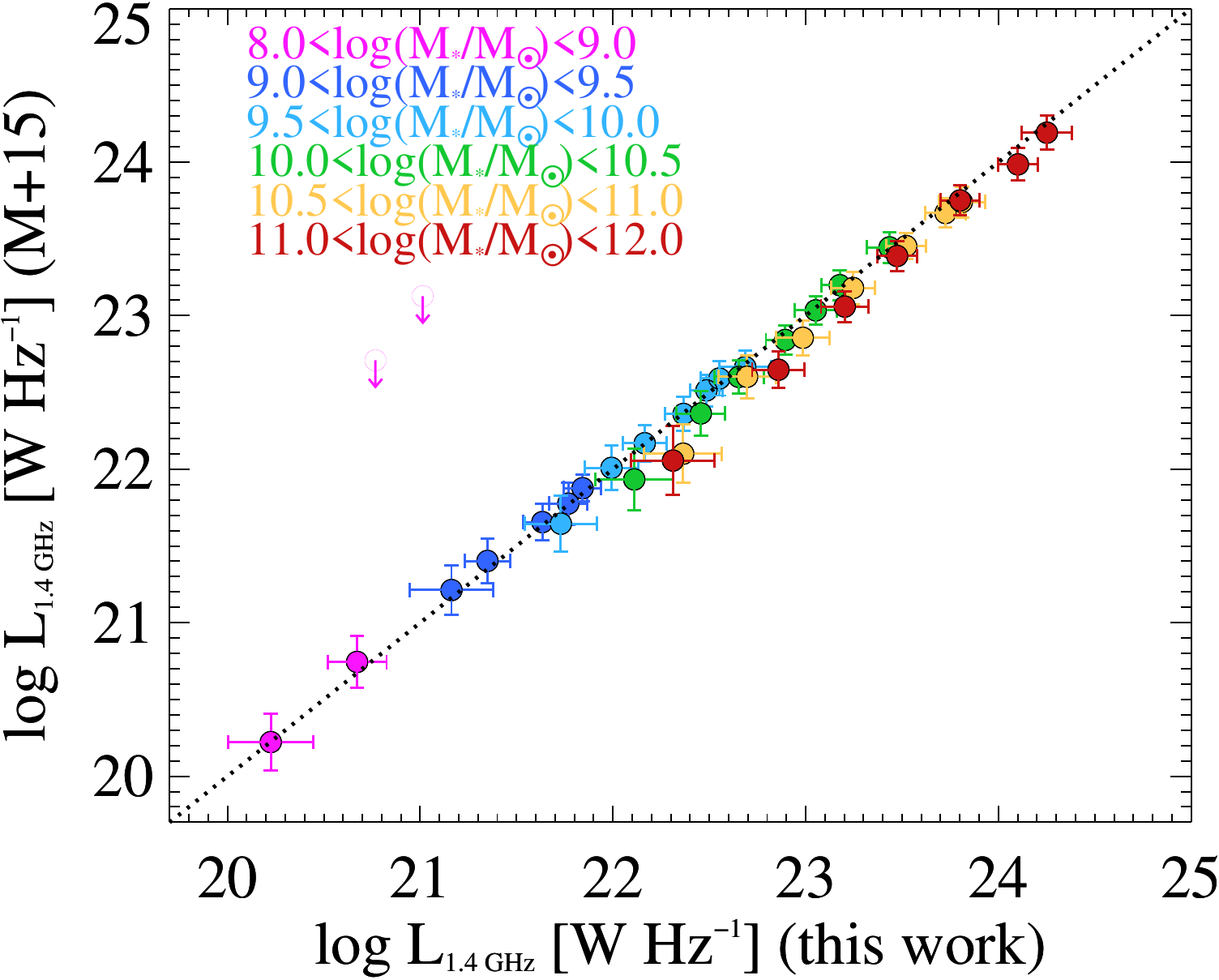}
 \caption{\small Comparison between AGN-corrected L$_{1.4~GHz}$ from this work (x-axis) and median L$_{1.4~GHz}$ obtained from stacking detections and non-detections together (\citealt{Magnelli+15}, y-axis). Different M$_{\star}$ ranges are colour-coded, while downward arrows mark 3$\sigma$ upper limits for 2/37 bins. Despite these different approaches, we notice a very good agreement in all bins, that strengthens the reliability of our recursive AGN subtraction.
 }
   \label{fig:testmagnelli}
\end{figure}

We compare our q$_{IR}$ with mock measurements obtained by following the stacking method of \citet{Magnelli+15}, but applied to the sample used in our work. Fig.~\ref{fig:testmagnelli} displays the final L$_{1.4~GHz}$ estimates that we obtained after removing radio AGN (x-axis) against those derived from median radio stacking (\citealt{Magnelli+15}, y-axis). We note that our L$_{IR}$ estimates and \citeauthor{Magnelli+15}'s were instead calculated through a fully consistent approach, therefore only a difference in L$_{1.4~GHz}$ might lead to systematics in the final q$_{IR}$ trends. The colour bar highlights the average M$_{\star}$ of each bin. Out of 37 bins analyzed in this work, 35 yield a S/N$>$3 from median 3~GHz stacking (circles), while 3$\sigma$ upper limits are shown for the remaining bins (downward arrows). This comparison clearly reveals a very good agreement between final 1.4~GHz luminosities, with all measurements being consistent within the uncertainties. Despite the different approaches, the agreement extends down to dwarf galaxies, supporting the AGN nature of most radio-detected sources (Sect.~\ref{agn_extrapolation}). A possible (though not significant) deviation of $\sim$0.1~dex might be present at the highest M$_{\star}$, with our measurements returning slightly higher L$_{1.4~GHz}$ measurements than those of \citeauthor{Magnelli+15}. This might be ascribed to the contribution of radio-detected SFGs to our weighted average L$_{1.4~GHz}$, since they make a substantial fraction of the M$_{\star}$-selected sample in that M$_{\star}$ bin ($\sim$45\%, Table~\ref{tab:fagn}). Therefore, this test proves our radio AGN subtraction broadly consistent with a totally independent approach.

%

\begin{figure}
\centering
     \includegraphics[width=\linewidth]{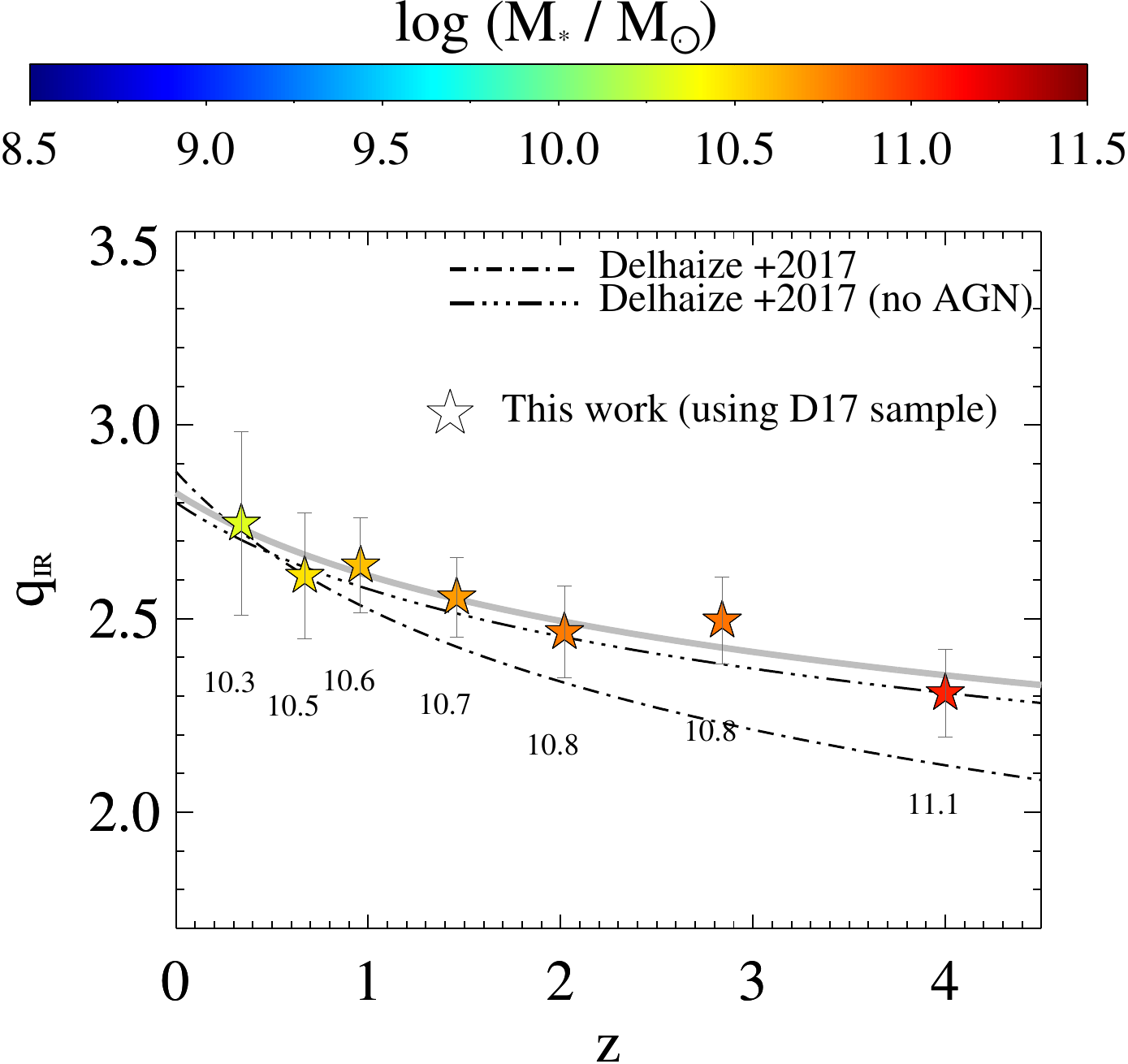}
 \caption{\small Median q$_{IR}$ as a function of redshift obtained by analysing the SFG sample of \citeauthor{Delhaize+17} (\citeyear{Delhaize+17}, stars). Black lines indicate the median q$_{IR}$--z trend of \citeauthor{Delhaize+17} before (dot-dashed) and after (triple dot-dashed) removing 2$\sigma$ outliers. The grey solid line marks the resulting best-fit q$_{IR}$ trend with redshift, that is highly consistent with that of \citet{Delhaize+17} after removing radio AGN. Numbers below each star denote the median M$_{\star}$ of the underlying sample.
 }
   \label{fig:testdelhaize}
\end{figure}
%

\subsection{Sample selection and binning}

An additional aspect worth testing is whether different sample selections lead to distinct IRRC trends. We started from an M$_{\star}$-selected sample of SFGs based on K$_s$-band priors, that typically reaches much deeper than any infrared or radio survey, compared to an average galaxy SED. A rare exception is represented by very high-redshift (z$>$4) or heavily dust-obscured systems, which are visible only in IRAC (e.g. \citealt{Davidzon+17}) or deep ALMA imaging (e.g. \citealt{Franco+20}). For this reason, studies that derived the IRRC based on exclusive or joint samples of radio/IR detections, are partly biased against low-M$_{\star}$ galaxies. For instance, the work of \citet{Delhaize+17} was based on a jointly-selected infrared (from \textit{Herschel}, with S/N$\geq$5 in at least one PACS or SPIRE band) and radio (VLA~3~GHz with S/N$\geq$5; \citealt{Smolcic+17}) sample of SFGs in the COSMOS field, out to z$\sim$5. By performing double-censored survival analysis to account for sources undetected at either radio or FIR wavelengths, they found an evolving $q_{IR}\propto$(1+z)$^{-0.19\pm0.01}$, which flattens to $q_{IR} \propto$(1+z)$^{-0.12\pm0.01}$ after removing 2$\sigma$ outliers (as reported in \citealt{Delvecchio+18}), particularly radio-excess AGN. We repeat our IR and radio stacking analysis using the same sample of SFGs from \citet{Delhaize+17} (9,575 sources), to demonstrate that our analysis leads to consistent results when matching the input sample.

We split the sample of \citet{Delhaize+17} among the same seven redshift bins analyzed in this work. For each, we perform median stacking of 3~GHz and IR images in all bands, combining both detections and non-detections. This approach should be comparable to the search for the median value carried out via survival analysis \citep{Delhaize+17}. Although we do not formally remove radio AGN in this check, we showed in Sect.~\ref{test_magnelli} that median radio stacking yields broadly consistent results (see \citealt{Magnelli+15}). Fig.~\ref{fig:testdelhaize} displays the median q$_{IR}$ obtained by stacking the SFG sample of \citet{Delhaize+17} in different redshift bins (stars). This yields a best-fitting q$_{IR}\propto$(1+z)$^{-0.11\pm0.05}$, that is fully consistent with the flatter trend of \citet{Delhaize+17} after removing 2$\sigma$ outliers (triple dot-dashed line). This check proves our technique solid against different sample selections from the literature. 

Fig.~\ref{fig:testdelhaize} also highlights the important role played by the binning grid in driving a declining IRRC with redshift. In particular, the colour-coded M$_{\star}$ clearly indicates how a joint IR and radio selection is sensitive to increasing galaxy M$_{\star}$ with redshift. Moreover, the scatter of the IRRC reported by \citet{Delhaize+17} is around 0.35~dex, while the dispersion that we measured at M$_{\star}>$10$^{10.5}$~M$_{\odot}$ (Sect.~\ref{ragn_highmass}) is only 0.21--0.22~dex. This is similar to the value reported by \citet{Bell03} (i.e. 0.26~dex) for nearby galaxies, recently narrowed down to 0.16~dex in \citet{Molnar+20}. A possible reason for the smaller than 0.35~dex dispersion in our study might be that we are splitting SFGs among different M$_{\star}$, each carrying an intrinsically smaller dispersion compared to the full SFG sample. Because of the decreasing q$_{IR}$ with M$_{\star}$, binning only as a function of redshift leads to a mixture of different galaxy M$_{\star}$ that results into a larger global dispersion.

 \end{appendix}

 \end{document}